\documentclass[iop,apj]{emulateapj}
\usepackage{amsmath,amssymb,amstext,float}

\usepackage[citecolor=black,linkcolor=black]{hyperref}
\usepackage{color,upgreek}

\usepackage[all]{hypcap} %Links go to figures; breaks on deluxetables (use \capstartfalse \capstarttrue to fix it)
\usepackage{aas_macros}
\usepackage{mathtools}
\usepackage{xspace}
\usepackage{lineno}
\usepackage[nabla,differential]{boldtensors}
\usepackage{perpage} %the perpage package
\MakePerPage{footnote} %the perpage package command
\shorttitle{Toroidal Flux Rope Emergence}
\shortauthors{Knizhnik et al.}
\newcommand{\beg}[1]{\begin{equation}\label{#1}}
\newcommand{\done}{\end{equation}}
\newcommand{\pd}[2]{\frac{\partial #1}{\partial #2}}

\newcommand{\vecB}{\boldsymbol{B}}
\newcommand{\vecJ}{\textbf{J}}

\newcommand{\veck}{\textbf{k}}

\newcommand{\vecv}{\boldsymbol{v}}

\newcommand{\unit}[1]{\hat{\textbf{#1}}}

\newcommand{\curl}[1]{\nabla\times{#1}}
\newcommand{\divv}[1]{\nabla\cdot{#1}}

\numberwithin{equation}{section}
\begin{document}

\title{The Rise and Emergence of Untwisted Toroidal Flux Ropes on the Sun}
%\author[0000-0002-2544-2927]{K. J. Knizhnik}
%\affiliation{Naval Research Laboratory, 4555 Overlook Avenue SW, Washington, DC 20375, USA}
%\author{J. E. Leake}
%\affiliation{Heliophysics Science Division, NASA Goddard Space Flight Center, 8800 Greenbelt Rd, Greenbelt MD 20771, USA}
%\author{M. G. Linton}
%\affiliation{Naval Research Laboratory, 4555 Overlook Avenue SW, Washington, DC 20375, USA}
%\author{S. Dacie}
%\affiliation{Max Planck Institute for Meteorology, Bundesstra{\ss}e 53, 20146 Hamburg, Germany}
\author{K. J. Knizhnik\altaffilmark{1}, J. E. Leake\altaffilmark{2}, M. G. Linton\altaffilmark{1}, and S. Dacie\altaffilmark{3}}
\altaffiltext{1}{Naval Research Laboratory, 4555 Overlook Ave SW, Washington, DC 20375}
\altaffiltext{2}{Heliophysics Science Division, NASA Goddard Space Flight Center, 8800 Greenbelt Rd, Greenbelt MD 20771, USA}
\altaffiltext{3}{Max Planck Institute for Meteorology, Bundesstra{\ss}e 53, 20146 Hamburg, Germany}

%\linenumbers
\begin{abstract}
Magnetic flux ropes (MFRs) rising buoyantly through the Sun's convection zone are thought to be subject to viscous forces preventing them from rising coherently. Numerous studies have suggested that MFRs require a minimum twist in order to remain coherent during their rise. Furthermore, even MFRs that get to the photosphere may be unable to successfully emerge into the corona unless they are at least moderately twisted, since the magnetic pressure gradient needs to overcome the weight of the photospheric plasma. To date, however, no lower limit has been placed on the critical minimum twist required for an MFR to rise coherently through the convection zone or emerge through the photosphere. In this paper, we simulate an untwisted toroidal MFR which is able to rise from the convection zone and emerge through the photosphere as an active region that \textcolor{black}{resembles} those observed on the Sun. We show that untwisted MFRs can remain coherent during their rise and then pile-up near the photosphere, triggering the undular instability, allowing the MFR to emerge through the photosphere. We propose that the toroidal geometry of our MFR is critical for its coherent rise. Upon emerging, a pair of lobes rises into the corona which interact and reconnect, resulting in a localized high speed jet. The resulting photospheric magnetogram displays the characteristic salt-and-pepper structure often seen in observations. Our major result is that MFRs need not be twisted to rise coherently through the convection zone and emerge through the photosphere. 
\end{abstract}

\maketitle

\section{Introduction}\label{sec:intro}
An important unsolved problem in solar physics is the origin of active regions on the photosphere and in the overlying corona. While it is generally agreed that photospheric flux distributions are due to the emergence of sub-photospheric magnetic fields (\citealt{Parker55}; see also review by \citealt{Cheung14}) that are generated by a dynamo process \citep{Parker55,Moffatt78,Nelson14}, the exact manner in which the magnetic fields make their way from deep in the solar interior to the solar surface is a matter of debate. \par
First, magnetic flux ropes need to be able to rise through the convection zone. One method that has been studied extensively in the literature is that of a buoyant rise of magnetic flux ropes \citep{Parker55}. In these models, a flux rope in the convection zone that is in pressure balance with the outside ambient plasma is either in gravitational or thermal equilibrium with its surroundings. Since the magnetic pressure of the flux rope lowers its internal plasma pressure, if the flux rope is in thermal equilibrium, the density inside the flux rope will be lower than the density outside the flux rope, while the temperature inside and outside will be the same. Conversely, if the flux rope is in gravitational equilibrium, the temperature inside the flux rope will be lower than temperature outside the flux rope, while the density inside and outside will be the same. In this case, and in the absence of convection, the flux rope can remain stable in the convection zone until thermal conduction brings it into thermal equilibrium, causing the density inside the flux rope to decrease (assuming it maintains pressure balance), making the flux rope buoyant and causing it to rise.  Thus, the fate of a pressure balanced flux rope is to be buoyantly unstable as a result of the density deficit caused by the added magnetic pressure. In this picture, therefore, the rise of magnetic flux ropes through the convection zone is a buoyantly driven process. This model has been used in numerous studies of flux emergence \citep{Dsilva93,Fan93,MI94,Schussler94,Galsgaard07,Archontis13,Knizhnik18b}. \par 

%A second method relies on the action of upflows driven by the convectively unstable convection zone. These flows push magnetic flux toward the photosphere \citep{Abbett07}, distributing and dispersing the field, creating localized concentrations of $\mathrm{kG}$ fields, such as are seen in active regions \citep{Lites13}. This model has also been used in numerous studies of flux emergence \citep{Magara05,Abbett07,Cheung07,Isobe08}, and flux emergence buffeted by granular scale flows have been observed \citep{DePontieu02,Centeno07} and simulations with magnetoconvection have been found to best reproduce many properties of observed active regions \citep{Dacie17}. \par

One issue with this model is that it has been thought that buoyantly rising flux ropes require a critical twist in order to maintain their integrity against viscosity-generated vortical flows that fragment the flux ropes during their rise \citep{Parker79,Schuessler79,Longcope96,MI96,Emonet98,Fan98a,Wissink00,Toriumi11}.  While \citet{Rempel14} showed that convection supports the integrity of even an untwisted flux rope against these forces so that it can rise coherently through the upper portion of the convection zone, in the absence of convection the twist that is required to withstand viscous fragmentation is unclear.
%In addition, the low pressure wake created by the rising flux rope generates a downward directed pressure gradient to counteract the flux rope's buoyancy, eventually decelerating it to its terminal velocity \citep{Longcope96}. 
Furthermore, if the flux tube breaks up into fragments, for flux tube vortices with circulation $\Gamma$ rising with velocity $v$, the Kutta-Zhukhovsky theorem \citep{Landau87,Batchelor00} predicts a force per unit length
\beg{KJ}
f = -\rho v\Gamma,
\done 
which acts perpendicular to the velocity to separate the fragments. The additional horizontal component of the velocity creates its own Kutta-Zhukhovsky force, this time directed downward, eventually bringing the rise to a complete stop. Thus, it seems critical that, for a coherent rise, a flux rope must maintain its integrity against viscous fragmentation, and therefore that weakly twisted flux ropes should not be able to rise to the photosphere. \par
In two dimensions, \citet{Emonet98} found that the critical twist necessary for the flux tube to maintain its coherence during its rise depended on a magnetic Weber number 
\beg{Weber}
We_m = \frac{E_{kin,rise}}{E_{mag,twist}} = \frac{\rho v^2/2}{B_{tw}^2/8\pi},
\done 
where $\rho$ is the density, $v$ is the tube's rise velocity, and $B_{tw}$ is the strength of the twist component of the flux tube's magnetic field. They found that the tube remained coherent if $We_m<1$, meaning that a significant twist was needed for a coherent rise. However, three dimensional studies have shown that the degree of curvature of the rising $\Omega$-loop can significantly reduce the strength of the critical twist field necessary for coherence \citep{Abbett00}. To this end, \citet{MacTaggart09} performed simulations of the emergence of weakly twisted toroidal flux ropes - with $qR=0.2$, where $R$ is the radius and $q$ is the twist wavenumber - that start out in the shape of a torus. They found that such flux ropes were able to coherently rise and emerge through the photosphere into the corona.
%, and that the evolution of a rising toroidal flux rope was different than the evolution of the more commonly used cylindrical flux rope models \citep[e.g.,][]{Fan01a,Archontis04,Toriumi14,Takasao15,Knizhnik18b}, 
%and that the curvature of the rising structure allows more efficient draining - relative to the cylindrical flux rope emergence - of the plasma down the flux rope legs, enabling the magnetic field in the center of the emerged polarity to be more vertical. 
%Furthermore, unlike in cylindrical flux rope emergence, where the distance between opposite polarities continually increased as the flux rope emerged, for toroidal flux emergence the distance between the opposite polarities was instead governed by the size of the original torus. 
However, the twist used by \citet{MacTaggart09}, though small, leaves open the question of whether magnetic flux ropes with even smaller twists would be able to rise coherently. \par 
There is a further problem for flux ropes that, having survived the rise through the convection zone, make it to the photosphere. It has been shown that a significant twist is needed in order to emerge through the photosphere without the aid of convection, since the emergence process is thought to occur through a magnetic buoyancy instability \citep{Parker55,Newcomb61,Thomas75,Acheson79,Murray06,Toriumi10,Archontis13,Leake13}. In this picture, the photosphere forms an interface separating low- and high-density plasma that is modified by the pile-up of flux just below the photosphere. Perturbations which bend the field lines can easily grow, since plasma drains from the concave down field lines, unburdening them and allowing them to rise further \citep{Fan01a}. \par
\citet{Murray06} found that flux ropes needed to be at least moderately twisted to be able to exceed the threshold for this instability, and that weakly twisted flux ropes were unable to emerge through the photosphere. At low twists, at values of about $qR=0.25$, they found that the gradient of the magnetic field was too small near the photosphere to overcome the weight of the plasma. Similarly, \citet{Toriumi11} performed a simulation of the rise of weakly twisted cylindrical flux ropes from deep in the solar interior. The found that at twists below $qR=0.25$, the flux ropes failed to emerge through the photosphere, but that at higher twists, the magnetic buoyancy instability was triggered and flux ropes were able to emerge. \par 
Subsequently, \citet{Archontis13} ran a simulation of a weakly twisted flux rope with the same twist ($qR=0.25$) as that used by \citet{Murray06} and \citet{Toriumi11} and found that it did, in fact, emerge through the photosphere if given enough time, and attributed the emergence to the magnetic buoyancy instability. They found that the emergence occurred quite differently than a typical moderately twisted flux rope. Instead of forming a simple arcade in the corona, such as that formed in the emergence of the toroidal flux rope of \citet{MacTaggart09}, the weakly twisted flux rope formed two distinct arcades that they termed `magnetic lobes', separated by a neutral line. The magnetic buoyancy instability allowed portions of the flux rope to emerge into the corona while keeping other parts of the flux rope submerged. The concave down portions of the field were able to emerge, whereas the concave up portions containing heavy plasma remained stuck below the photosphere. Additional plasma draining down from the emerging loops onto the submerged loops exacerbated the situation. Further, they found that the two lobes interacted and reconnected, causing a high speed jet and a reconfiguration of the field. \par 
The contrasting results of \citet{Murray06}, \citet{Toriumi11} and \citet{Archontis13} for the same twist begs the question of whether there is, in fact, a critical twist that determines whether flux ropes will emerge or not, or whether there are additional considerations in whether emergence occurs. \citet{Toriumi11} and \citet{Archontis13} use different field strengths and depth-to-radius ratios: \citet{Toriumi11} use a field strength of $15\;\mathrm{kG}$ and a depth to radius ratio of $20$ whereas \citet{Archontis13} used a field strength of $2.8\;\mathrm{kG}$ and depth to radius ratio of $<5$. In addition, the flux rope of \citet{Toriumi11} was initially placed about $10$ times deeper than the flux rope of \citet{Archontis13}. As a result, a direct comparison is difficult to perform and there are indications that the flux ropes of \citet{Toriumi11} did, in fact, emerge, or would have had the simulation been run out for longer (see, for example, Figure 10 of \citealt{Toriumi11}). As a result, the question of just how weakly twisted a flux rope can be before emergence is suppressed remains open. \par
The picture that ``\emph{emerges}" therefore, is one in which a critical twist is posited to be necessary for magnetic flux ropes to rise coherently through the convection zone, and a (perhaps different) critical twist is thought to be required for magnetic flux ropes to emerge through the photosphere. While some numerical studies, described above, have constrained the value of these twists, no lower limit has yet been set. Furthermore, the critical twist cannot be constrained observationally, since there exists an apparent discrepancy between observed twists and theoretically estimated values. Several authors have measured the twist in active regions to be of order $0.01\;\mathrm{Mm^{-1}}$ \citep{Pevtsov94,Pevtsov95,Pevtsov97,Longcope98,Longcope99}. On the other hand, \citet{Toriumi11} struggled to emerge flux ropes with twists as `low' as $0.25/R$. For their flux rope of radius $0.43\;\mathrm{Mm}$, this converts to a twist of $0.59\;\mathrm{Mm^{-1}}$, which is more than an order of magnitude higher than observed. Thus, there is already a huge gap between the observed twists and flux ropes that were too weakly twisted to emerge. This discrepancy between observations and theory has been discussed previously in \citet{Knizhnik18b}. They argued that the `twist' as measured in the photosphere of their simulations has little physical correspondence to the twist as measured by the rotation of field lines per unit length in the subsurface flux rope. Since there is expected to be significant reconnection, diffusion, and possibly kinking of the flux rope as it rises, the twist at the photosphere would consequently be reduced. In addition, significant twist may remain trapped below the photosphere \citep{Fan09}. Thus, twists measured at the photosphere are unlikely to constrain the subsurface twist of rising flux ropes.\par 
In this study, we perform a simulation of a completely untwisted toroidal flux rope rising from inside the convection zone, and show that it is able to emerge, implying that in the abscence of twist, strong axial curvature is a necessary property for magnetic flux ropes to both rise coherently through the convection zone or emerge through the photosphere. This paper is organized as follows. In \S \ref{sec:model} we describe our numerical model. In \S \ref{sec:results} we show the results of the rise and emergence of an untwisted toroidal flux rope. In \S \ref{sec:Conclusions} we summarize our results and compare them to the previous studies.

\section{Numerical Model}\label{sec:model}
\subsection{MHD Equations}
Our numerical model uses the visco-resistive MHD Lagrangian-remap code Lare3D \citep{Arber01}. It solves the equations of MHD having the Lagrangian form:
\beg{mass}
\frac{D\rho}{Dt} = -\rho \divv{\vecv}
\done
\beg{momentum}
\frac{D\vecv}{Dt} = -\frac{1}{\rho}\Big(\mu_0^{-1}(\curl{\vecB})\times\vecB+\nabla P +\rho\textbf{g}+ \divv{S_{ij}}\Big)
\done 
\beg{energy}
\frac{De}{Dt} = \frac{1}{\rho}\Big(-P\divv{\vecv} + \Xi_{ij}S_{ij}+\frac{\eta}{\mu_0^2}(\curl{\vecB})^2\Big)
\done 
\beg{induction}
\frac{D\vecB}{Dt} = (\vecB\cdot\nabla)\vecv - \vecB(\divv{\vecv}) - \eta\nabla^2\vecB.
\done 
In these equations, $\rho$ is the mass density, $\vecv$ is the velocity, $\vecB$ is the magnetic field, and $e$ is the specific energy density. $\mu_0$ is the permeability of free space, which is used to relate the magnetic field to the current density $\mu_0\vecJ=\curl{\vecB}$. $\eta=16\;\mathrm{\Omega\;m}$ is the resistivity, $\textbf{g} = -274\;\mathrm{ms^{-2}}\unit{z}$ is the gravity at the photosphere, and \textbf{S} is the stress tensor with components given by
\beg{stresstensor}
S_{ij} = \nu \Big(\Xi_{ij}-\frac{1}{3}\delta_{ij}\divv{\vecv}\Big)
\done 
The viscosity $\nu = 3.35\times10^3\;\mathrm{kg}\;\mathrm{m^{-1}}\;\mathrm{s^{-1}}$, $\delta_{ij}$ is the Kronecker delta function, and $\Xi$ can be written in terms of the components of the Jacobian matrix $\nabla\vecv$ as 
\beg{upzeta}
%\upzeta_{ij} = \frac{1}{2}\Big(v^i_{\;,j} + v^j_{\;  ,i}\Big).
\Xi_{ij} = \frac{1}{2}\Big(\pd{v_i}{x_j} + \pd{v_j}{x_i}\Big).
\done 
The gas pressure is defined through the mass density, Boltzmann constant $k_B$, temperature $T$ and reduced particle mass $\mu_m$ as
\beg{idealgas}
p = \frac{k_B\rho T}{\mu_m},
\done 
where $\mu_m=1.25m_p$, with $m_p$ the proton mass. This choice of mass is discussed in \citet{Leake13} and \citet{Leake13b}. The pressure is related to the specific energy density via
\beg{eandP}
e = \frac{p}{\rho(\gamma-1)},
\done 
where $\gamma=5/3$ is the ratio of specific heats. Lare3D defines the plasma variables $e$ and $\rho$ on cell centers, magnetic field components at cell faces, and the velocity components at the cell vertices, and preserves $\divv\vecB$.
\subsection{Normalization}
The MHD equations are scaled by choosing three quantities to normalize the physical parameters. We choose a normalizing field strength 
\beg{magscl} 
B_0 = 0.13 \mathrm{T},
\done 
length scale 
\beg{lenscl}
L_0 = 1.7\times10^5\;\mathrm{m},
\done 
(which is the photospheric pressure scale height), and gravitational acceleration 
\beg{gravscl}
g_0=g_{sun}=274\;\mathrm{ms^{-2}}.
\done
These choices constrain the normalizing values for the gas pressure
\beg{prescl}
P_0=\frac{B_0^2}{\mu_0} = 1.3\times10^4 \;\mathrm{Pa},
\done 
mass density 
\beg{rhoscl}
\rho_0 = \frac{B_0^2}{\mu_0L_0g_0} = 3\times10^{-4}\; \mathrm{kg\;m^{-3}},
\done
velocity 
\beg{velscl}
v_0 = \sqrt{L_0g_0} = 6.8\times10^3\;\mathrm{m\; s^{-1}},
\done
time
\beg{timescl}
t_0 = \sqrt{\frac{L_0}{g_0}} = 24.9\; \mathrm{s},
\done 
temperature
\beg{temscl}
T_0 = \frac{m_pL_0g_0}{k_B} = 5.6\times10^3\;\mathrm{K},
\done 
current density
\beg{curscl}
J_0 = \frac{B_0}{L_0\mu_0} = 0.6 \;\mathrm{A\;m^{-2}},
\done 
viscosity
\beg{viscscl}
\nu_0 = P_0t_0 = 3.35\times10^5\;\mathrm{kg}\;\mathrm{m^{-1}}\;\mathrm{s^{-1}},
\done 
and resistivity
\beg{resistscl}
\eta_0 = \mu_0v_0L_0 = 1.6\times10^3\;\mathrm{\Omega}\;\mathrm{m}.
\done 
\subsection{Domain and Boundary Conditions}
\begin{figure*}
\includegraphics[width=\linewidth]{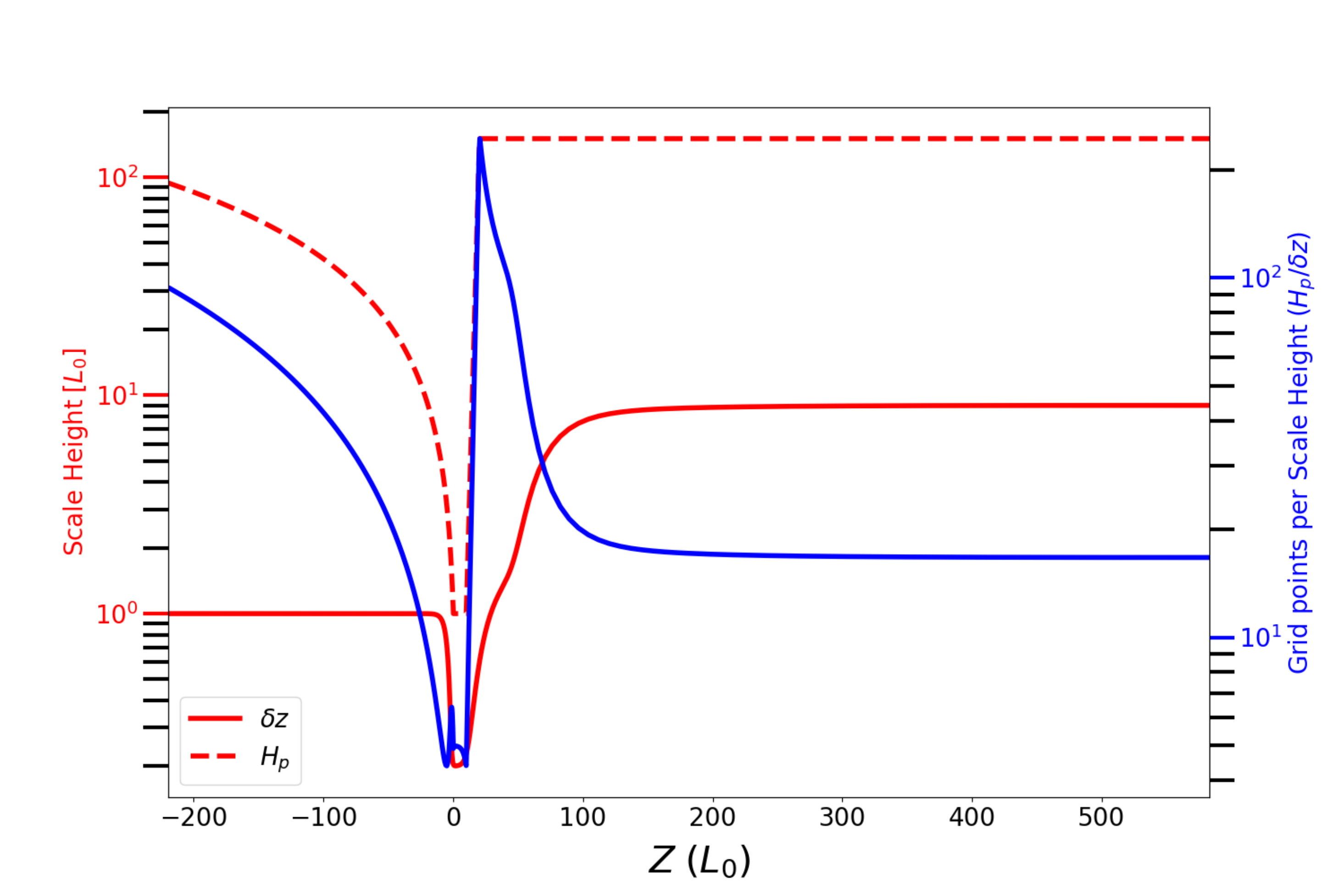}
\caption{Profiles with height of grid spacing $\delta z$ (red solid line), pressure scale height (red dashed line), and grid points per local scale height (blue line).\label{fig:grid}}
\end{figure*}

The simulation domain has extents $X\times Y\times Z = [-492,492]L_0\times[-492,492]L_0\times[-220,598]L_0$. In the $x$- and $y$-directions, we use $1024$ grid points, for a resolution of $\delta x=\delta y\approx L_0$, while in the $z$-direction, $400$ grid points are used, but the grid is stretched so that the grid spacing has the form:
\beg{stretchz}
\delta z = L_0\frac{1 + f_0[1+\tanh({\frac{z-L_v}{w}})]}{1+f_1[\tanh(\frac{z+f_2}{f_3})-\tanh(\frac{z+f_4}{f_5})]}.
\done 
Here $f_0 = 4.0$, $f_1 = 2.0$, $f_2=-2.0L_0$, $f_3=5.0L_0$, $f_4=-117.0L_0$, $f_5=5.0L_0$, $L_v = 100L_0$, and $w=L_z/20$. \autoref{fig:grid} shows the profile of $\delta z$, along with the local scale height and number of grid points per local scale height. 
Near the photosphere, $\delta z=0.25L_0$, so that a minimum of four grid points resolve the photospheric pressure scale height, while farther away from the photosphere, the resolution coarsens, though the number of grid points per scale height increases.\par
At the boundaries, all velocity components and normal gradients of the magnetic field, plasma density and specific energy density are set to zero. As in \citet{Leake13}, the resistivity is smoothly decreased to zero close to the bottom boundary to reduce diffusion of the magnetic field. In addition, a damping region is applied to the velocity at all four side boundaries, as described in \citet{Leake13}. This prevents reflected waves from interacting with the solution in the interior. As a result of these boundary conditions, the field lines are line-tied at the bottom boundary ($v_h=0$ and $\eta=0$). 

\subsection{Initial Conditions}
The simulations are initialized with a hydrostatic background atmosphere with a convection zone ($-220L_0<z<0$), a photosphere/chromosphere ($0<z<10L_0$), transition region ($10L_0<z<20L_0$) and corona ($z>20L_0$). \par 
In the convection zone, the atmosphere is polytropic, with the temperature following
\beg{temperature}
T_{cz}(z) = T_0 \Big(1-\frac{z}{l_c}\Big),
\done 
where $l_c = (g\mu_m/k_B)(\gamma-1)/\gamma$. In the photosphere, $T(z) = T_0$, $\rho(z) = \rho_0$ and $P(z) = P_0$. In the corona, the temperature is 
\beg{Tcorona}
T_{cor}(z) = 150 T_0.
\done
The photosphere and corona are connected by the transition region, where the temperature increases from $T_{ph}$ to $T_{cor}$ via
\beg{Ttr}
T_{tr}(z) = T_0\Big(\frac{T_{cor}}{T_0}\Big)^{(z-10L_0)/10L_0}
\done
We then numerically integrate $\nabla P = -\rho g$ and the ideal gas law, Equation \ref{idealgas}, to obtain the density and pressure profiles throughout the atmosphere.
\par 
We place a toroidal flux rope in the convection zone. In spherical coordinates ($r,\theta,\phi$), the flux rope has the form:
\beg{Btotal}
\vecB = \nabla\times \Big(A(r,\theta)\hat{\phi}\Big) + B_\phi(r,\theta)\hat{\phi},
\done 
where 
\beg{Bphi}
B_\phi(r,\theta) = \frac{aB_t}{r\sin\theta}\exp\Big(-\varpi^2/a^2\Big),
\done 
and
\beg{Artheta}
A(r,\theta) = \frac{1}{2}\zeta a r\sin\theta B_\phi(r,\theta),
\done 
and
\beg{pomega}
\varpi^2 = r^2+r_0^2-2rr_0\sin\theta.
\done 
Note that a more general form of this profile is given in \citet{Fan08}. In these equations, $r_0=120\;L_0$ and $a=20\; L_0$ represent the major and minor radii of the flux rope, $B_t = 100\;\mathrm{B_0}$ is the initial field strength, and $\zeta$ is a dimensionless number representing the flux rope's twist. In terms of the Cartesian coordinates ($x,y,z$), 
\beg{defr}
r^2=x^2+(z-z_0)^2,
\done 
\beg{deftheta}
\theta = \arccos{\Big(\frac{x}{r}\Big)},
\done 
and
\beg{defphi}
\phi = \arctan{\Big(\frac{z-z_0}{y}\Big)}
\done 
are the radial, azimuthal, and toroidal coordinates, respectively, and $z_0=-220\;L_0$. Thus, our initial flux rope depth-to-radius ratio is similar to \citet{MacTaggart09}, who use $z_0/a = 10.0$ and $z_0/r_0 = 1.7$, compared to our values of $11.0$ and $1.8$, respectively, though our flux rope is placed about $9$ times deeper. \par 
We perturb the background plasma and energy densities by 
\beg{rhoperturb}
\rho(r,z) = \rho_b(z)\Big(1-\frac{p_1(r)}{p_0(z)}\Big)
\done 
and
\beg{eperturb}
e(r,z) = \frac{p_0(z)-p_1(r)}{\rho(r,z)(\gamma-1)},
\done
where $p_1$ is given by
\beg{p1}
p_1(r) = \frac{1}{2\mu_0}B^2_\phi(r,\theta).
\done 
In this way, the entire flux rope is made buoyant, but the largest buoyancy is near the center at $x=y=0$ and $z=z_0+r_0$.
 \par 
%Note that this flux rope is both larger and deeper than the flux rope used by \citet{MacTaggart09}, who used $r_0=2.5\;L_0$ and $z_0=-25\;L_0$. \par

%In Figure \ref{fig:buoyancy}, we plot $\delta\rho/\rho_{fr}$, where $\delta\rho$ is the difference between the flux rope density and the ambient density. The buoyancy is distributed throughout the entire flux rope, but it is most buoyant near its center at $y=0$. 
%\begin{figure*}
%\includegraphics[width=\linewidth]{Buoyancy.pdf}
%\caption{Buoyancy profile $\delta\rho/\rho$ used to initialize the flux rope's rise.\label{fig:buoyancy}}
%\end{figure*}
In the present study, we set $\zeta=0$, so that the only nonzero component of $\vecB$ is $B_\phi$ (Equation \ref{Bphi}), corresponding to a completely untwisted flux rope that is bent into the shape of a torus.

\section{Results}\label{sec:results}
\subsection{Rise Through the Convection Zone}
\begin{figure*}
\includegraphics[width=0.50\linewidth,trim={0cm 8cm 0cm 0cm},clip]{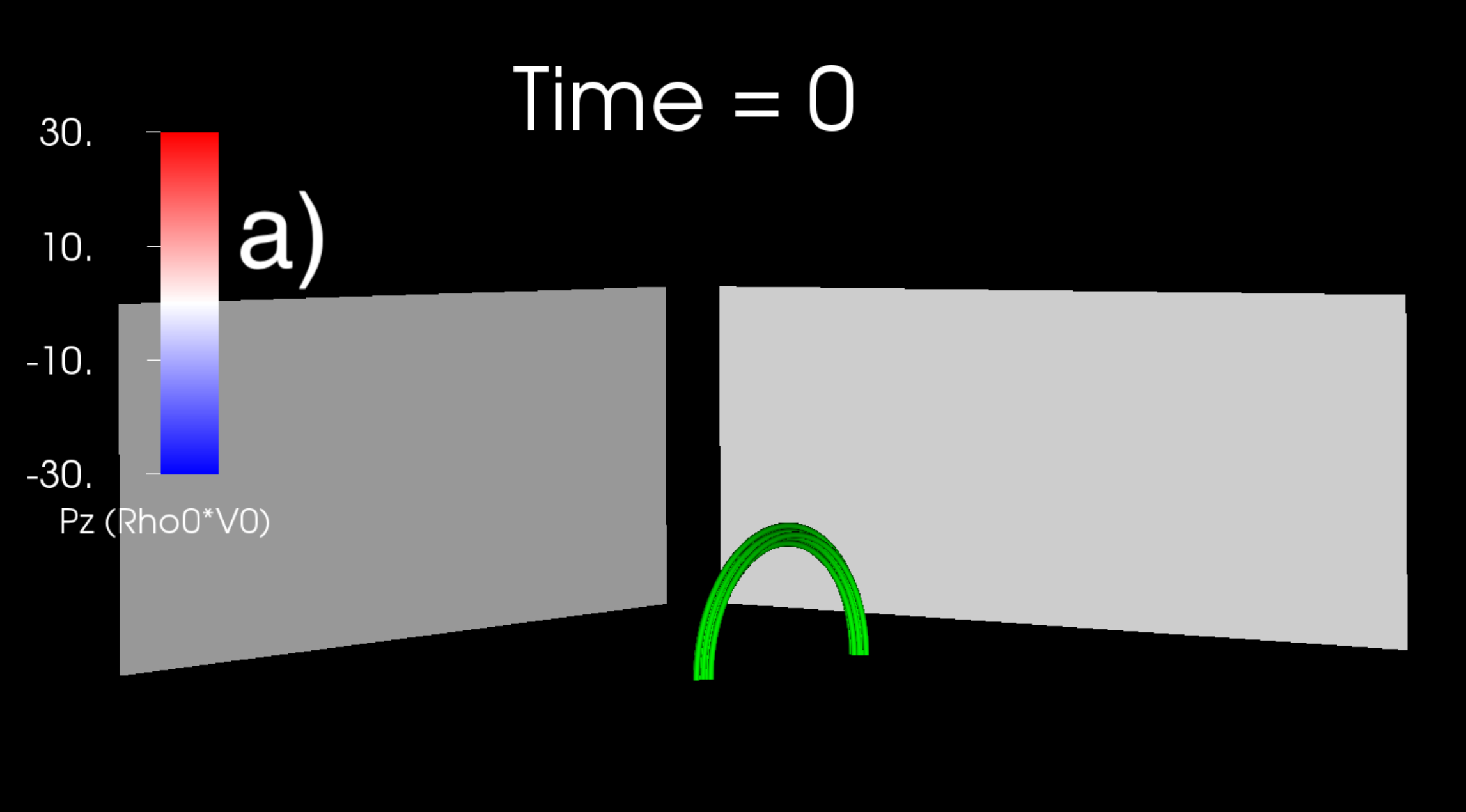}
\includegraphics[width=0.50\linewidth,trim={0cm 8cm 0cm 0cm},clip]{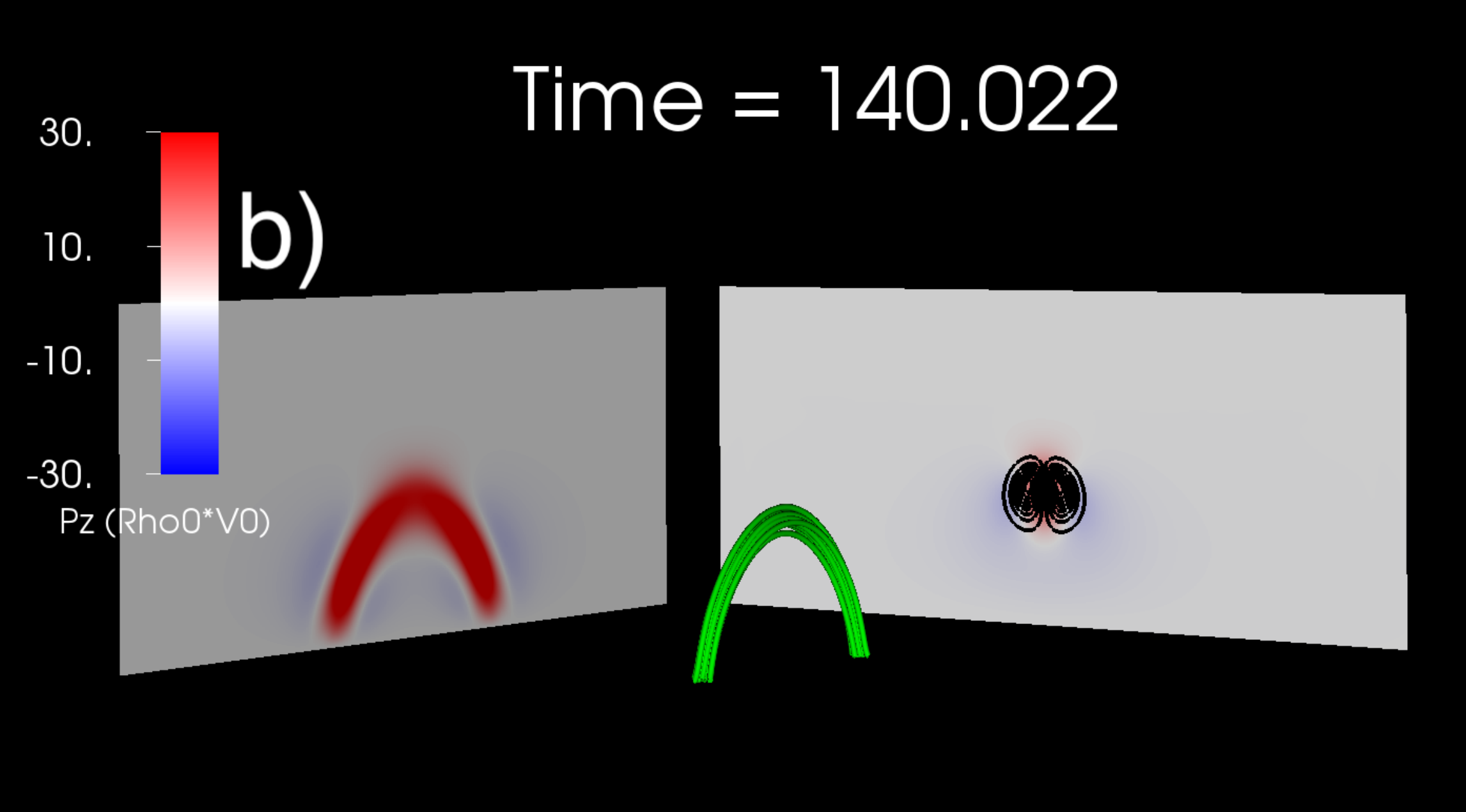}
\includegraphics[width=0.50\linewidth,trim={0cm 8cm 0cm 0cm},clip]{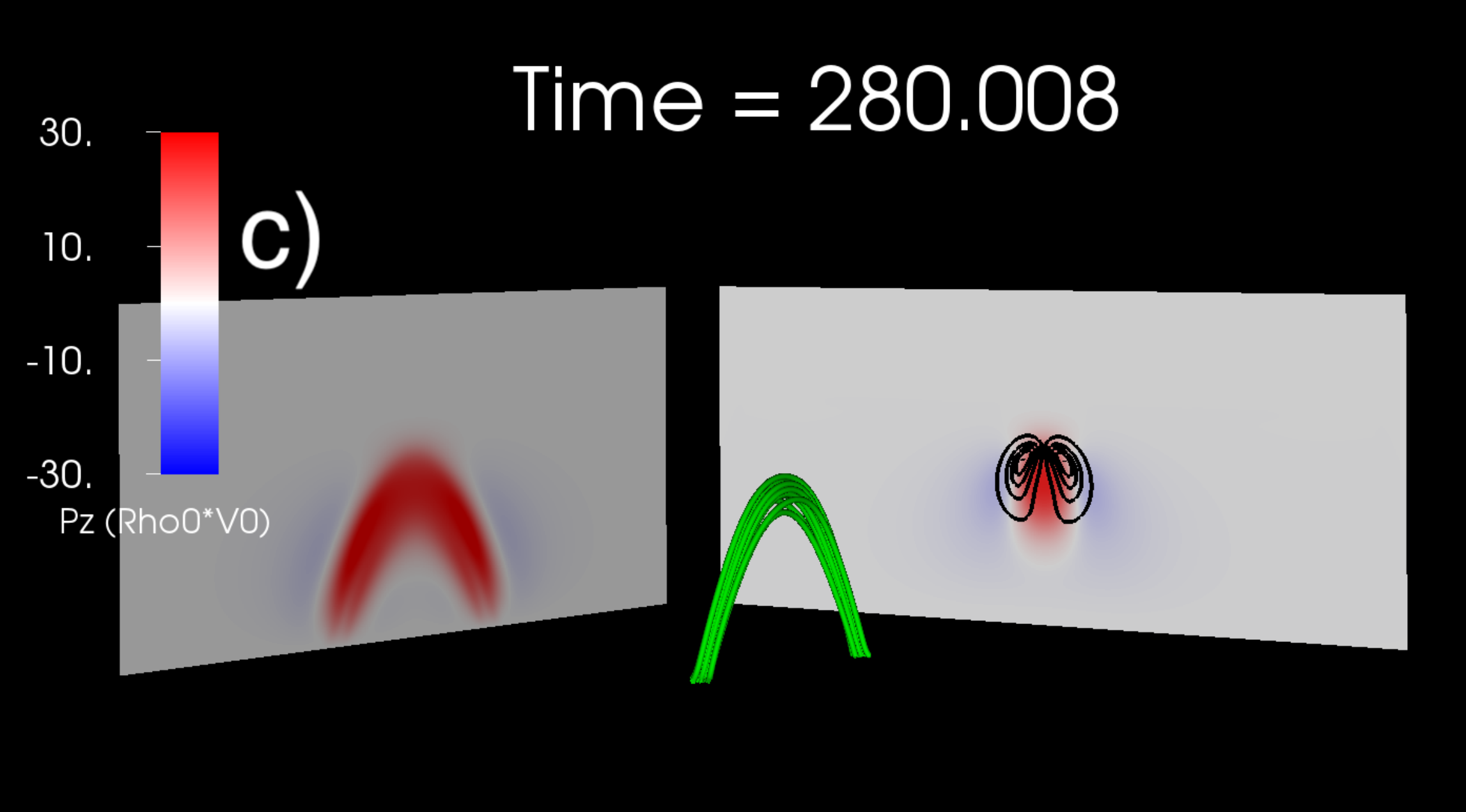}
\includegraphics[width=0.50\linewidth,trim={0cm 8cm 0cm 0cm},clip]{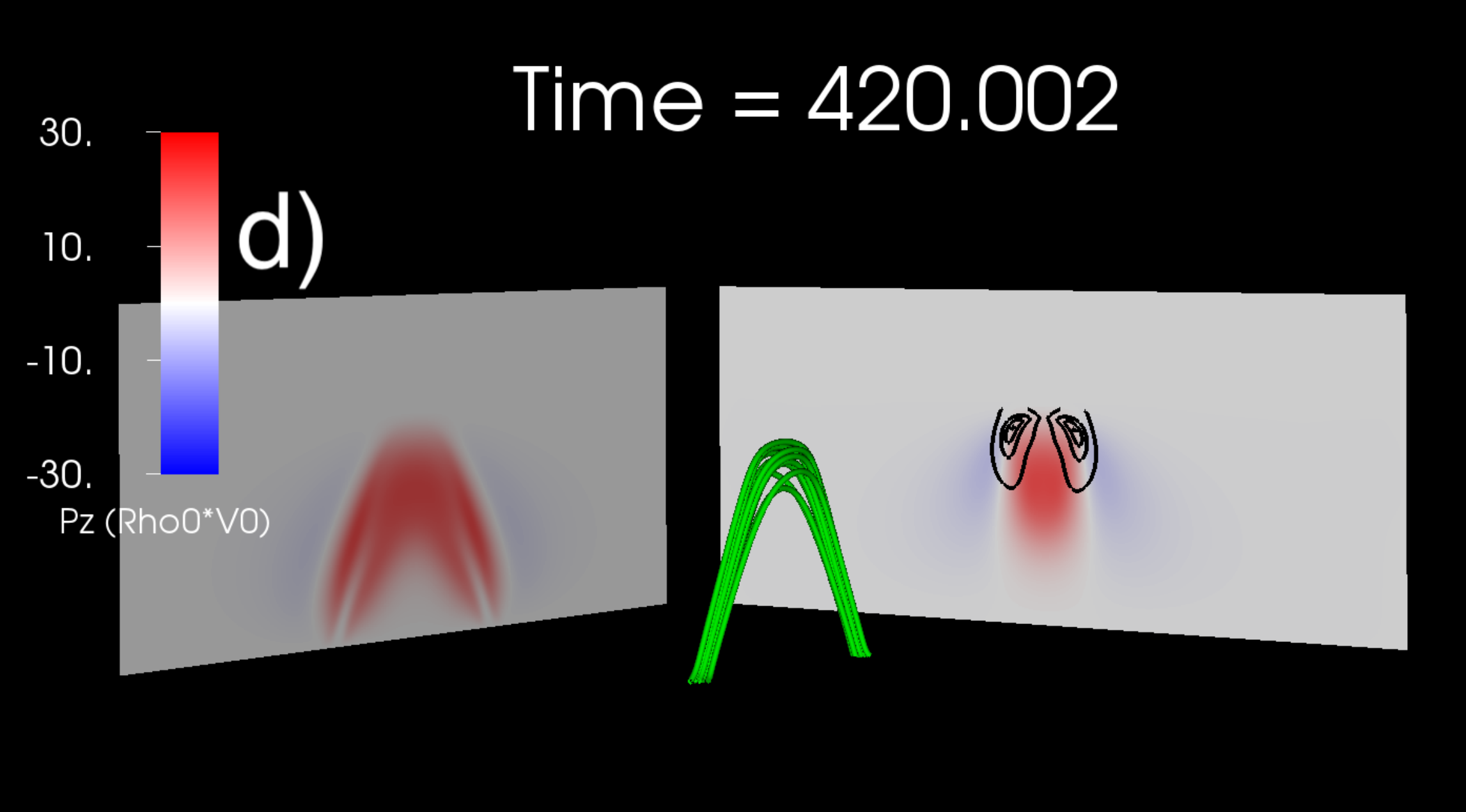}
\caption{Field lines in the torus at different times during the simulation, traced from the bottom boundary. Color shading shows vertical momentum $p_z$, and contours show out of plane vorticity. The two vertical cuts are taken in the $x=0$ and $y=0$ mid-planes, respectively, but plotted on the boundaries for clarity. Only the lower part of the domain is shown, from the bottom boundary to about $120\;L_0$ above the photosphere.\label{fig:fieldlines_vz}}
\end{figure*}

\begin{figure*}
\includegraphics[width=0.5\linewidth,trim={0cm 6cm 0cm 4cm},clip]{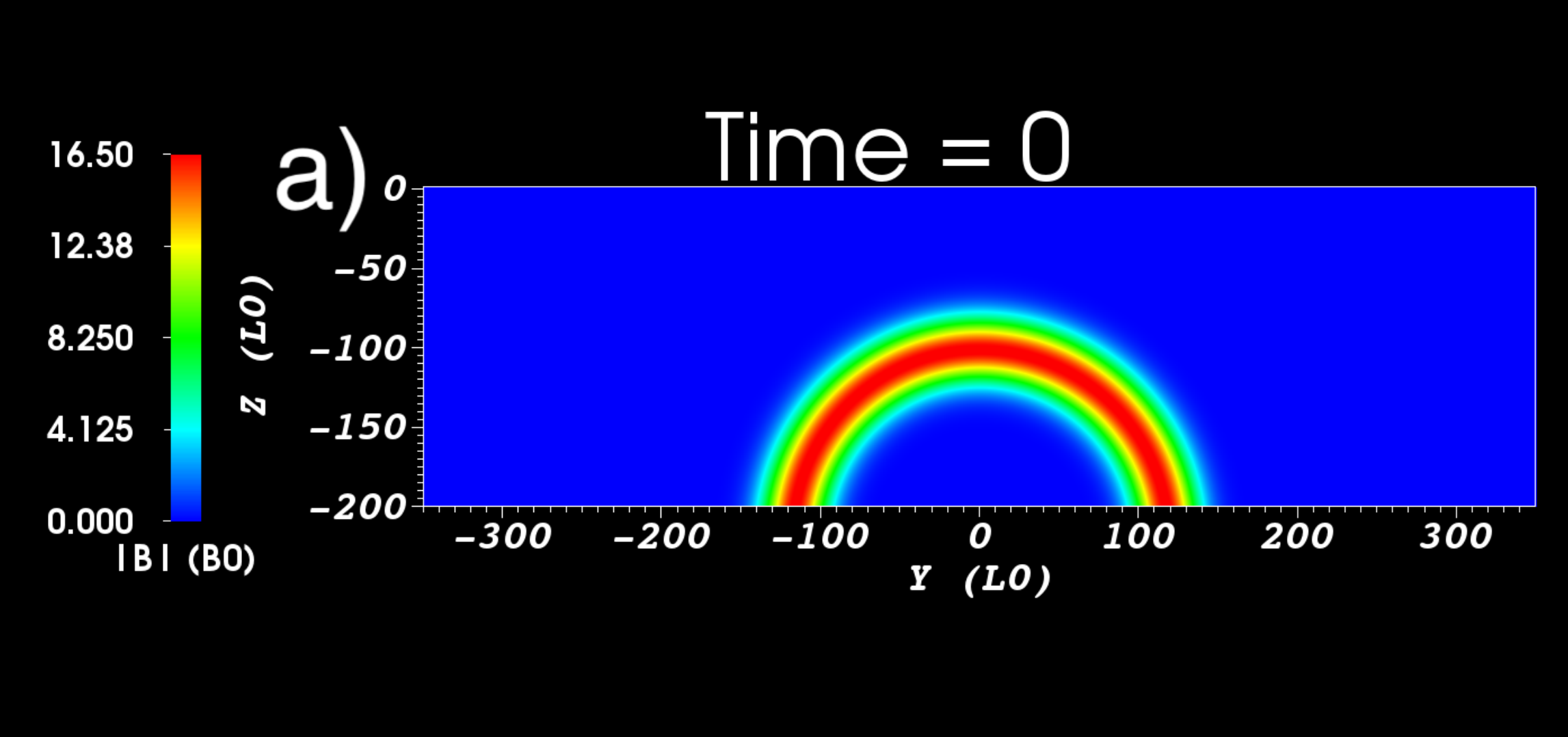}
\includegraphics[width=0.5\linewidth,trim={0cm 6cm 0cm 4cm},clip]{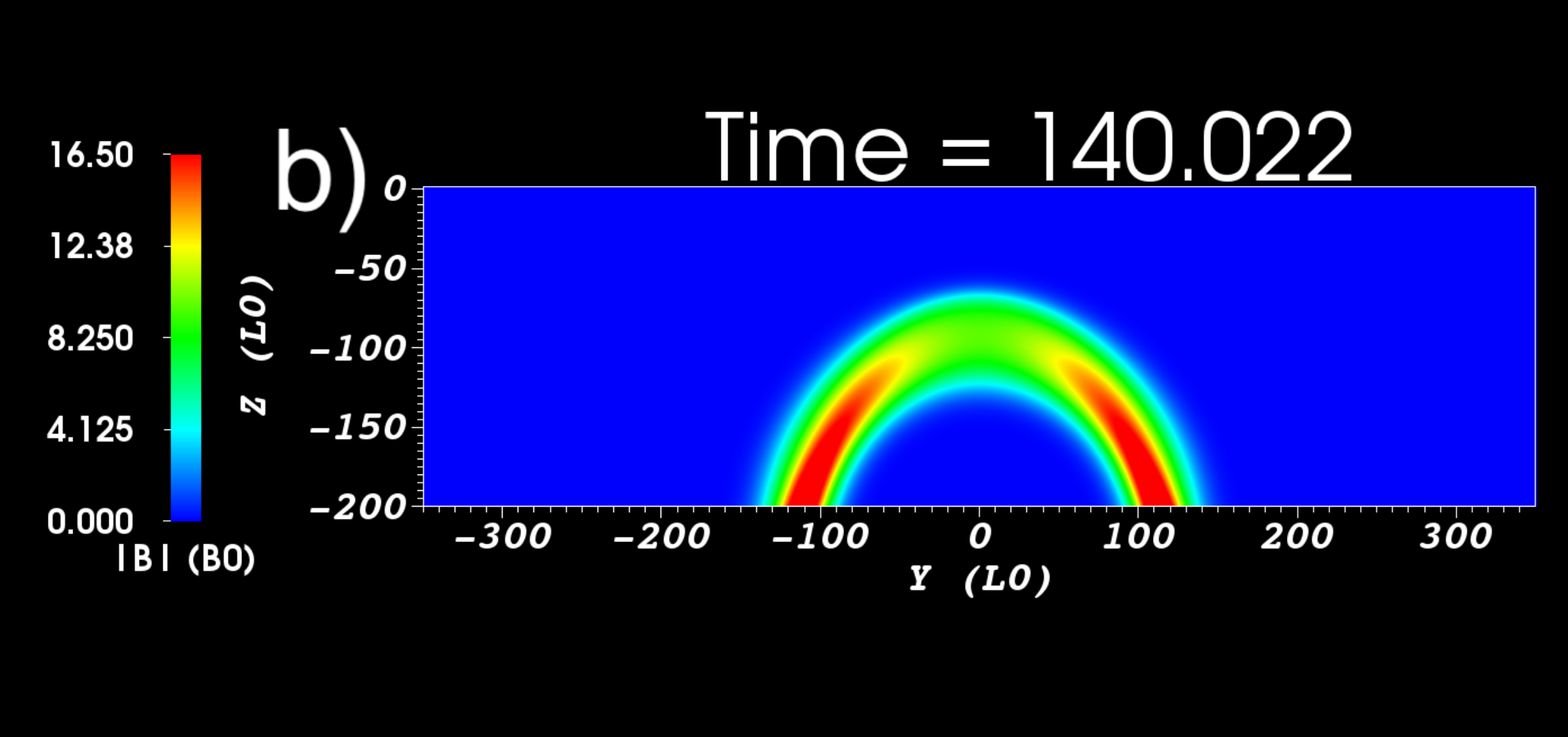}
\newline
\includegraphics[width=0.5\linewidth,trim={0cm 6cm 0cm 4cm},clip]{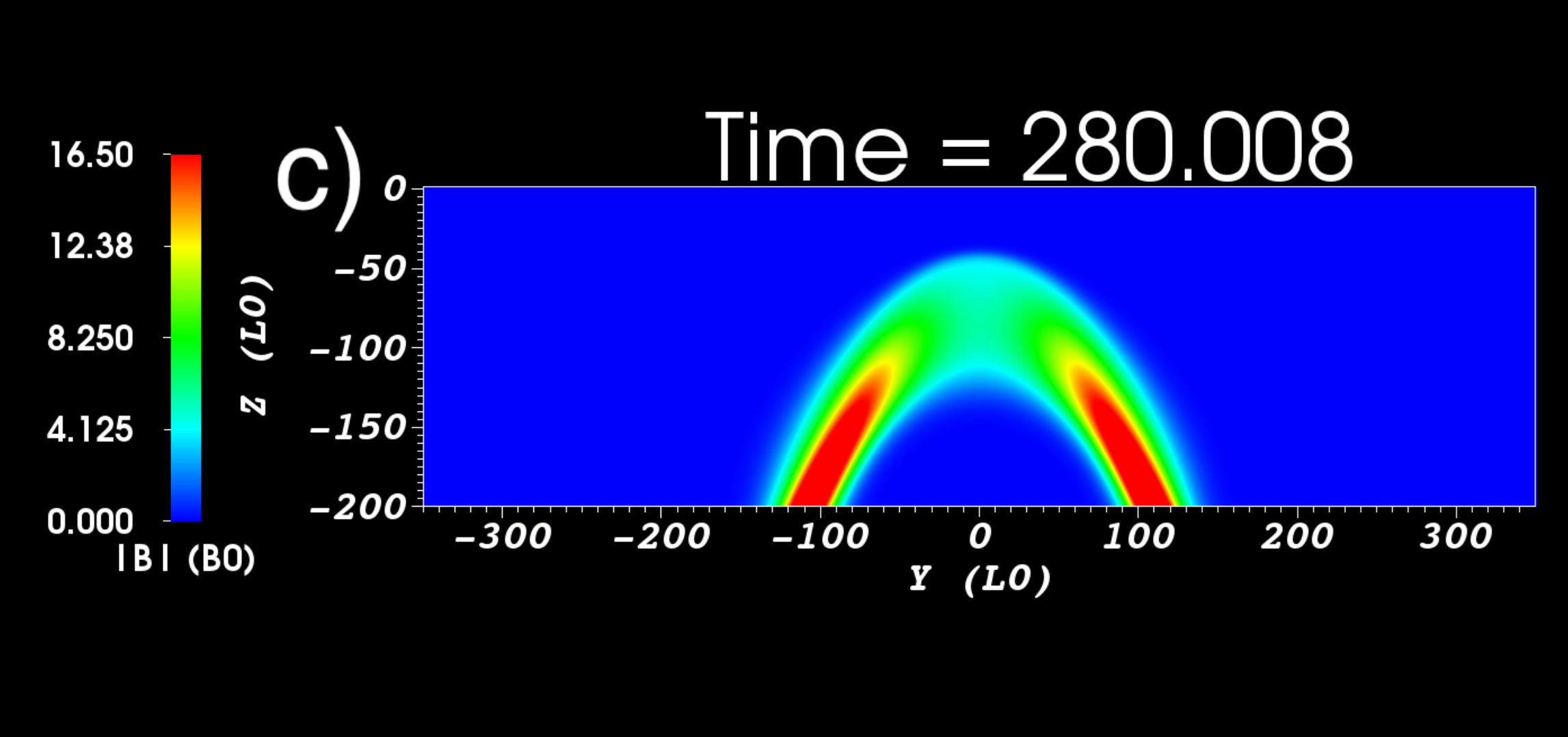}
\includegraphics[width=0.5\linewidth,trim={0cm 6cm 0cm 4cm},clip]{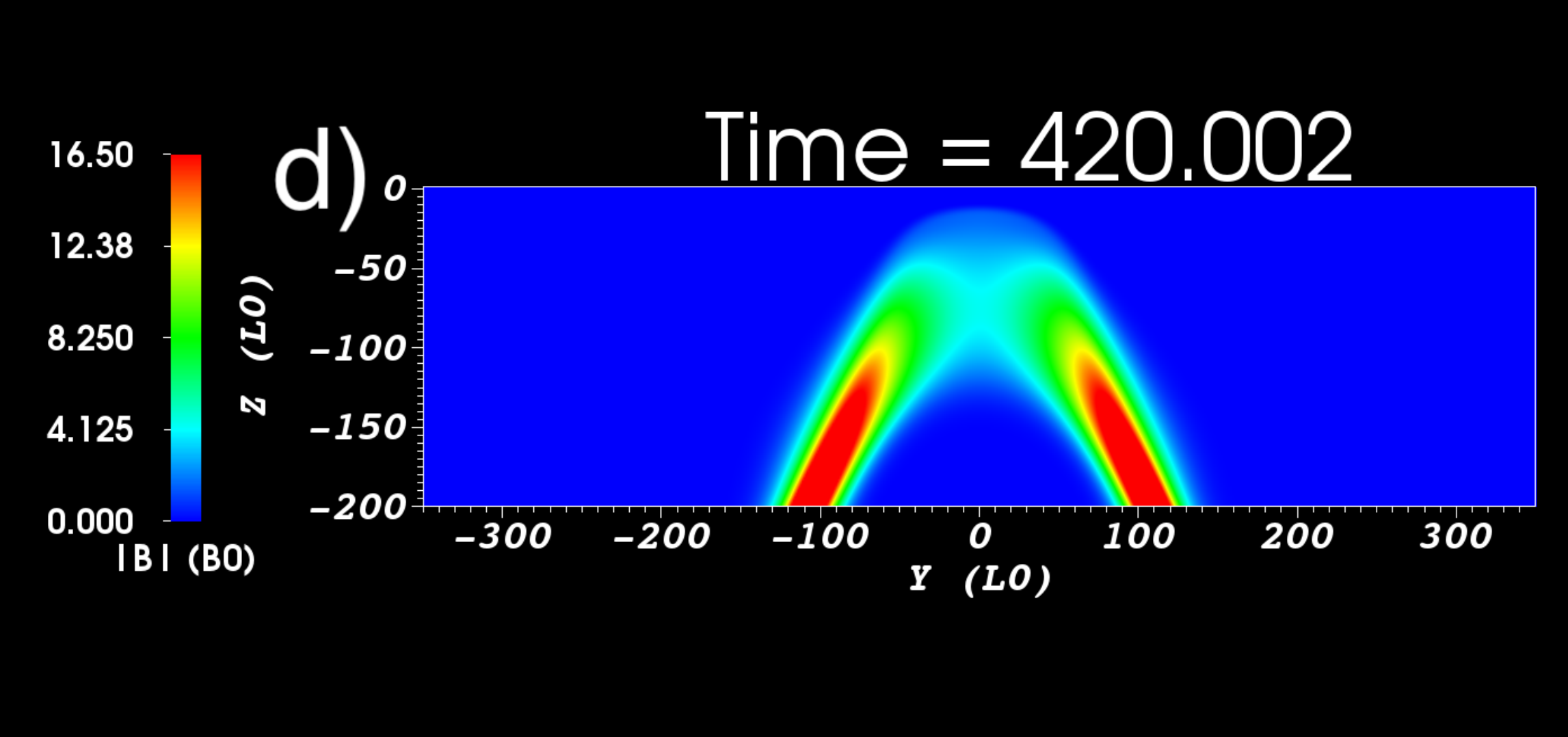}
\newline
\includegraphics[width=0.5\linewidth,trim={0cm 6cm 0cm 4cm},clip]{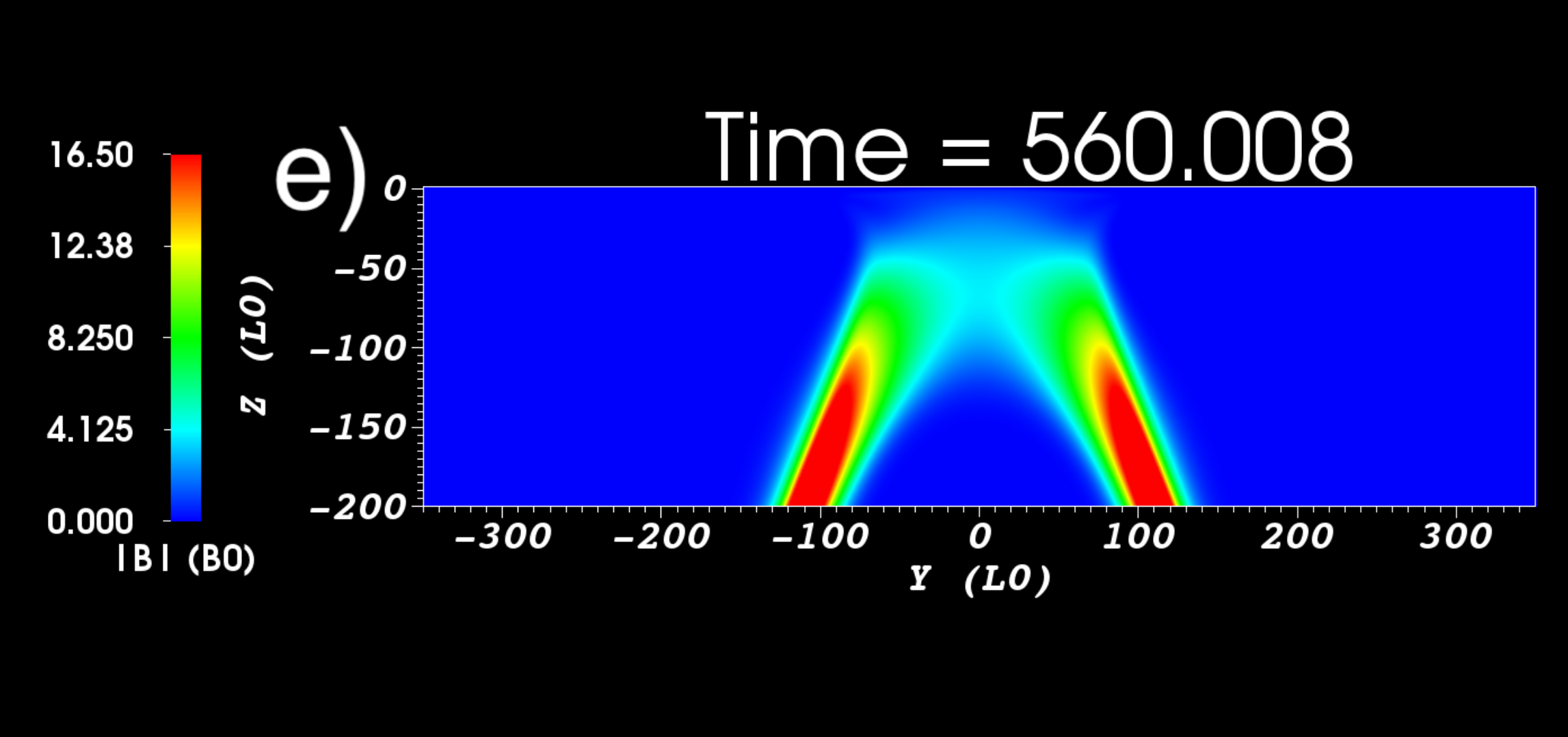}
\includegraphics[width=0.5\linewidth,trim={0cm 6cm 0cm 4cm},clip]{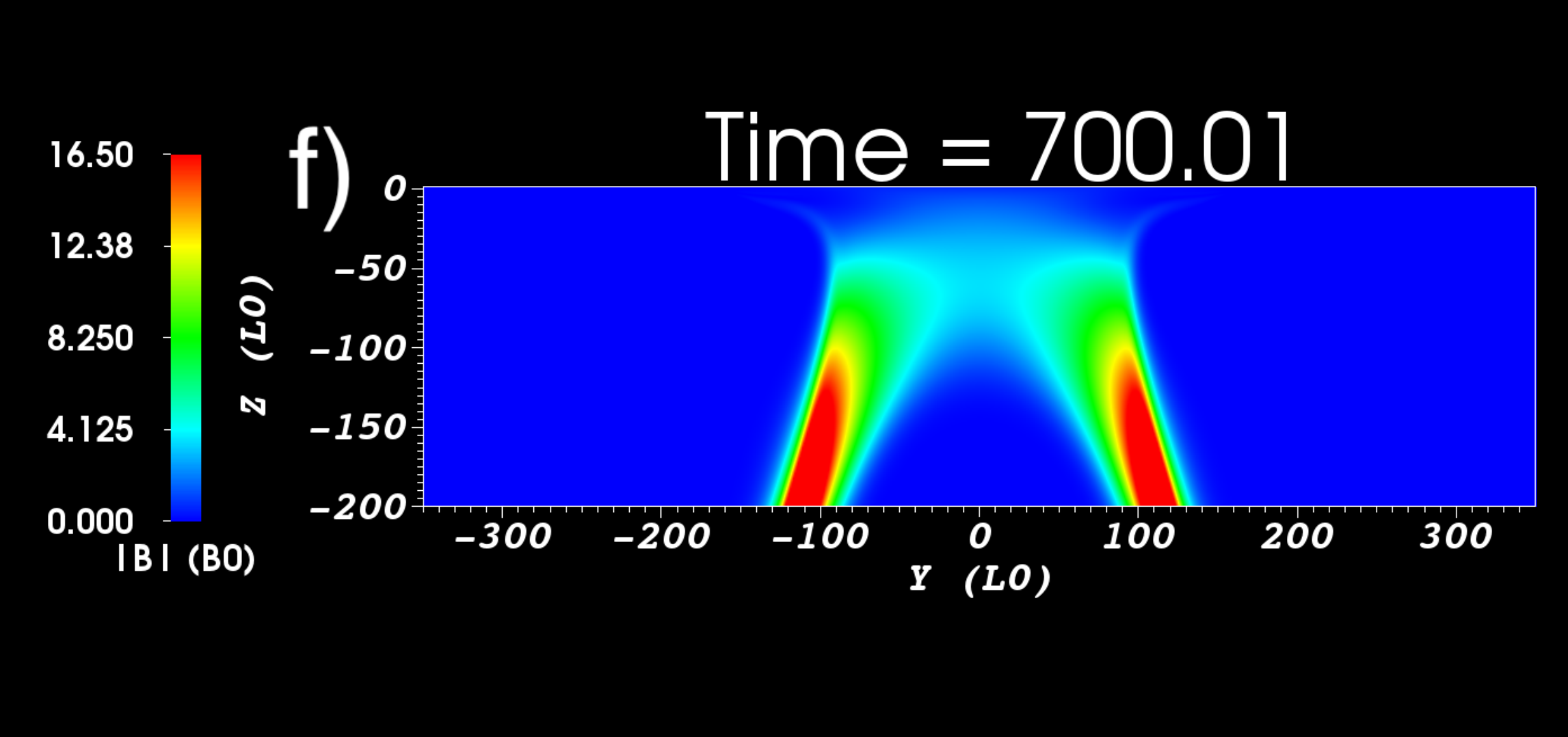}
\caption{Contours of $|B|$ at different times during the simulation in the $y-z$ plane. \label{fig:bmag0p0}}
\end{figure*}

\begin{figure*}
\includegraphics[width=0.5\linewidth,trim={0cm 6cm 0cm 4cm},clip]{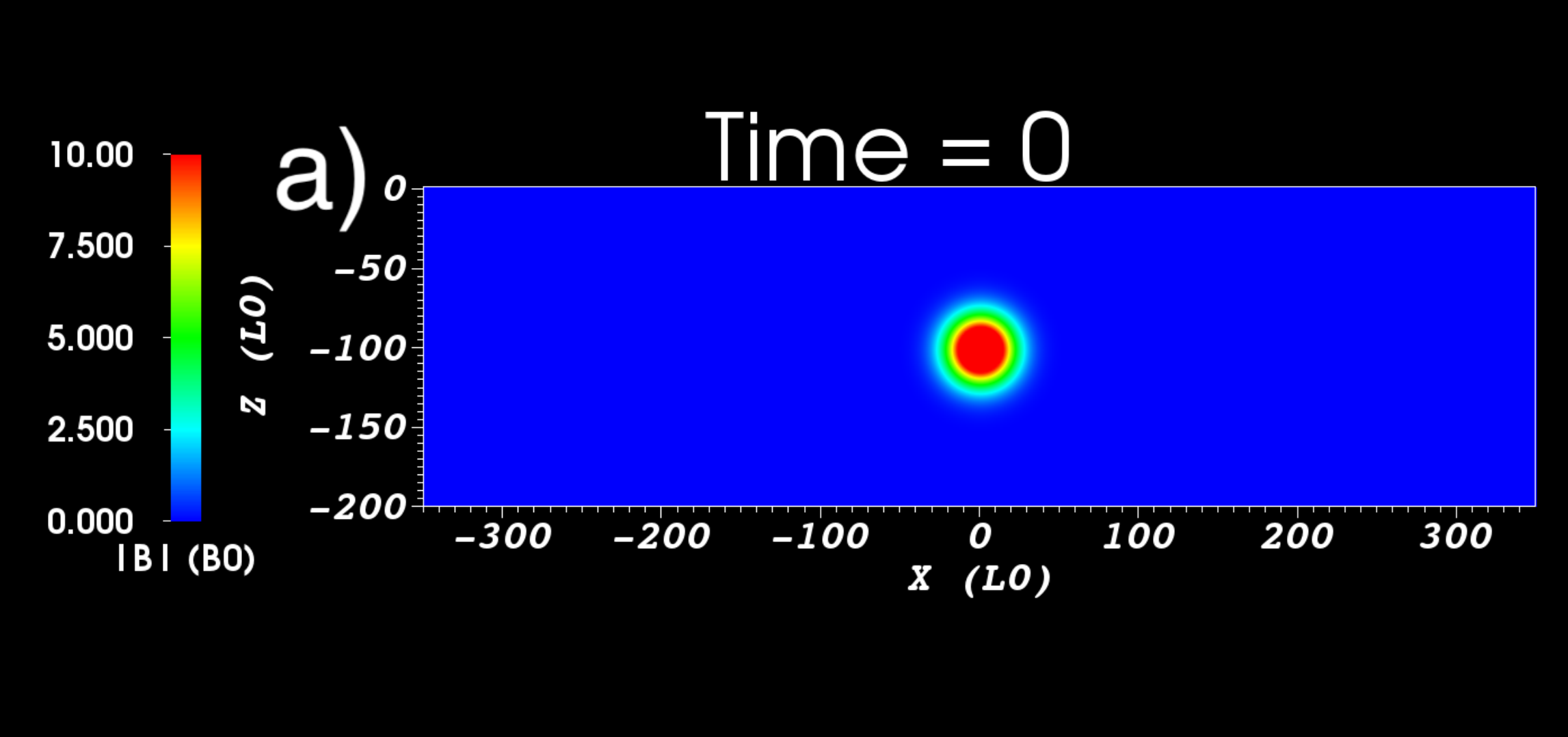}
\includegraphics[width=0.5\linewidth,trim={0cm 6cm 0cm 4cm},clip]{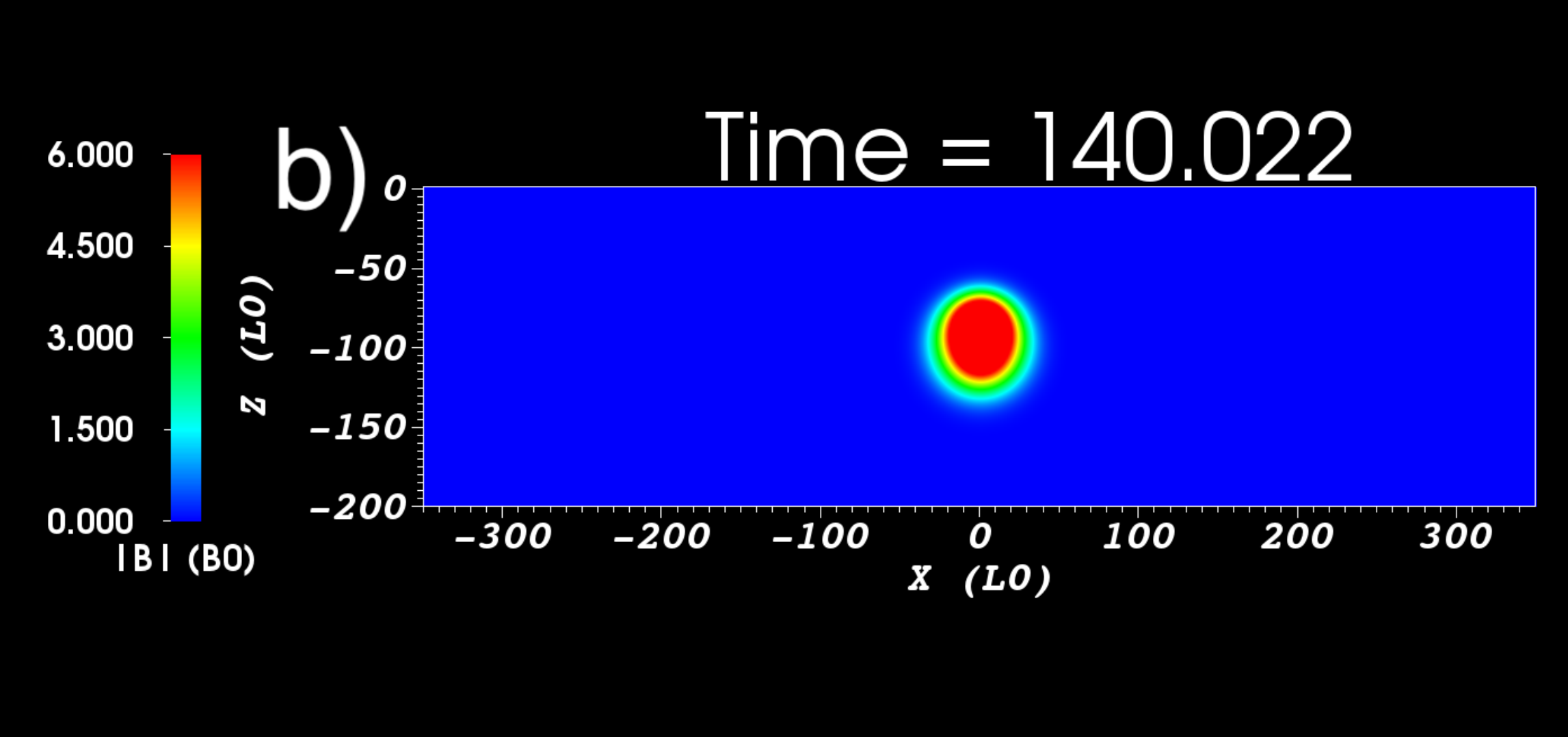}
\newline
\includegraphics[width=0.5\linewidth,trim={0cm 6cm 0cm 4cm},clip]{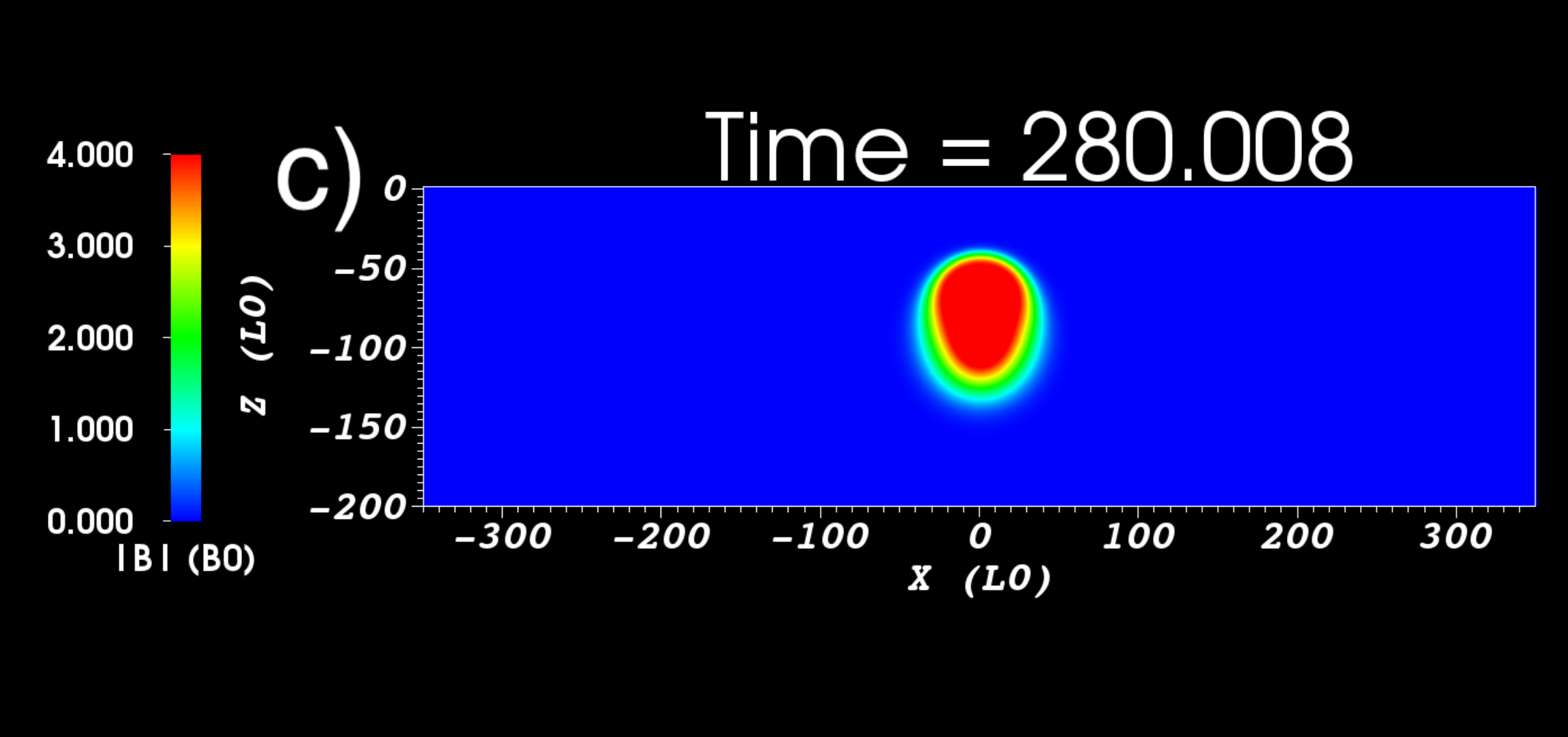}
\includegraphics[width=0.5\linewidth,trim={0cm 6cm 0cm 4cm},clip]{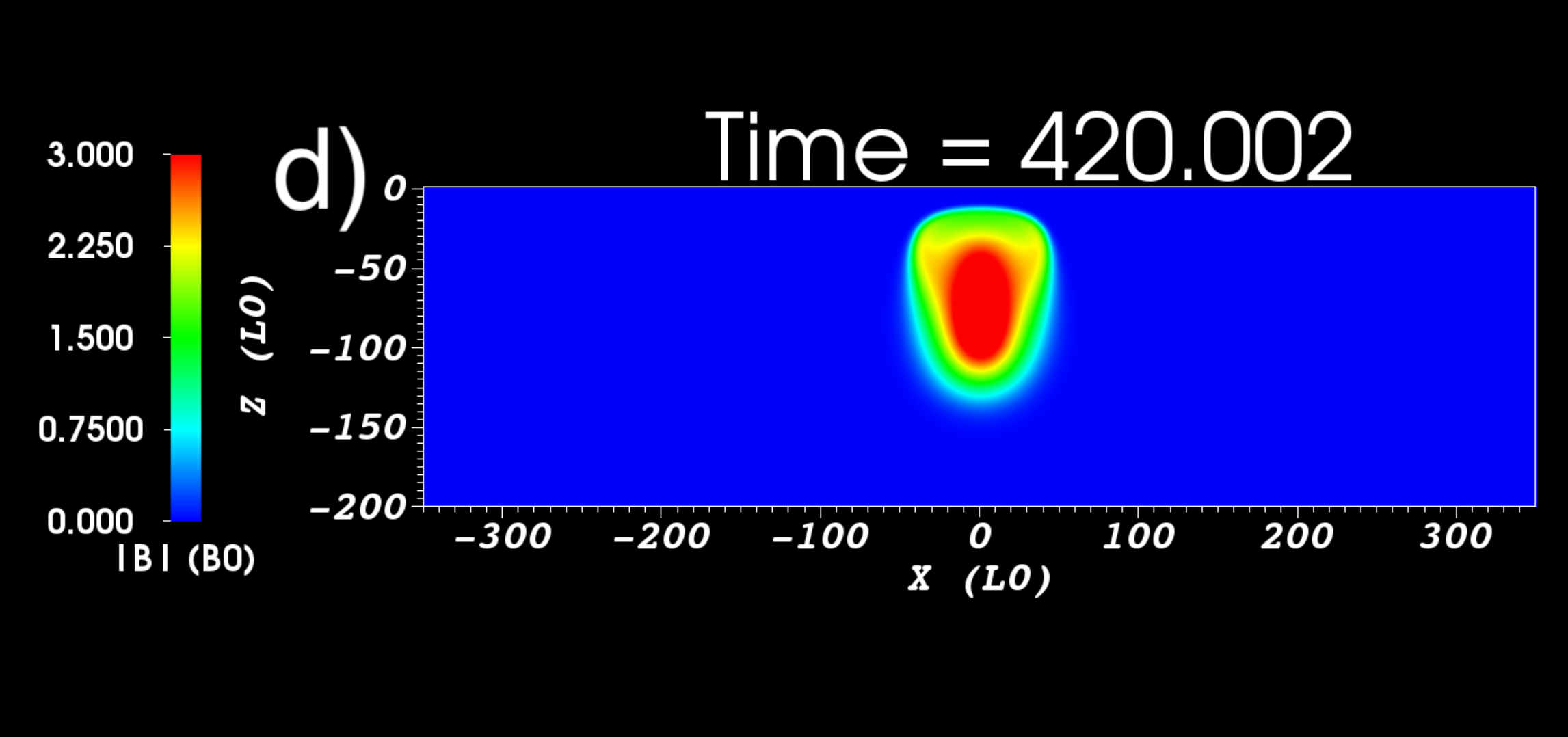}
\newline
\includegraphics[width=0.5\linewidth,trim={0cm 6cm 0cm 4cm},clip]{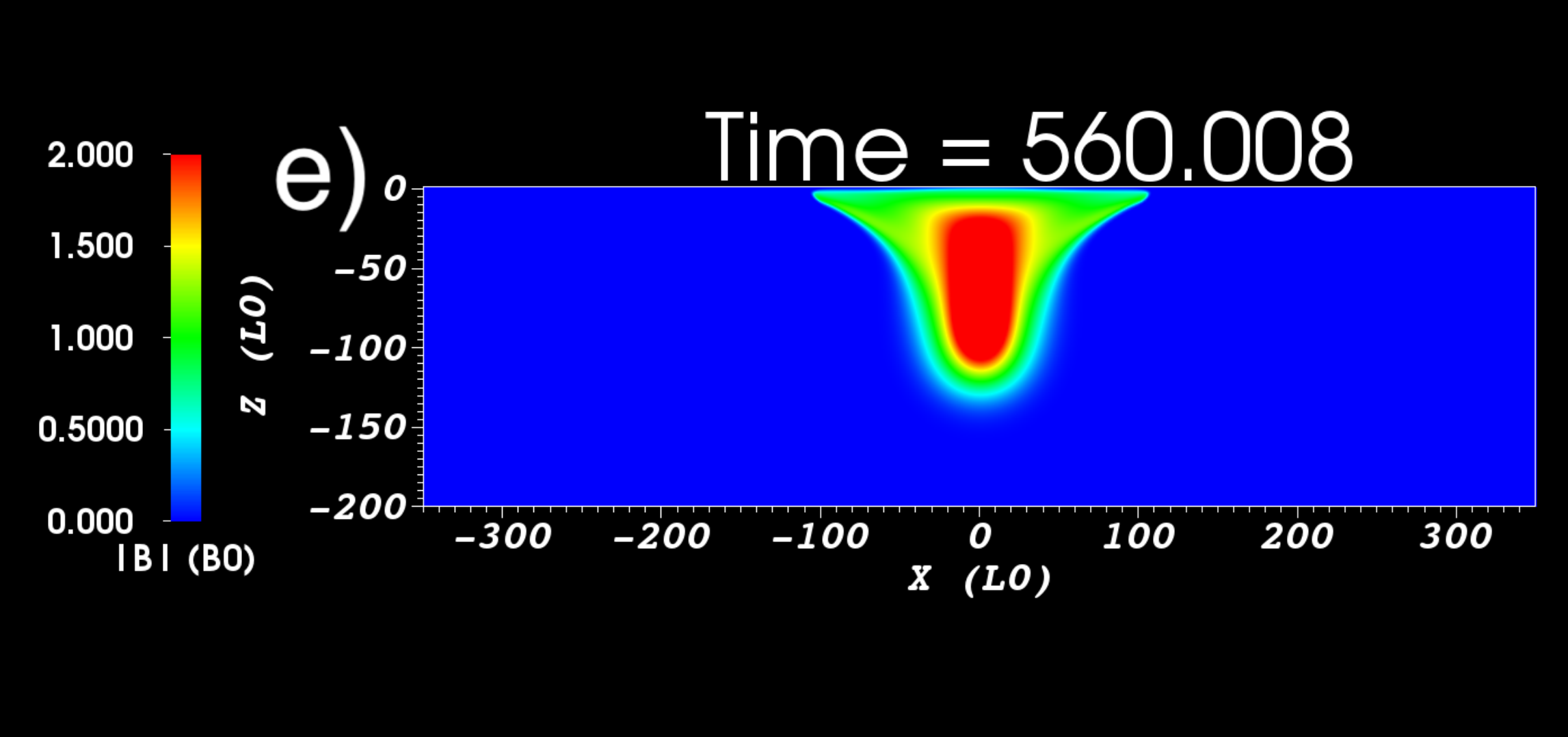}
\includegraphics[width=0.5\linewidth,trim={0cm 6cm 0cm 4cm},clip]{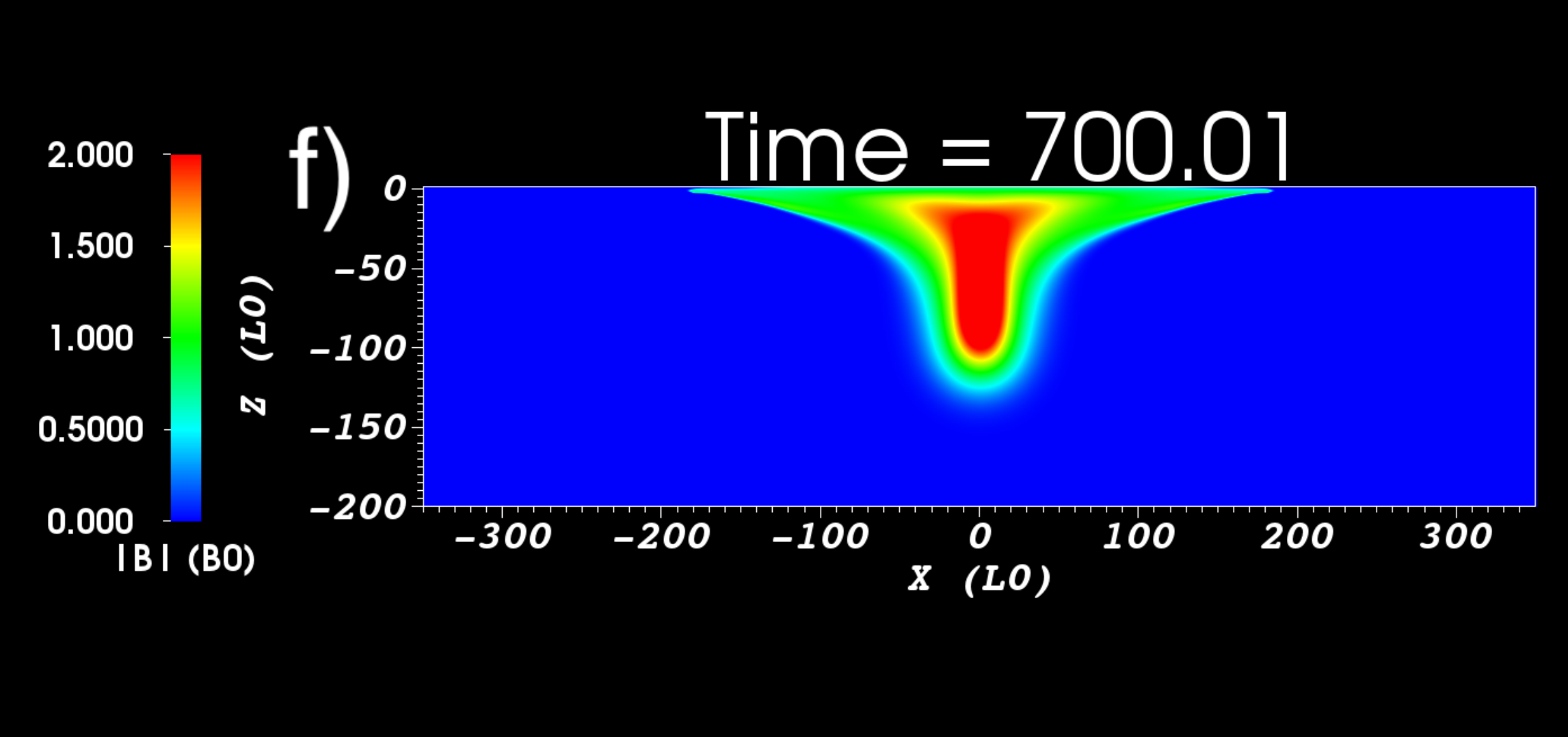}
\caption{Contours of $|B|$ at different times during the simulation in the $x-z$ plane. \label{fig:bmagxz0p0}}
\end{figure*}

\autoref{fig:fieldlines_vz} shows field lines comprising the torus traced from the fixed bottom boundary at different times during the simulation, where the background color shading shows vertical momentum $p_z$, and the contours on the $y=0$ plane represent the out-of-plane ($y$-directed) vorticity. Both vertical cuts, taken at $x=0$ and $y=0$, respectively, are plotted on the side boundaries of the simulation ($x=-492\;\mathrm{L_0}$ and $y=-492\;\mathrm{L_0}$) for clarity. Only the lower part of the domain is shown, from the bottom boundary to just above the photosphere. The rise of the flux rope manifests itself as a straightening of the legs of the torus, as its bottom end is anchored at the bottom boundary. The semi-circular nature of the torus gradually changes to a more diagonally oriented pair of legs. Meanwhile, vortices are generated by the rising flux rope below the photosphere, visible in the $y=0$ cuts in panels b-d. These vortices have no observable effect on the evolution of $|B|$ shown in \autoref{fig:bmag0p0}-\autoref{fig:bmagxz0p0}, which show the field strength of the flux rope as a function of time in the $y-z$ and $x-z$ planes. \autoref{fig:bmag0p0} shows that the field strength at the apex of the rising flux rope decreases as it rises, and the semi-circle of the torus axis is gradually transformed into more of a triangular shape. Meanwhile, \autoref{fig:bmagxz0p0} shows that the cross section gradually deforms from the circular shape to a more elongated, oval-like shape, before spreading out as the flux rope approaches the surface. Although the field weakens with time (note the changing color table), there is no hint of fragmenting of the flux rope, and no evidence of vortical magnetic structures breaking off the rope, as is often seen in weakly twisted 2D flux ropes \citep{Schuessler79,Longcope96}. Evidently, this flux rope is able to maintain its coherence throughout its rise, despite the generation of velocity vortices, implying that the curvature of the axis plays an important role in maintaining the integrity of the flux rope \citep{Abbett00}. 

\subsection{Photospheric Evolution}
\begin{figure*}
\includegraphics[width=0.5\linewidth,trim={0cm 1cm 0cm 0cm},clip]{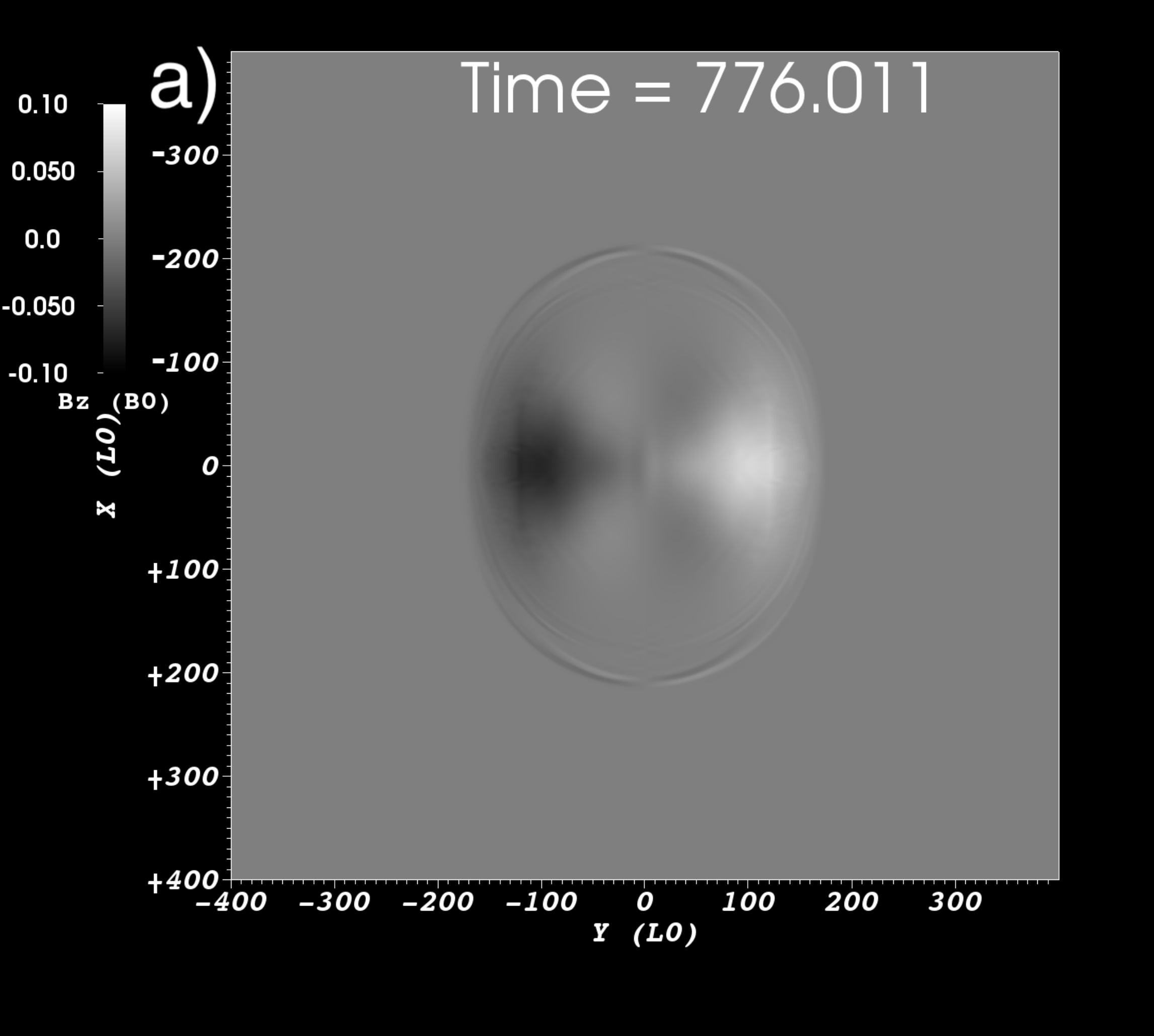}
\includegraphics[width=0.5\linewidth,trim={0cm 1cm 0cm 0cm},clip]{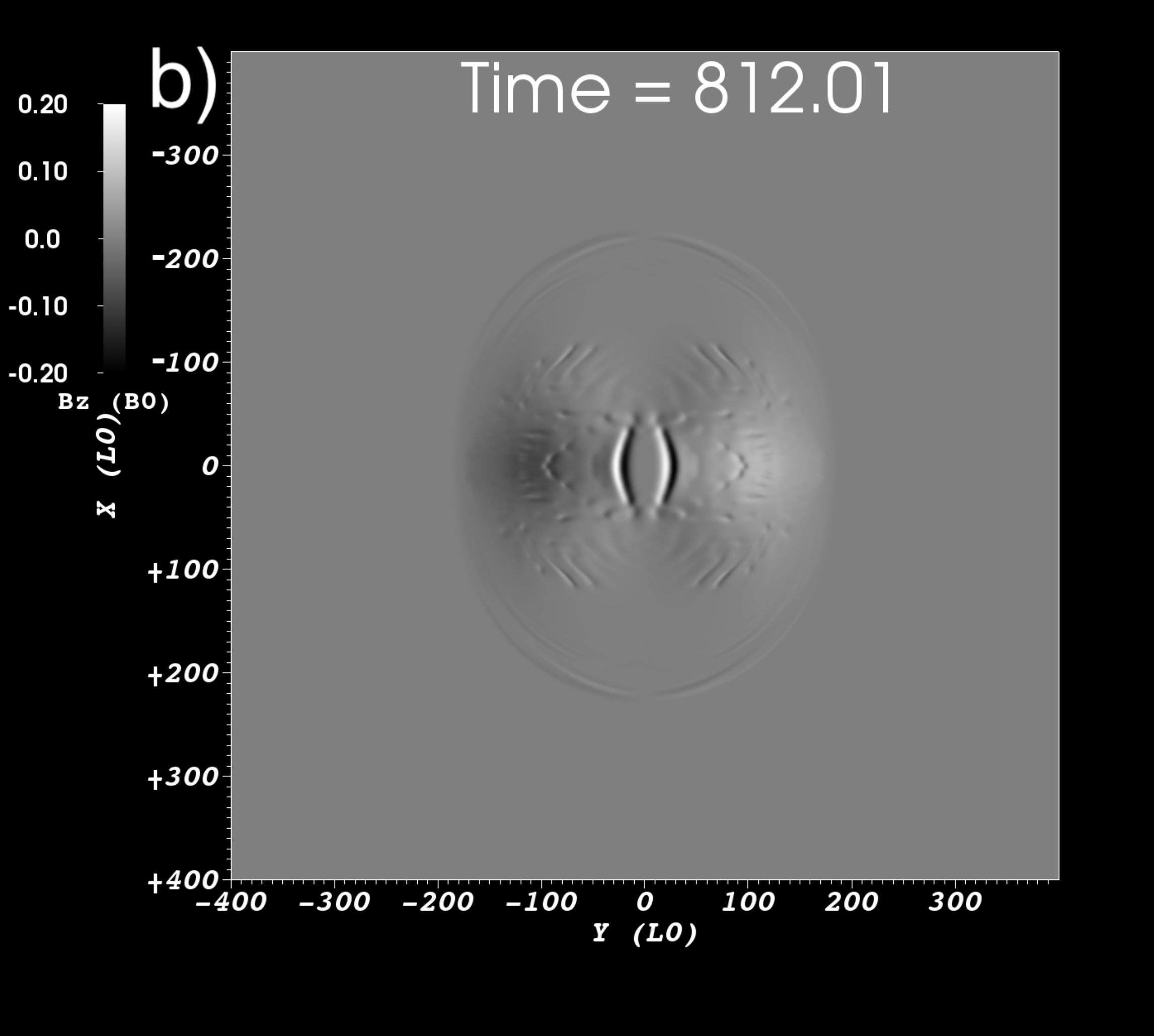}
\includegraphics[width=0.5\linewidth,trim={0cm 1cm 0cm 0cm},clip]{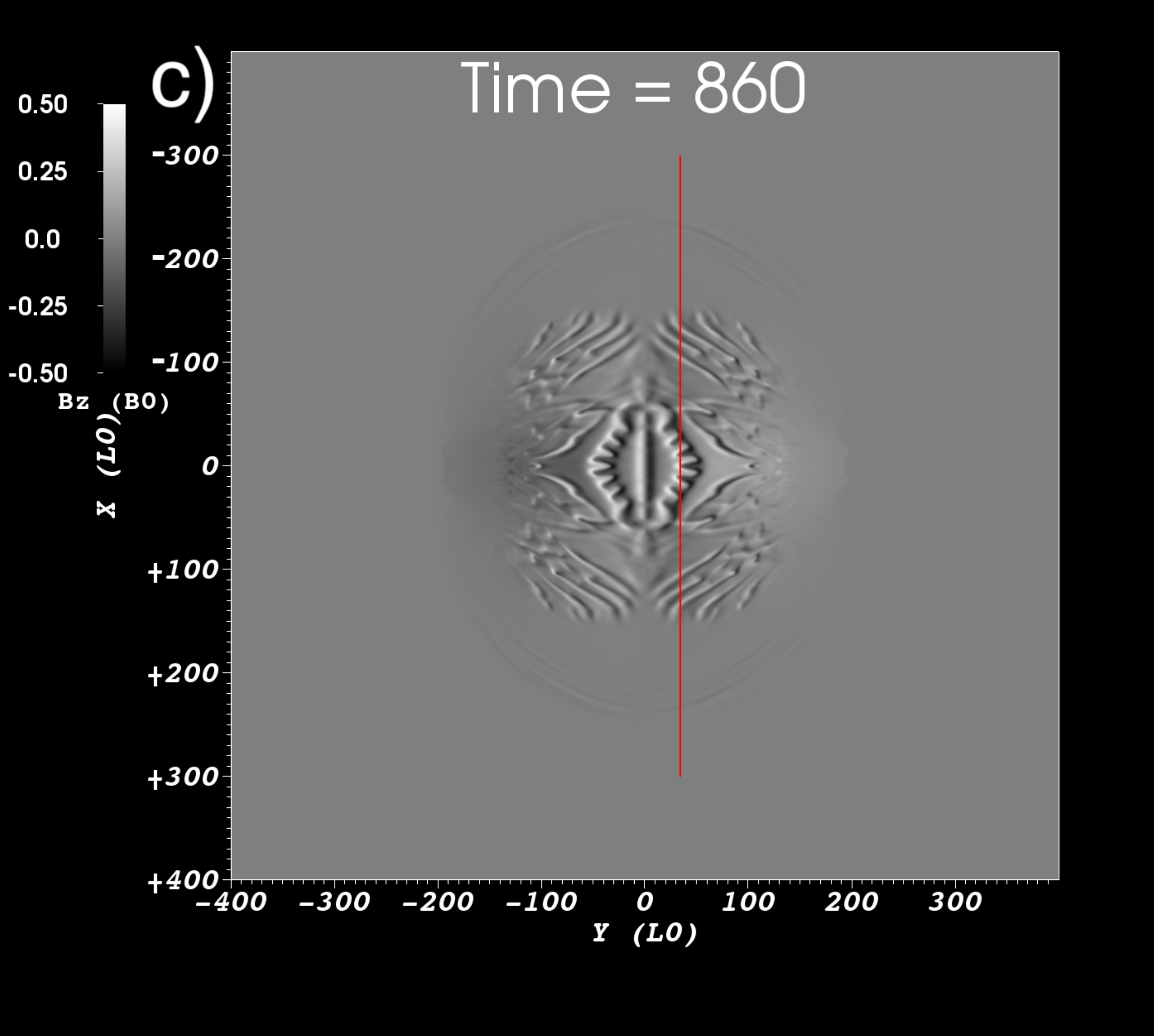}
\includegraphics[width=0.5\linewidth,trim={0cm 1cm 0cm 0cm},clip]{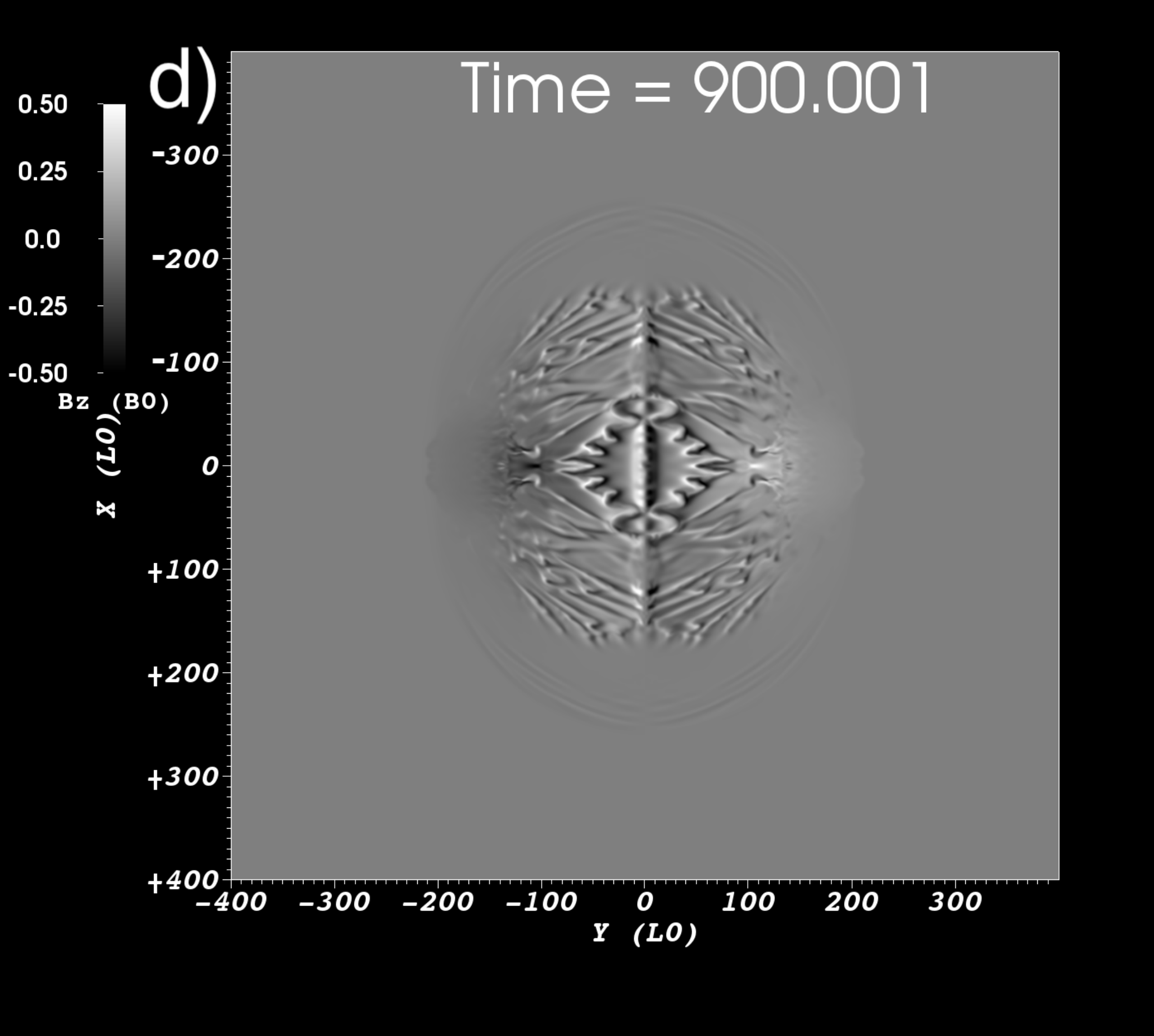}
\includegraphics[width=0.5\linewidth,trim={0cm 1cm 0cm 0cm},clip]{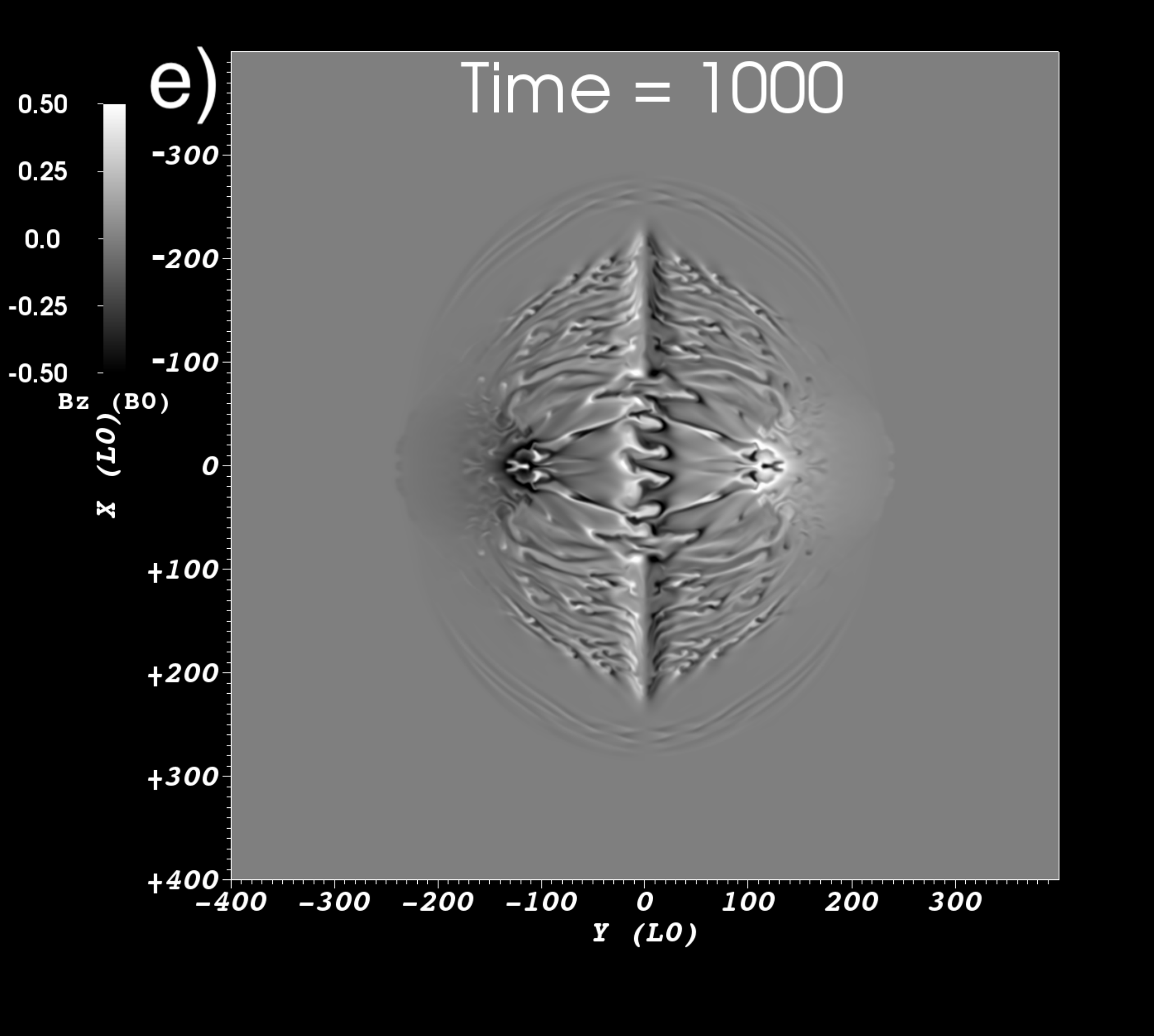}
\includegraphics[width=0.5\linewidth,trim={0cm 1cm 0cm 0cm},clip]{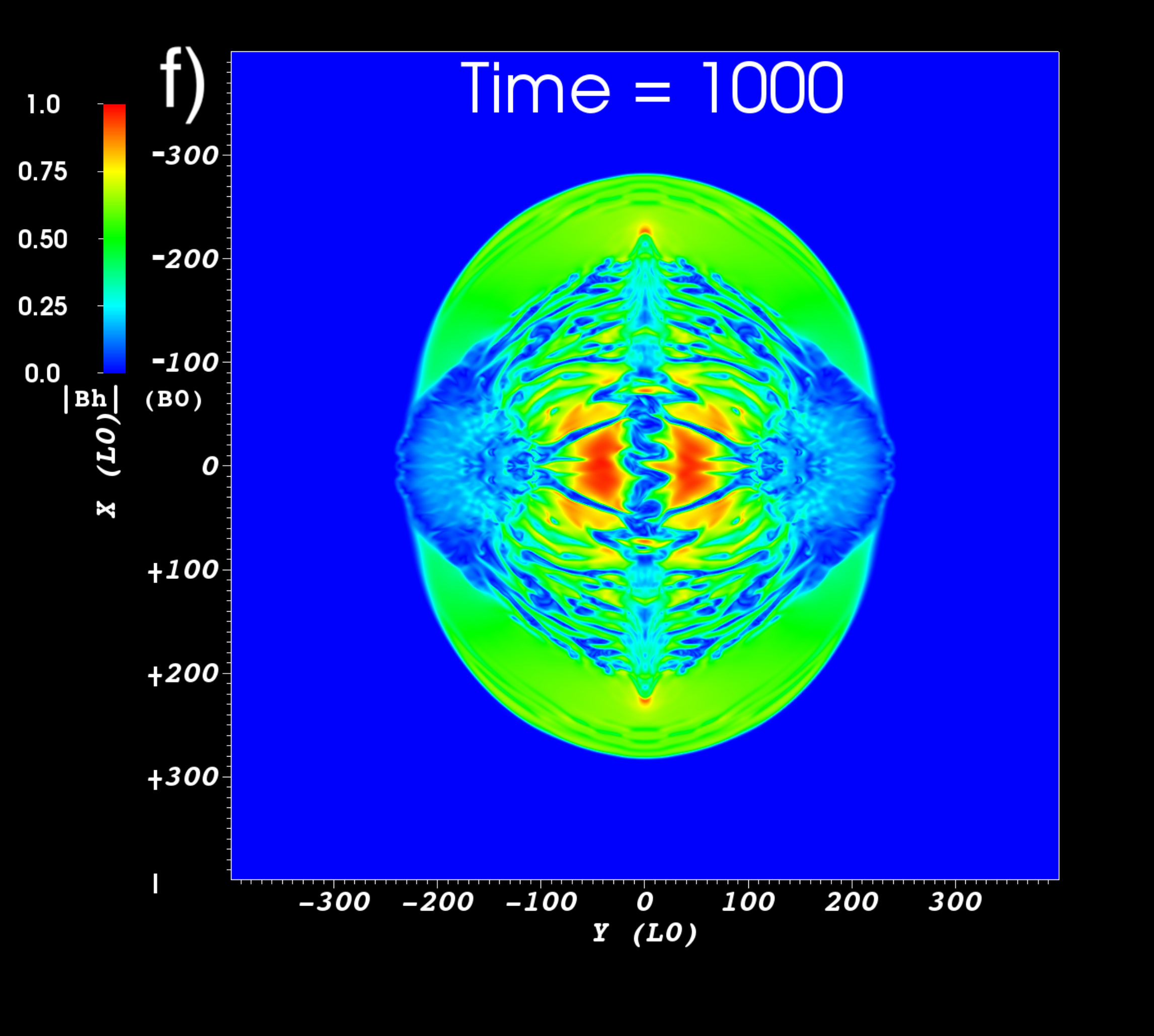}
\caption{Photospheric $B_z$ at different times throughout the simulation. The red line in panel c) shows the location of the cut analyzed in Figure \ref{fig:perpbzcut}. The last panel shows the magnitude of the in-plane magnetic field \label{fig:photosphere0}.}
\end{figure*}

\autoref{fig:photosphere0} shows the photospheric magnetic field distribution at several times during the simulation. Around $t=776$, the flux rope first appears at the photosphere as a pair of triangular shaped opposite polarity regions, enclosed by a circular band of flux. By $t=812$, these polarities are in the process of fragmenting, and two small perturbations have developed between them, resembling a pair of crescents in the photosphere. Around $t=860$, the entire region is overwhelmed by narrow lanes of adjacent positive and negative flux. The primary polarities are basically no longer visible at this time. The narrow lanes, in turn, grow significantly, forming many extended lanes on the photosphere. Many of these lanes are oriented at $\pm 45^\circ$ to the $x-y$ axes. The main polarities are now located around  $(x,y)$ = $(0,\pm r_0)$, but are difficult to distinguish. In other words, the distance between these two primary polarities is approximately equal to the diameter of the initial toroidal flux rope. The photospheric field shows perturbations growing over the next $\approx 140\;t_0$, with the separation of the two primary polarities merely increasing slightly, but becoming more pronounced. The horizontal magnetic field magnitude $|B_h|=\sqrt{B_x^2+B_y^2}$, shown in \autoref{fig:photosphere0}f, shows that the strongest horizontal field is near the central region where the field is mostly trapped just below the surface (see Section \ref{sec:CoronalEvolution}). \par 
\subsection{Coronal Evolution}\label{sec:CoronalEvolution}
\begin{figure*}
\includegraphics[width=0.5\linewidth]{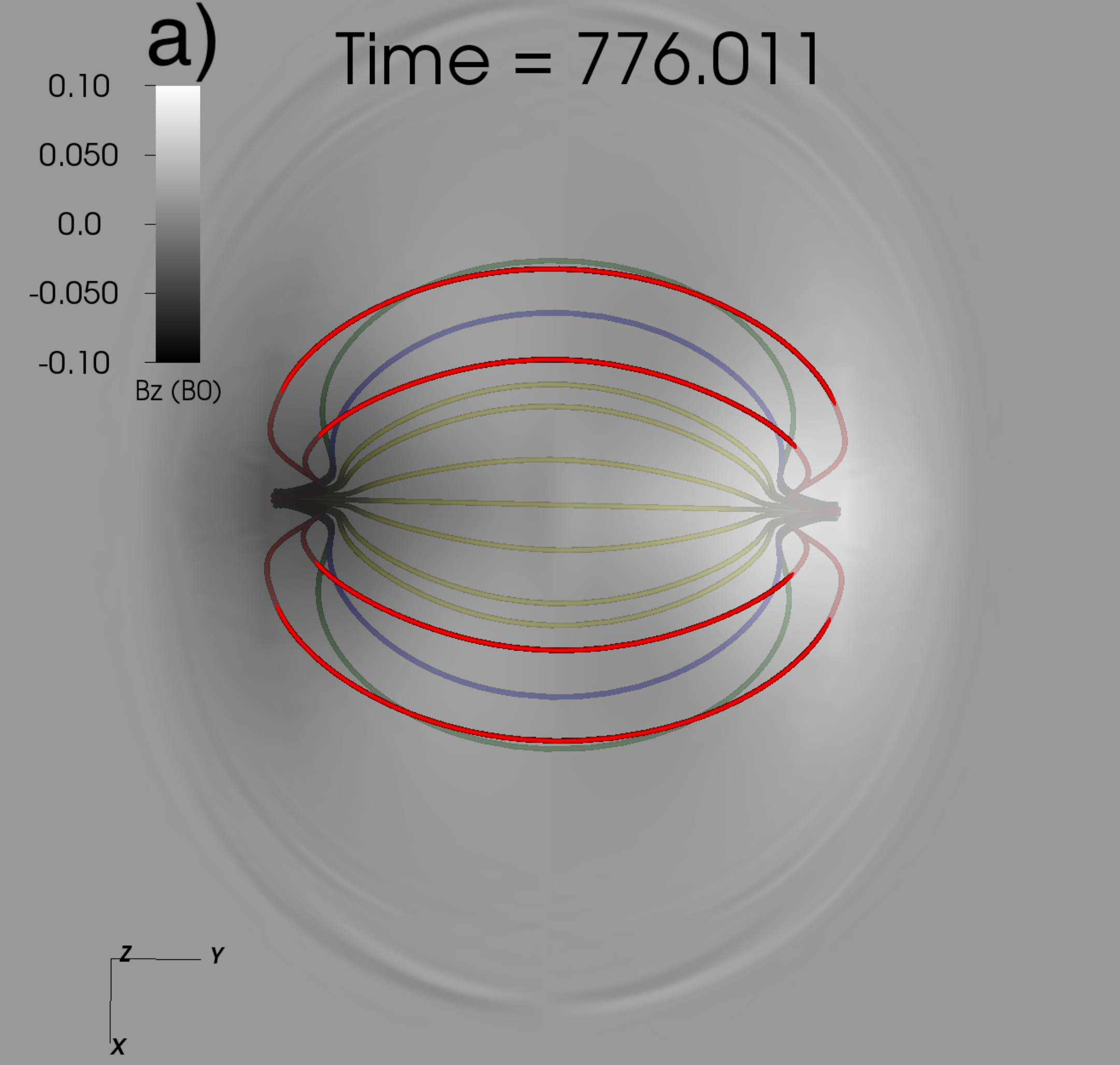}
\includegraphics[width=0.5\linewidth]{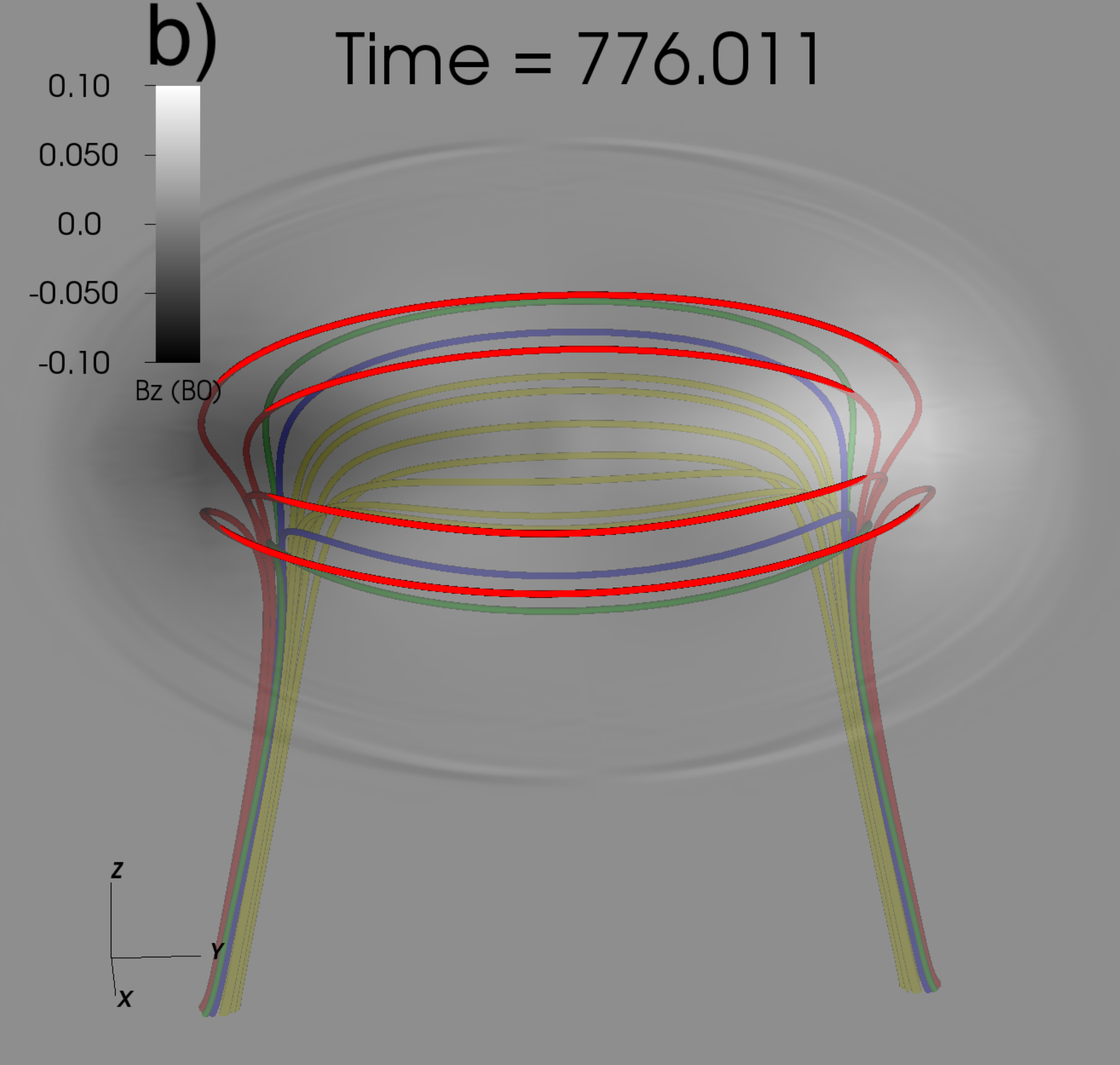}
\newline
\includegraphics[width=0.5\linewidth]{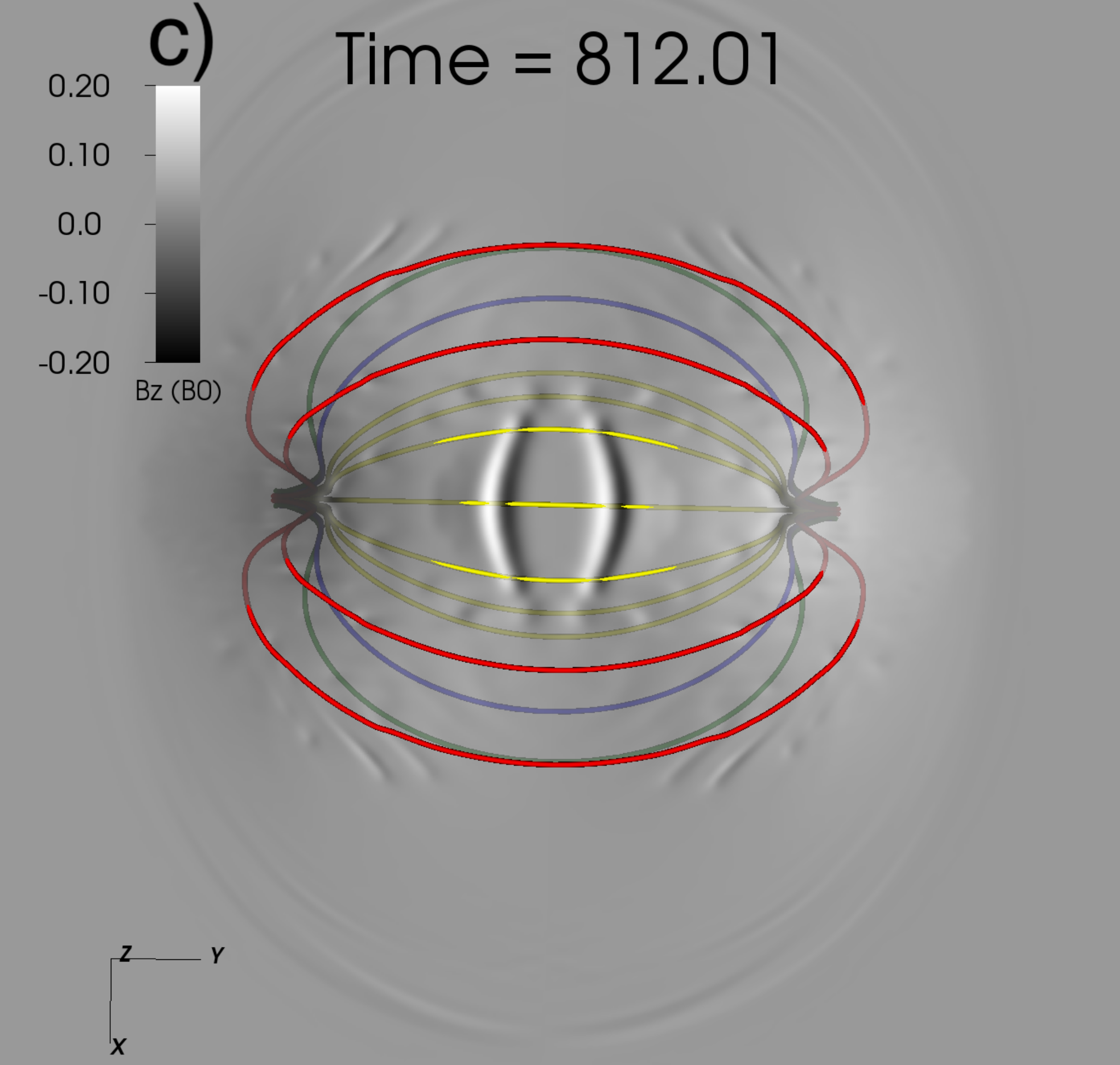}
\includegraphics[width=0.5\linewidth]{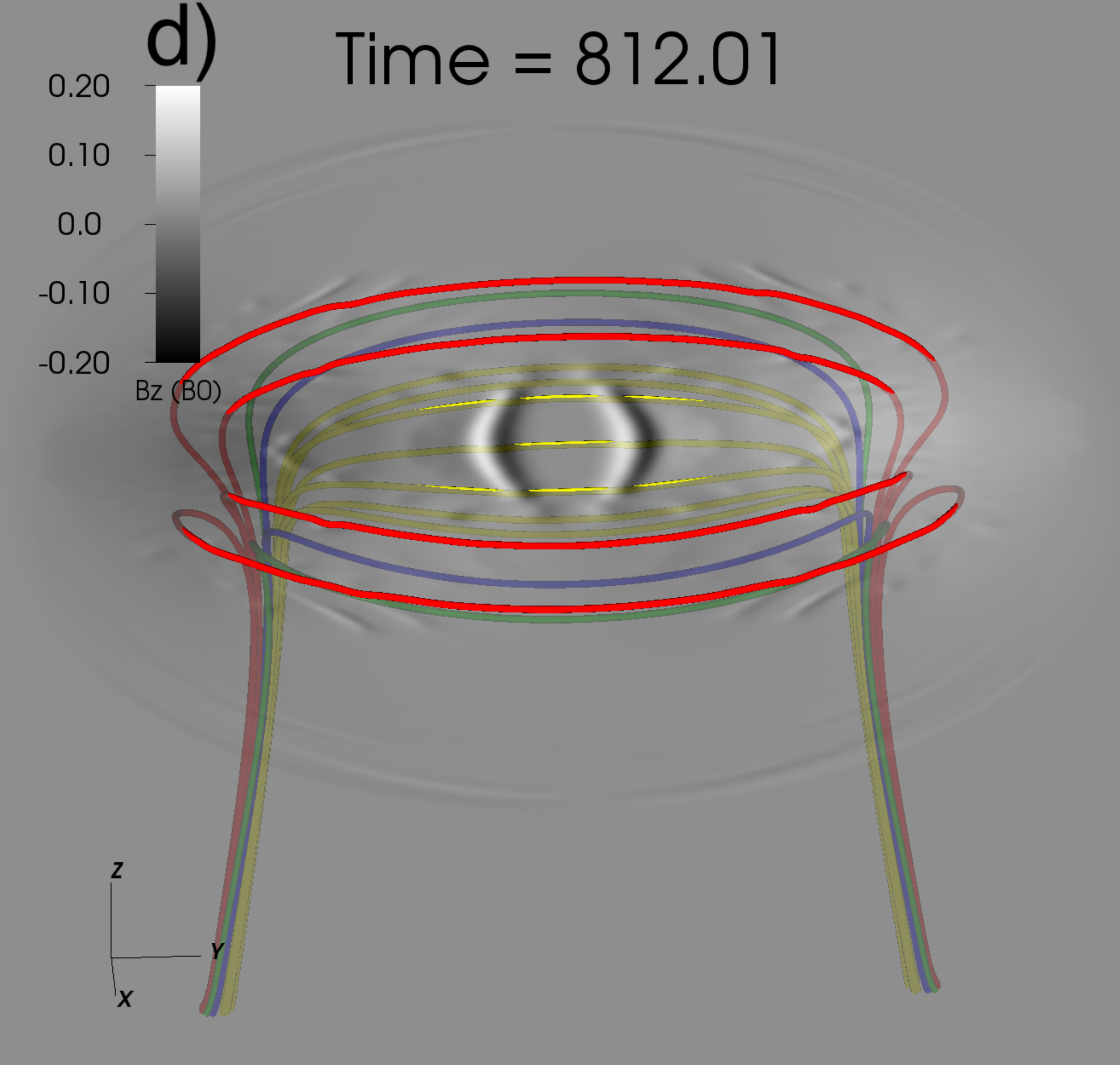}
\caption{Field lines, overplotted on photospheric magnetograms, at various stages early in the simulation. The seed points for these field lines are the same as in \autoref{fig:fieldlines0before2}. \label{fig:fieldlines0before1}}
\end{figure*}

\begin{figure*}
\includegraphics[width=0.5\linewidth]{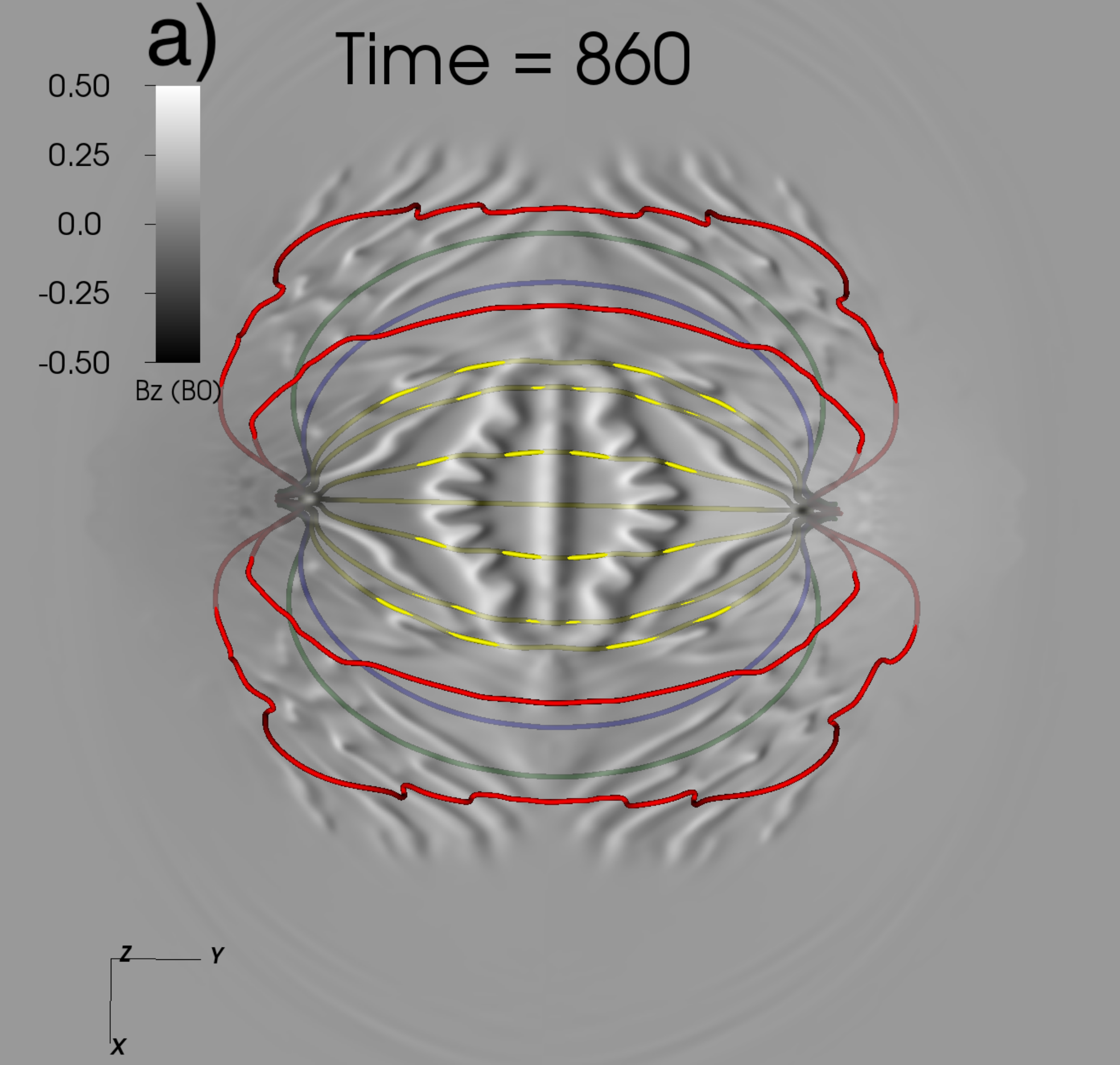}
\includegraphics[width=0.5\linewidth]{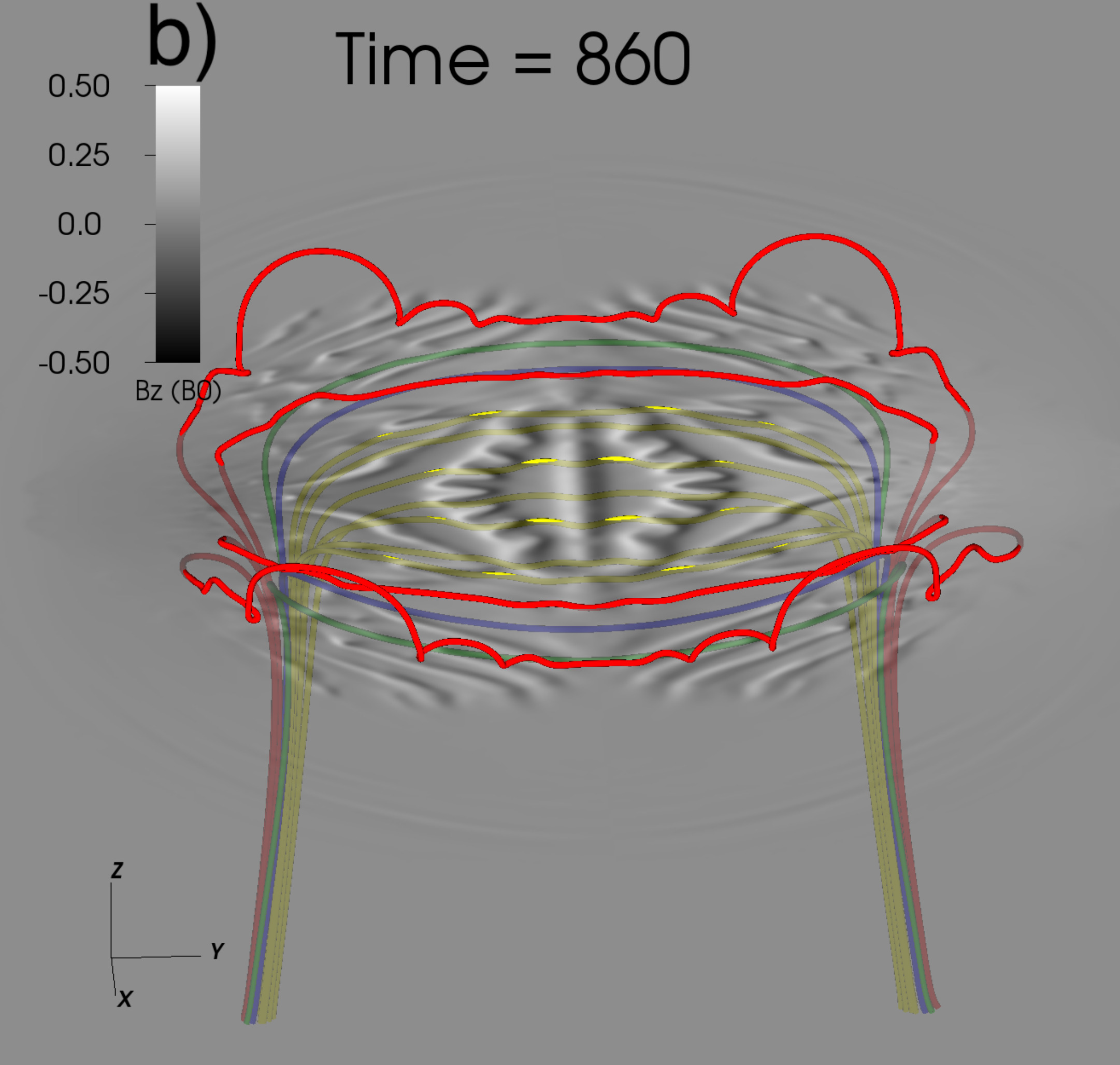}
\newline
\includegraphics[width=0.5\linewidth]{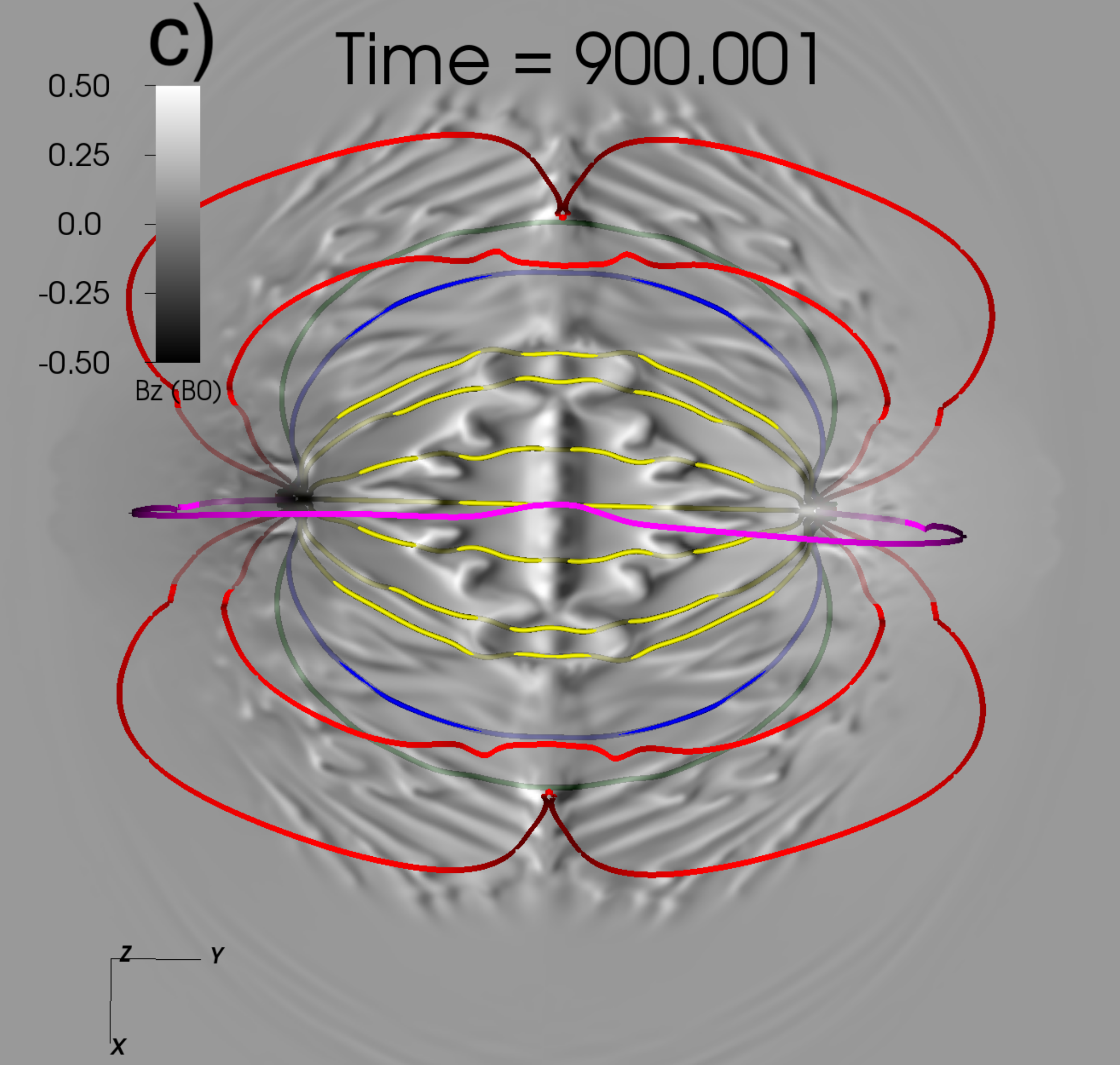}
\includegraphics[width=0.5\linewidth]{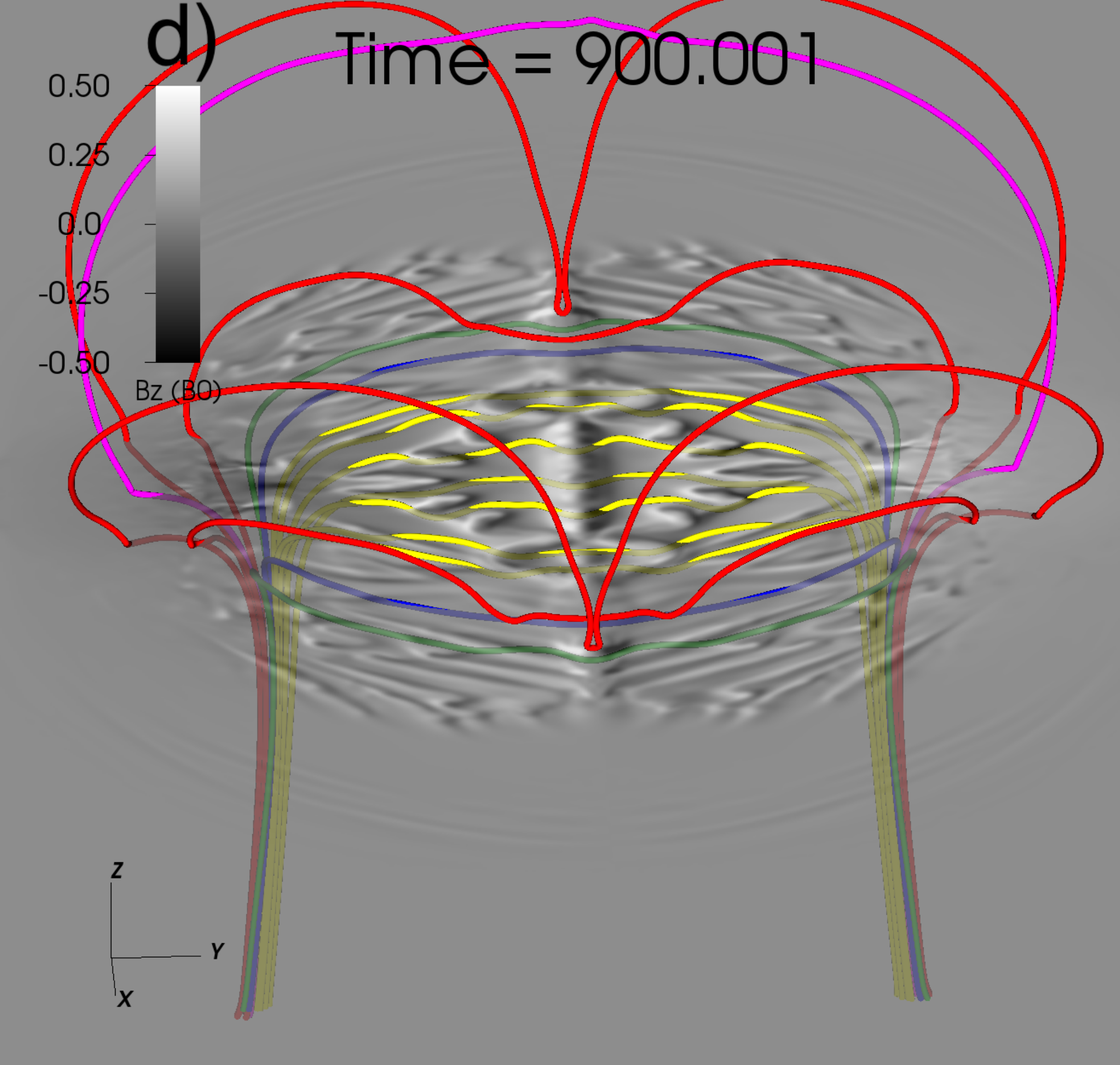}
\caption{Field lines, overplotted on photospheric magnetograms, at various stages in the middle of the simulation. The seed points for these field lines are the same as in \autoref{fig:fieldlines0before1}. \label{fig:fieldlines0before2}}
\end{figure*}

Figures \ref{fig:fieldlines0before1}-\ref{fig:fieldlines0before2} show field lines in the emerging flux rope as it rises through the convection zone and photosphere. In these figures, field lines are traced from inside each of the legs of the torus on the bottom line tied boundary. The color of the field line denotes the different behaviors evident in the Figure, with the seed points being identical for each panel. In these figures, the left panels are viewed from nearly directly above, while the right panels are viewed close to side-on. When the flux rope initially emerges (panels a-b of \autoref{fig:fieldlines0before1}; $t=776\;\mathrm{t_0}$), the red field lines are so curved that they are able to rise just above the photosphere, forming a bipolar active region. Blue and green field lines flatten out sideways as the flux rope approaches the photosphere, and the yellow field lines are coming up after them. In panels c-d ($t=812\;\mathrm{t_0}$), the red field lines have expanded to form lobes in the corona, and the blue and green field lines have still not emerged but have, instead, piled up near the photosphere. The yellow field lines undulate in and out of the photosphere, forming the crescents that are seen in the photospheric field. \par 
In panels a-b of \autoref{fig:fieldlines0before2} ($t=860\;\mathrm{t_0}$), the yellow field lines in the middle of the flux rope undulate, in a serpentine manner, into and out of the photosphere while the red field lines do the same just above the photosphere. These are highly reminiscent of the sea-serpent topology thought to be responsible for Ellerman bombs \citep{Pariat04,Pariat06,Isobe07,Pariat09,Archontis09b,Danilovic17}, which are associated with emergence resulting from the undular instability \citep{Isobe07}. In panels c-d ($t=900\;\mathrm{t_0}$), the emergence continues via the formation of a pair of magnetic lobes, evident in the red field lines, which manage to rise higher than the other field lines, possibly due to a series of reconnection events. Reconnection between these lobes, discussed below in \autoref{sec:interaction}, results in the formation of overlying (pink) field. Note that the side-on view shows that red field lines are really close to the photosphere at $t=860\;\mathrm{t_0}$ (\autoref{fig:fieldlines0before2}b), but reach much higher up at $t=900\;\mathrm{t_0}$ (\autoref{fig:fieldlines0before2}d). The central portions of these field lines are prevented from rising, as they are supporting heavy photospheric plasma, which is unable to drain.  Meanwhile, the blue and green field lines have remained flattened sideways. \par 
The evolution of this simulation is markedly different from that shown for the smaller scale, weakly twisted flux rope shown in \citet{MacTaggart09}, where two well-defined, large polarities form at the photosphere, with very little to no structure in between. At the coronal level, \citet{MacTaggart09} find a simple sheared arcade between the two polarities, whereas this simulation displays a series of $\mathcal{M}$ shaped loops connecting the two polarities that correspond to the flux rope's legs. We postulate below that this difference is due to the relative horizontal extent of the simulated emerging regions compared to the $170\;\mathrm{km}$ photospheric pressure scale height.\par  
\par 
\begin{figure*}
\includegraphics[width=0.5\linewidth]{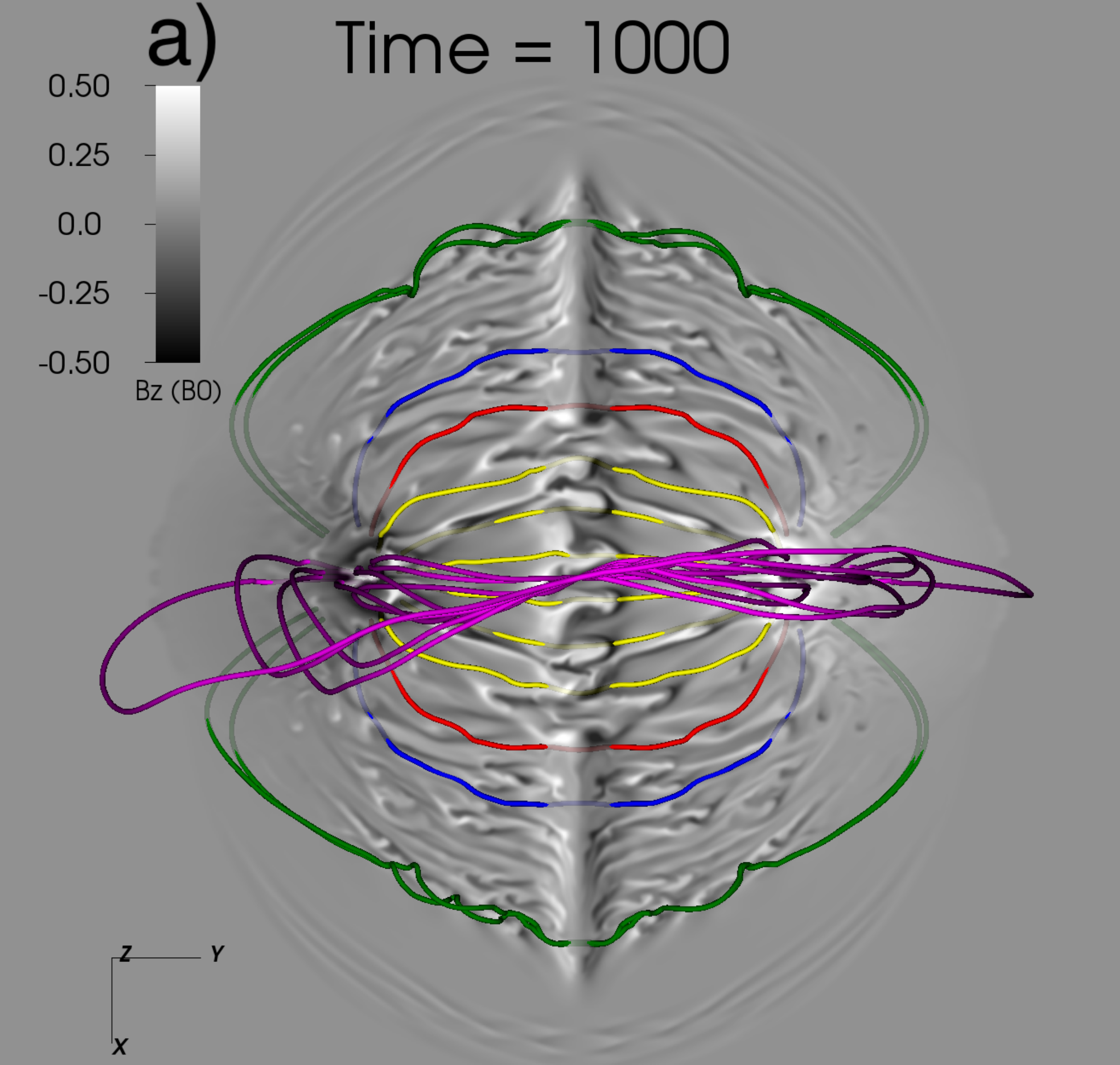}
\includegraphics[width=0.5\linewidth]{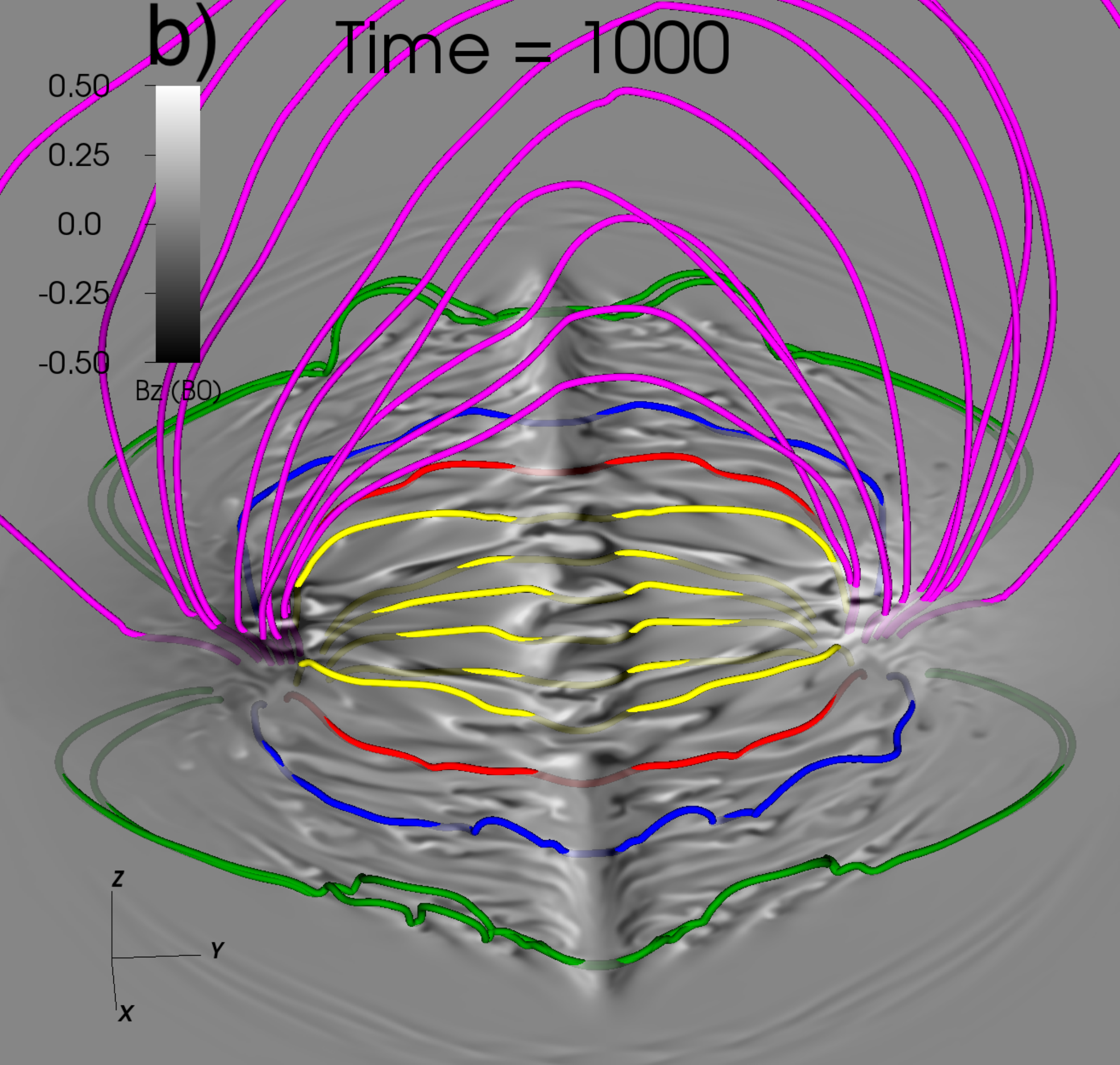}
\caption{Field lines, overplotted on semi-transparent photospheric magnetograms at the end of the simulation, after the emerged lobes haves interacted (cf. \ref{sec:interaction}).\label{fig:fieldlines0after}}
\end{figure*}

\autoref{fig:fieldlines0after} shows the photospheric field and sample field lines at the end of the simulation ($t=1000\;\mathrm{t_0}$). As will be described in \S \ref{sec:interaction}, by this point the magnetic lobes seen in panels c-d of \autoref{fig:fieldlines0before2} will have reconnected with each other, causing a reorganization of the magnetic field. Thus, the original field lines plotted above have changed their connectivities. To best represent the coronal field, the field lines shown here are different than the ones shown in \autoref{fig:fieldlines0before1}-\autoref{fig:fieldlines0before2}. At this stage, some remnants of the pre-interaction behavior can still be observed. The yellow, red, green and blue field lines are spread outwards, undulating in and out of the photosphere, connecting the two primary polarities. Overlying the entire flux system are a set of pink field lines, which result from reconnection of the magnetic lobes (described below, in \autoref{sec:interaction}). \par

\subsection{Comparison With Observations}
An interesting feature of this simulation, seen in \autoref{fig:photosphere0}, is the salt-and-pepper nature of the photospheric magnetogram. It is reminiscent of observational signatures, which are often attributed to the presence of convection breaking up emerging flux bundles \citep{Kitiashvili15}, which then coalesce back into large scale polarities \citep{Dacie16}. Such salt-and-pepper features have been observed in convectively unstable simulations \citep{Rempel14}, but, to date, have not been seen in convectively stable simulations \citep[\emph{e.g.,} ][]{Leake13}, which typically form coherent polarity concentrations. A characteristic quantitative feature of such salt-and-pepper magnetograms is that the flux strength distribution obeys a power law with a slope in the range $[-1.6,-1.5]$ between $10^1$-$10^3\;\mathrm{G}$ early in the active region's lifecycle, though slopes as steep as $-2.0$ and as shallow as $-1.3$ were also measured \citep{Dacie16}.
In view of these results, \citet{Dacie17} analyzed multiple simulations of emerging flux ropes and found that the convectively-stable simulations from \citet{Leake13} and \citet{MacTaggart09}, with buoyant twisted cylindrical and toroidal flux ropes, respectively, produced flux distribution slopes in the correct range only during the later phases of emergence. Those simulations did not show the signatures of the undular instability observed here. Meanwhile, the convective simulation of \citet{Rempel14} was able to generate a power law over the range $10^1-10^3\;\mathrm{G}$ in field strength, with slopes in the range $[-1.4,-1.0]$ over the entire duration of the emergence. Thus it seems that convective simulations produced the proper range of large and small field strengths while buoyant twisted flux ropes without convection could produce the proper range of small and large field strengths only during the later phases of emergence, when the primary polarities had become coherent. Since our simulation shows qualitative differences from the coherent emergence of \citet{Leake13} and \citet{MacTaggart09}, an interesting question is what is the photospheric flux distribution in our simulation? \par
\begin{figure*}
\includegraphics[width=0.5\linewidth]{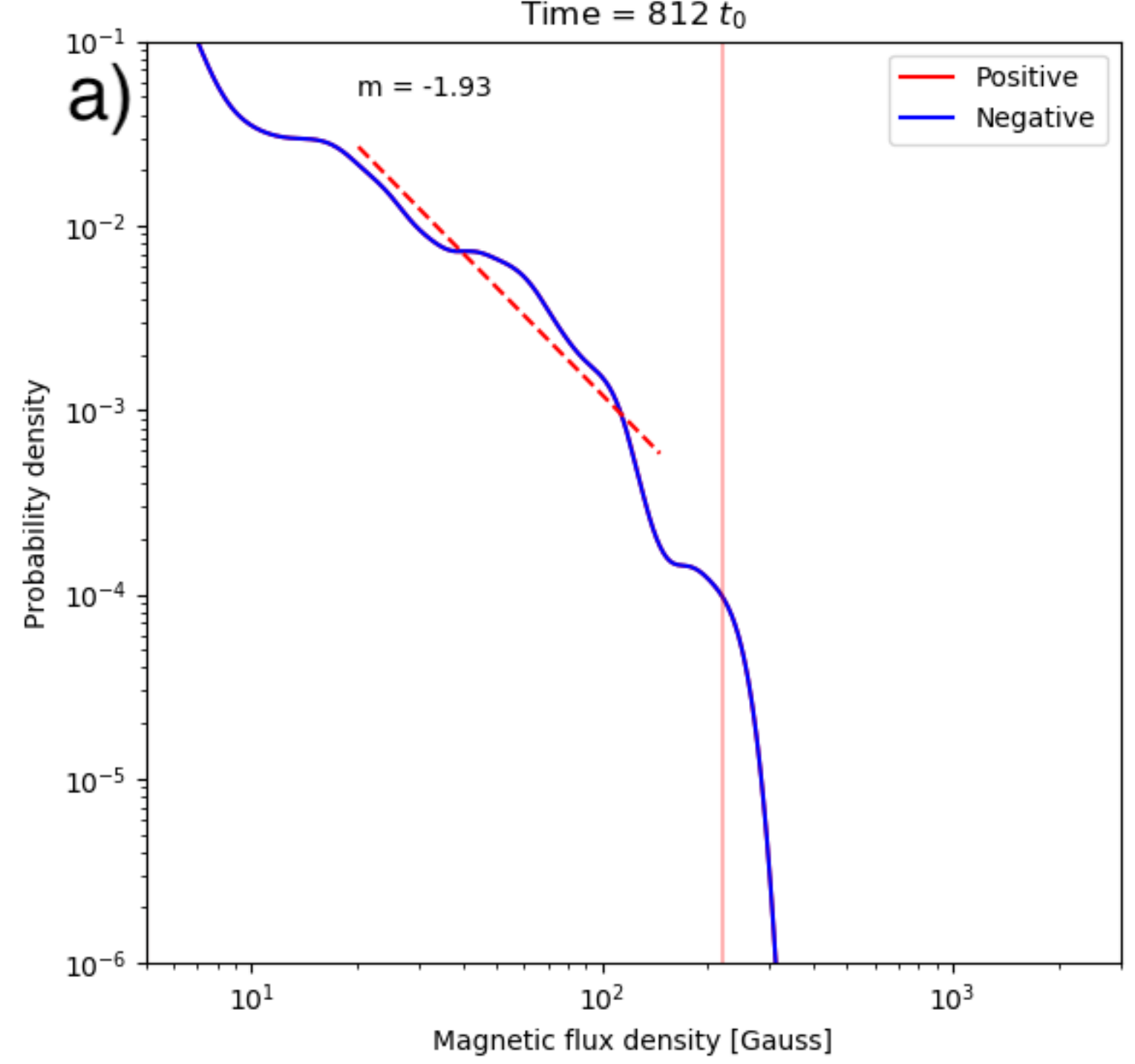}
\includegraphics[width=0.5\linewidth]{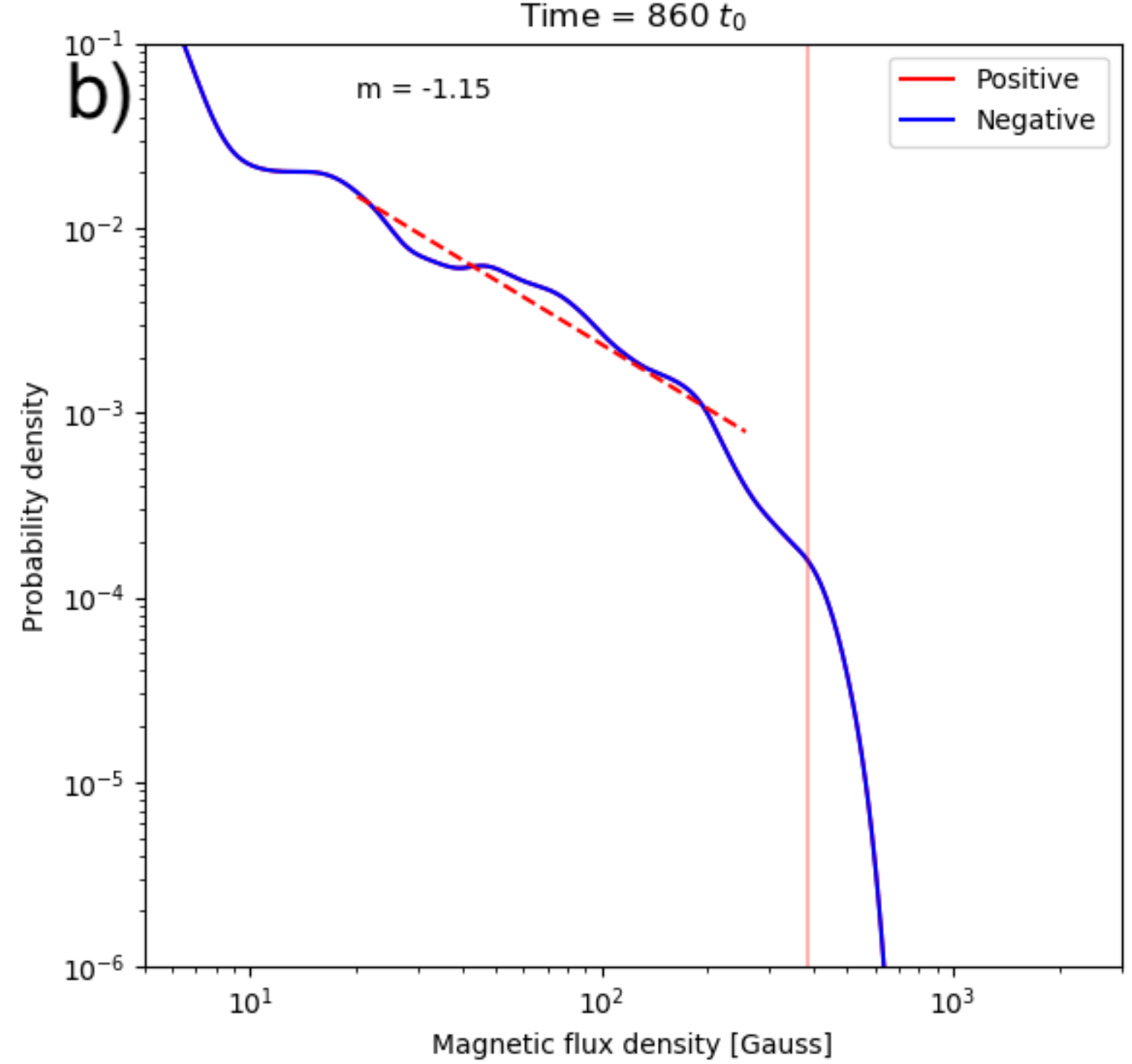}
\includegraphics[width=0.5\linewidth]{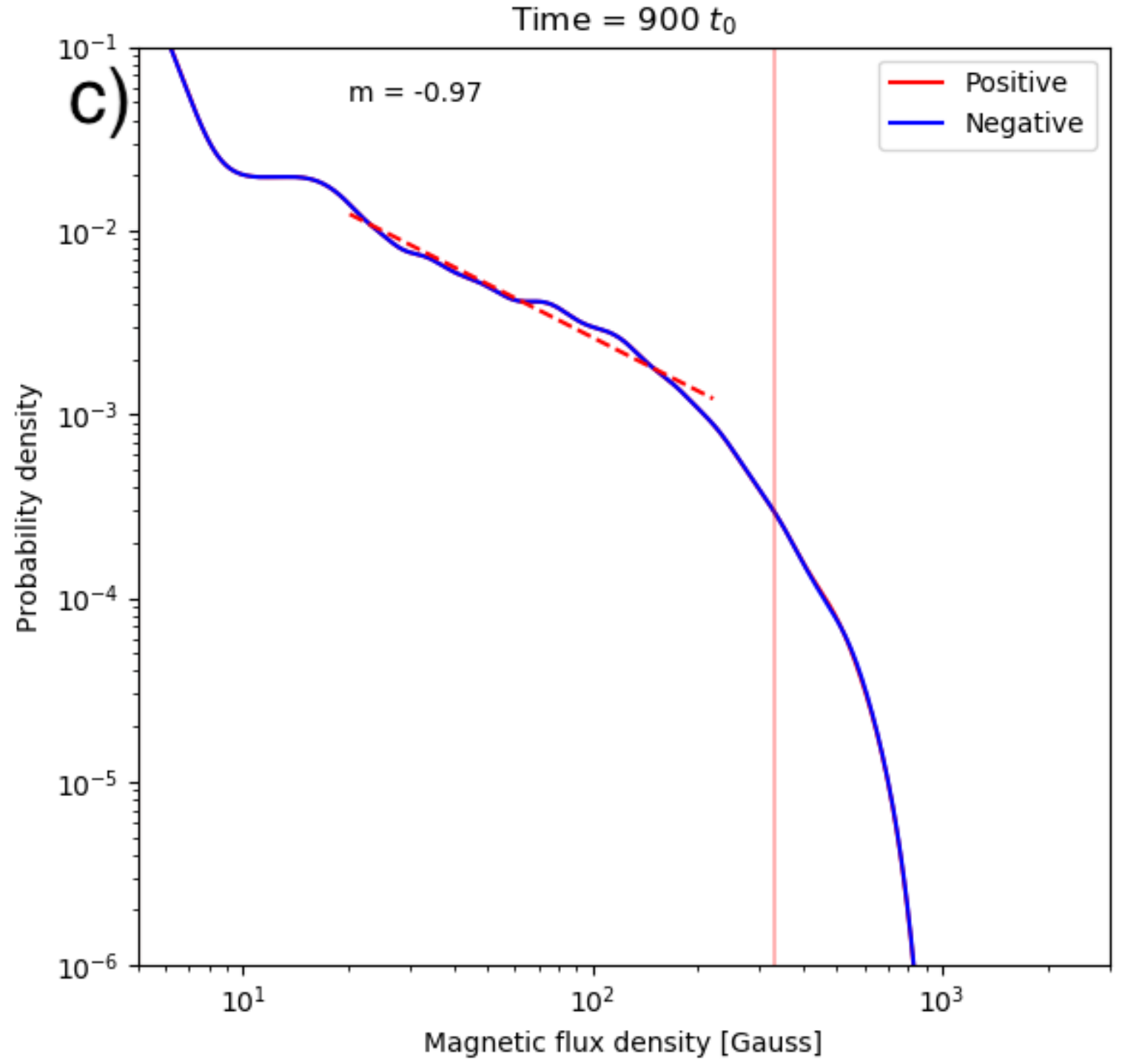}
\includegraphics[width=0.5\linewidth]{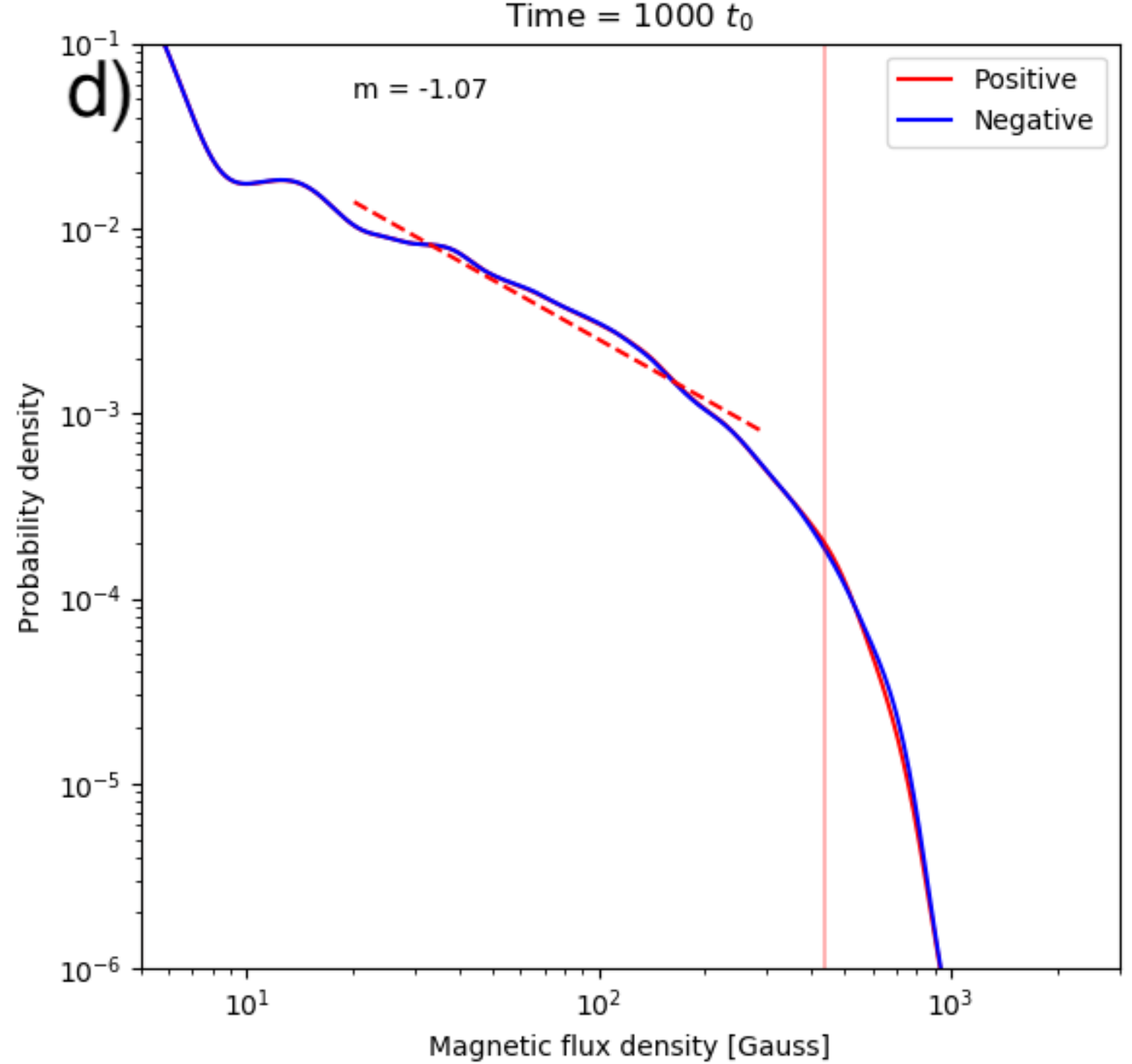}
\caption{Following \citet{Dacie17}, the probability distribution of magnetic flux at several times during the simulation. The vertical red line shows the value of the magnetic flux density $B_{-3}$ where the derivative of the probability density curve reached $-3$, and the slope was fitted between $20\;\mathrm{G}$ and $\frac{2}{3}B_{-3}$.} \label{fig:Dacie}
\end{figure*}

\begin{figure}
\includegraphics[width=\linewidth]{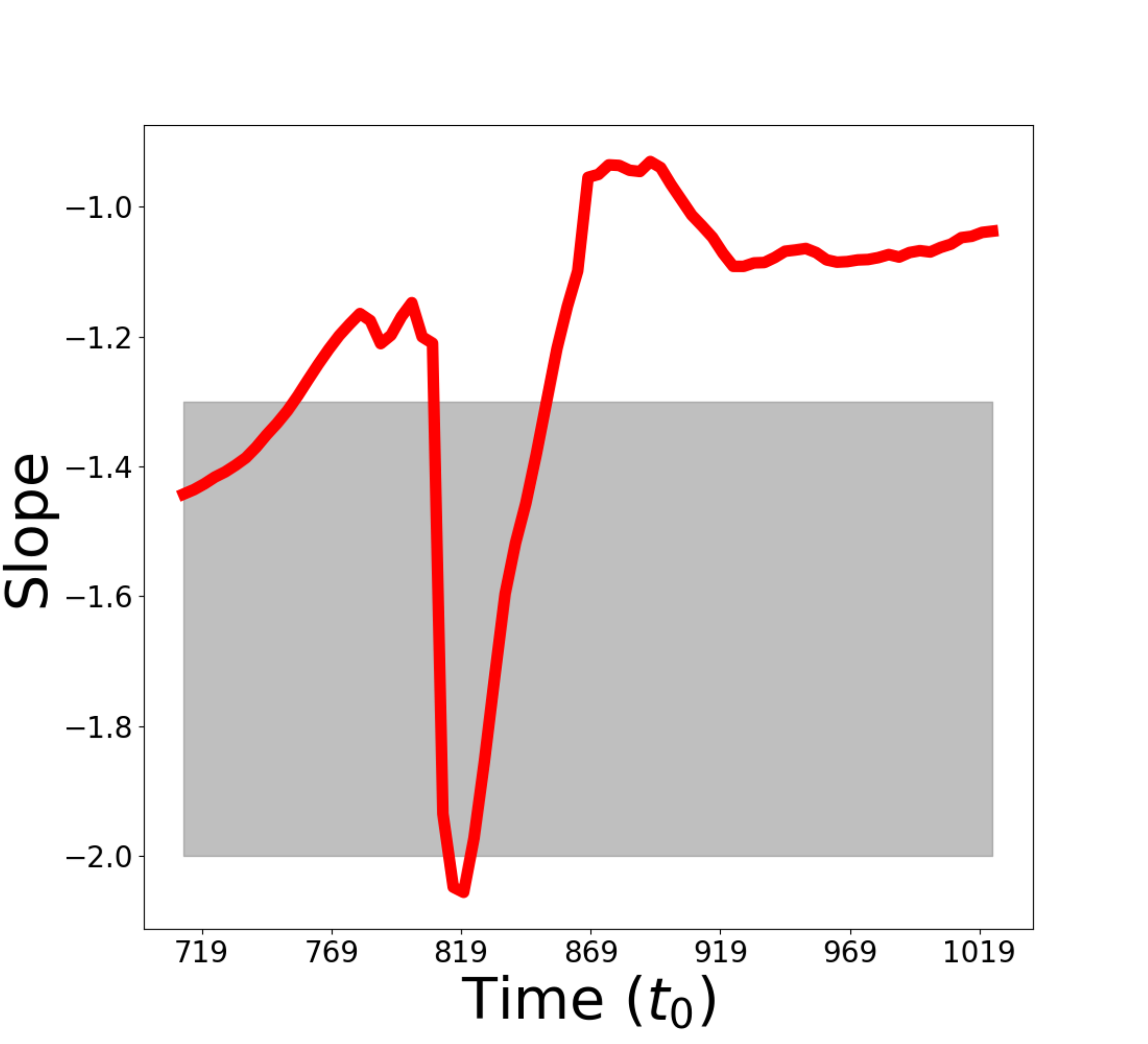}
\caption{Slope as a function of time for the magnetic field distribution during the simulation. Several snapshots of the distribution are shown in \autoref{fig:Dacie}. The grey shaded region is the approximate range of slopes seen in the observational analysis of \citet[cf. Figure 2]{Dacie17}.} \label{fig:Dacie_slope}
\end{figure}

 
We performed a kernel density estimation \citep{Dacie16,Dacie17} on this simulation by converting our magnetic field values to Gauss (cf. Equation \ref{magscl}), and we plot the results in \autoref{fig:Dacie}-\autoref{fig:Dacie_slope}. At each time, we fit a line to the distribution between $20\;\mathrm{G}$ and $\frac{2}{3}B_{-3}$, where $B_{-3}$ is defined as the magnetic field strength where the slope of the distribution curve exceeds $-3$ (vertical red line in \autoref{fig:Dacie}; i.e., the dashed red line is fit up to $2/3$ of the value at which there is an approximate knee in the distribution). The temporal profile of the slope of this distribution is shown in \autoref{fig:Dacie_slope}. The grey area denotes the range of slopes observed by \citet{Dacie17}.  
%At $t=548\;t_0$, during the emergence phase, the power law with slope of $-0.75$ extends only over a couple of Gauss, and would probably have been steeper if it were calculated from $20\;\mathrm{G}$. 
As can be seen from these figures, the slope of the distribution is relatively steep only during the very early time period, around $t=800\;t_0-830\;t_0$, corresponding approximately to the linear phase of the undular instability (cf. Section \ref{sec:MBI}). After this early phase, the slope rapidly becomes less steep, getting to values of approximately $-1$ for the rest of the emergence. 
%The large scale fields that are measured in this simulation appear only at the later stages of emergence when the flux rope legs are able to appear at the photosphere. 
Thus we find that even in the absence of convection, small-scale salt-and-pepper structure that resemble observations can be produced during the linear phases of emergence via the undular instability. 
%In addition, our simulation is also able to reproduce the large magnetic field strengths seen in observations and - previously - in numerical simulations with either twist or convection \citep{MacTaggart09,Leake13,Dacie17}. This is due to the fact that, at later times in our simulation, the legs of the flux rope also emerge through the photosphere, resulting in these large field strengths. Our results, however, demonstrate that both the small- and large-scale magnetic fields seen in observations during the emergence of an untwisted flux rope can be produced without convection via the growth of the undular instability.

\subsection{The Undular Instability}\label{sec:MBI}
The above results demonstrate that the photospheric signatures seen in our simulation exhibit small scale, salt-and-pepper structure, arising as a result of field lines dipping in and out of the photosphere. This behavior is largely due to the fact that the emergence process simulated here occurs as a result of the undular instability, also known as the magnetic buoyancy instability (MBI), the theory of which has been studied by numerous authors \citep{Parker55,Chandra61,Newcomb61,Gough66,Parker66,Parker69,Gilman70,Tayler73,Acheson78a,Acheson78b,Acheson79,Spiegel82,Spruit82,Hughes85,Hillier16,Hillier18}. 
%In contrast, the hydrodynamic interchange mode of the MBI satisfies $\veck\cdot\vecB=0$. 
The MBI mixes plasma above and below the photosphere, and keeps large portions of the magnetic field weighed down with plasma. The instability criterion \citep{Acheson78a,Acheson78b,Acheson79,Murray06,Archontis13} for the undular instability is given by\footnote{Note that other authors \citep{Newcomb61,Gilman70,Tayler73,Fan01a} give a different form of the instability criterion. However, \citet{Acheson78b} argue that these alternate forms of the criterion are equivalent. \citet{Chandra61} derives a different criterion in the Boussinesq approximation, and \citet{Spruit82} find a slightly simplified form of the instability criterion using the thin tube approximation. Neither the Boussinesq nor the thin tube approximations are valid in our simulation.}

\beg{fullinstabilitycriterion}
-H_p\frac{1}{|\vecB|}\pd{|\vecB|}{z} > -\frac{1}{2}\beta\gamma\delta + H_p^2k_{||}^2\Big(1+\frac{k_z^2}{k_\perp^2}\Big),
\done 
Here\footnote{The second term on the right hand side of \autoref{fullinstabilitycriterion} is given incorrectly in \citet{Archontis04} and \citet{Archontis13}. See discussion in \citet{Hood12}. } $H_p=RT_0/g$ is the photospheric pressure scale height, $|\vecB|$ is the magnitude of the magnetic field, $\beta$ is the ratio of plasma- to magnetic pressure, $\gamma=5/3$ is the ratio of specific heats, and 
\beg{defdelta}
\delta = \frac{d\log T}{d\log P} - 1 + \frac{1}{\gamma}
\done 
is the superadiabatic excess, and is approximately $-0.4$ in the photosphere. Perturbations have wave-vector $\veck$, where $k_{||}$ and $k_\perp$ are the horizontal components parallel and perpendicular to the magnetic field and $k_z$ is the vertical component. Both \citet{Murray06} and \citet{Archontis13} find that the second term on the right hand side is negligible compared to the the first term, so that \autoref{fullinstabilitycriterion} can be simplified and written \citep[see, e.g.,][]{Leake13b}:
\beg{MBI}
\frac{\xi}{\chi} > 1,
\done 
where 
\beg{MBIxi}
\xi = - \frac{H_p}{|\vecB|}\pd{|\vecB|}{z},
\done 
and
\beg{MBIchi}
\chi = -\frac{1}{2}\beta\gamma\delta.
\done 
\par 
%If any concave down portion of the magnetic field is able to emerge through the photosphere, it will enable plasma to drain from the top of the concave down portion of the loop, and its weight will prevent the field below from emerging. At the same time, the buoyancy at the top of the loop increases, causing the instability to grow \citep{Acheson79}.

%This reflects the fact that for this instability, a field free plasma is being supported against gravity by a plasma with $|\vecB|\ne0$, which provides the upward pressure gradient necessary to balance against gravity. However, if the system is in both gravitational and thermal equilibrium, then the density inside the flux rope will drop.  
%If the vertical gradient of $|\vecB|$ is large enough, then the density of that plasma would not be large enough to support the more dense plasma above it against the downward pressure gradient, and the two plasmas will interchange, and the magnetic field will appear at the photosphere \citep{Acheson79}.

\citet{Archontis13} found that a weakly twisted, initially cylindrical flux rope emerges via the appearance of a pair of lobes of magnetic field when the instability criterion in Equation \ref{MBI} is satisfied at a height of $\approx0.3\;\mathrm{Mm}$ ($1.76L_0$) above the photosphere. In that simulation, magnetic flux rose from the convection zone and piled up near the photosphere until there was sufficient flux to exceed the instability criterion.  \par
\begin{figure*}
\includegraphics[width=\linewidth]{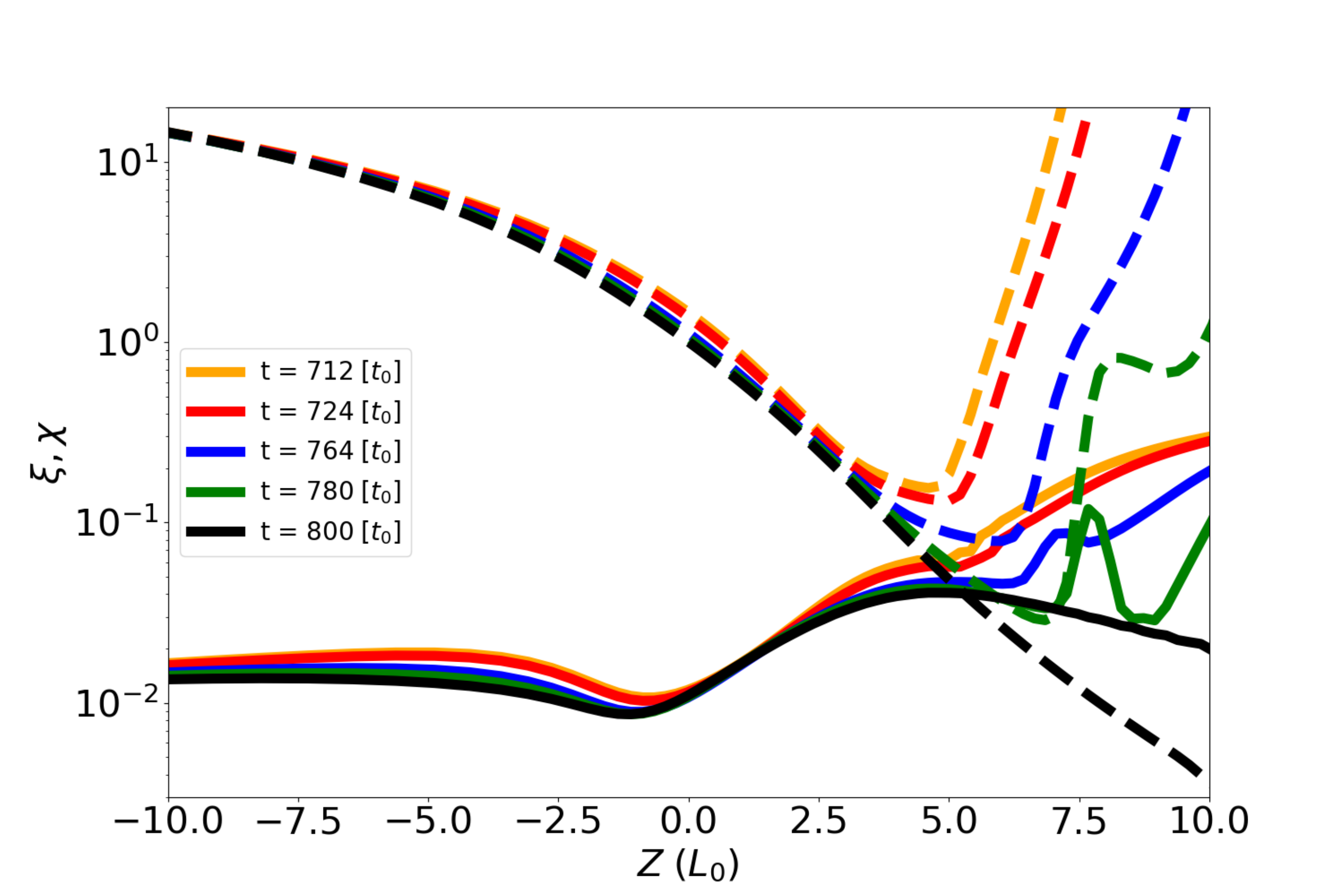}
\caption{The instability ($\xi$, solid) and stability ($\chi$, dashed) terms in the magnetic buoyancy relation at several times during the simulation near when the peak photospheric field strength first exceeds $0.05B_0$. The undular mode is unstable when $\xi$ exceeds $\chi$.}\label{fig:Parkerterms}
\end{figure*}

%This is a very different emergence scenario than is often assumed for moderately twisted flux ropes: namely, that the buoyant rise of the flux rope dynamically overshoots the photosphere, causing the active region to appear on the photosphere.\par 
For our simulation, we calculate the criterion in Equation \ref{MBI} near the time when the peak field strength on the photosphere exceeds $0.05B_0$, as well as at a later time, for reference. In \autoref{fig:Parkerterms}, we plot the destabilizing term $\xi$ and the stabilizing term $\chi$ at $x=y=0$ for $z$ near the photosphere around the time when undulations first appear at the photosphere. It is clear that $\xi/\chi$ goes from below $1$ to above $1$ near $t=800\;\mathrm{t_0}$ (black curve), at a height of around $z=z_i=5L_0$ ($0.85\;\mathrm{Mm}$). Here $z_i$ is defined as the height at which the ratio $\xi/\chi$ first reaches unity at $t=780\;\mathrm{t_0}$.\par 

\begin{figure*}
\includegraphics[width=0.33\linewidth]{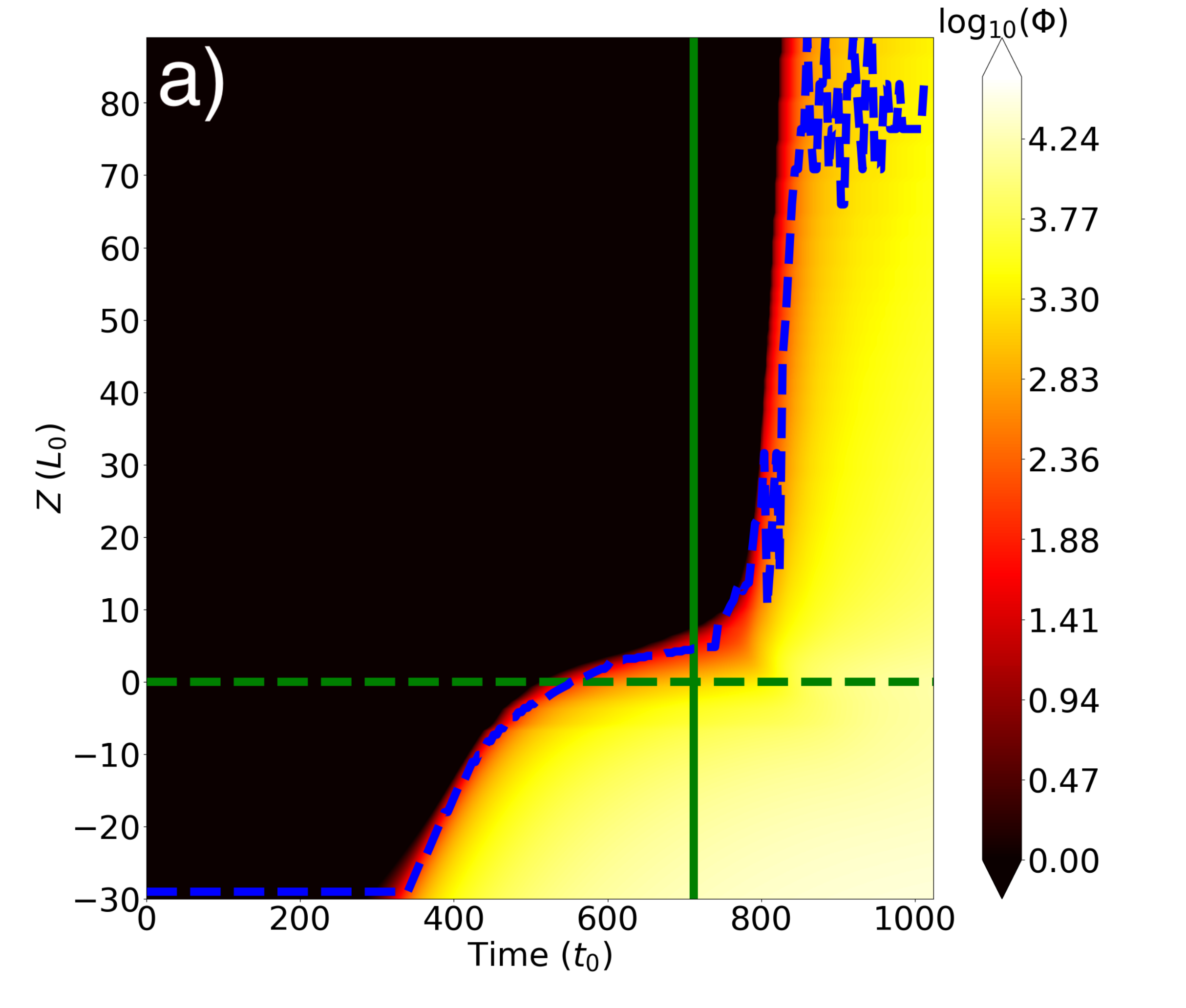}
\includegraphics[width=0.33\linewidth]{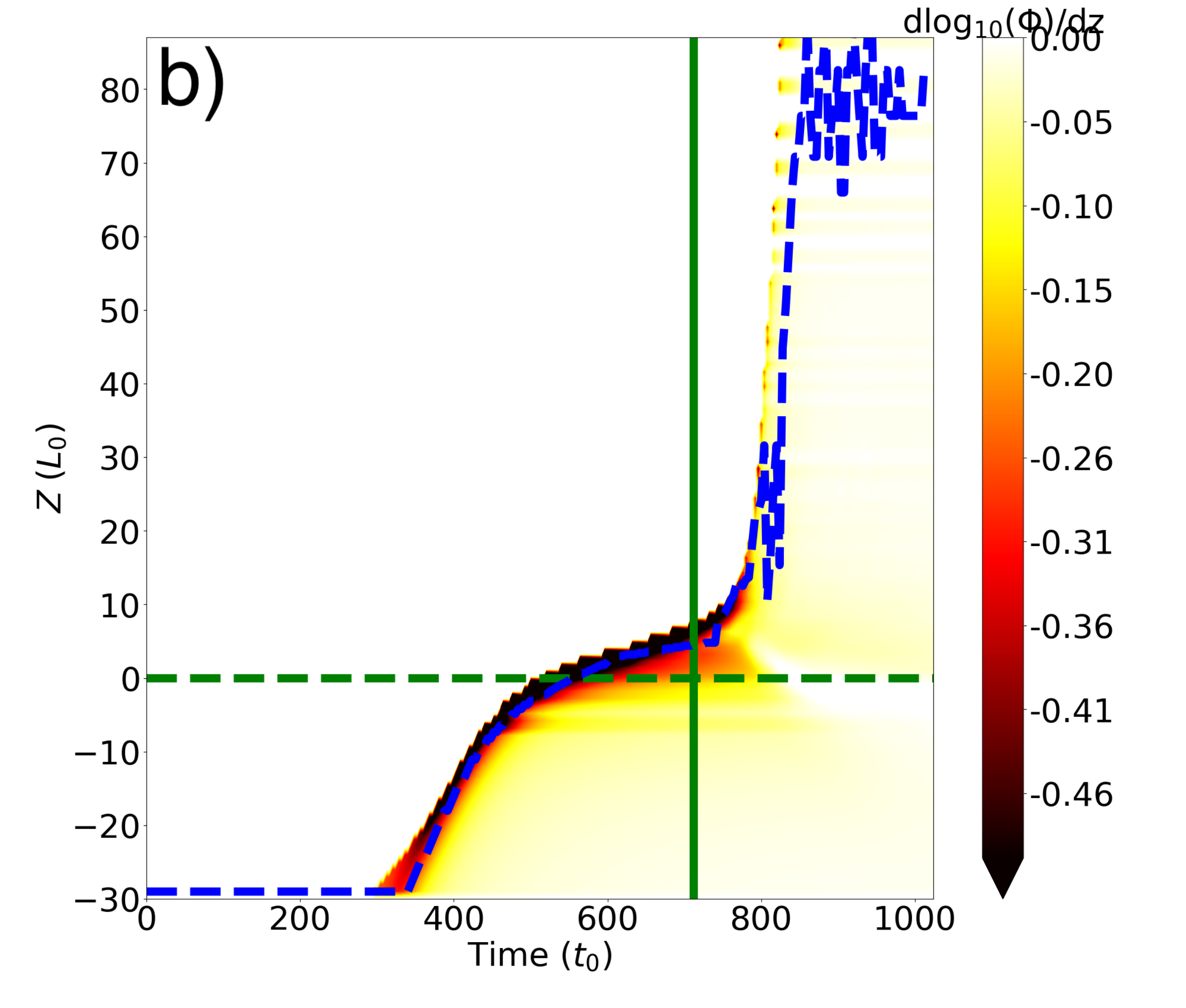}
\includegraphics[width=0.33\linewidth]{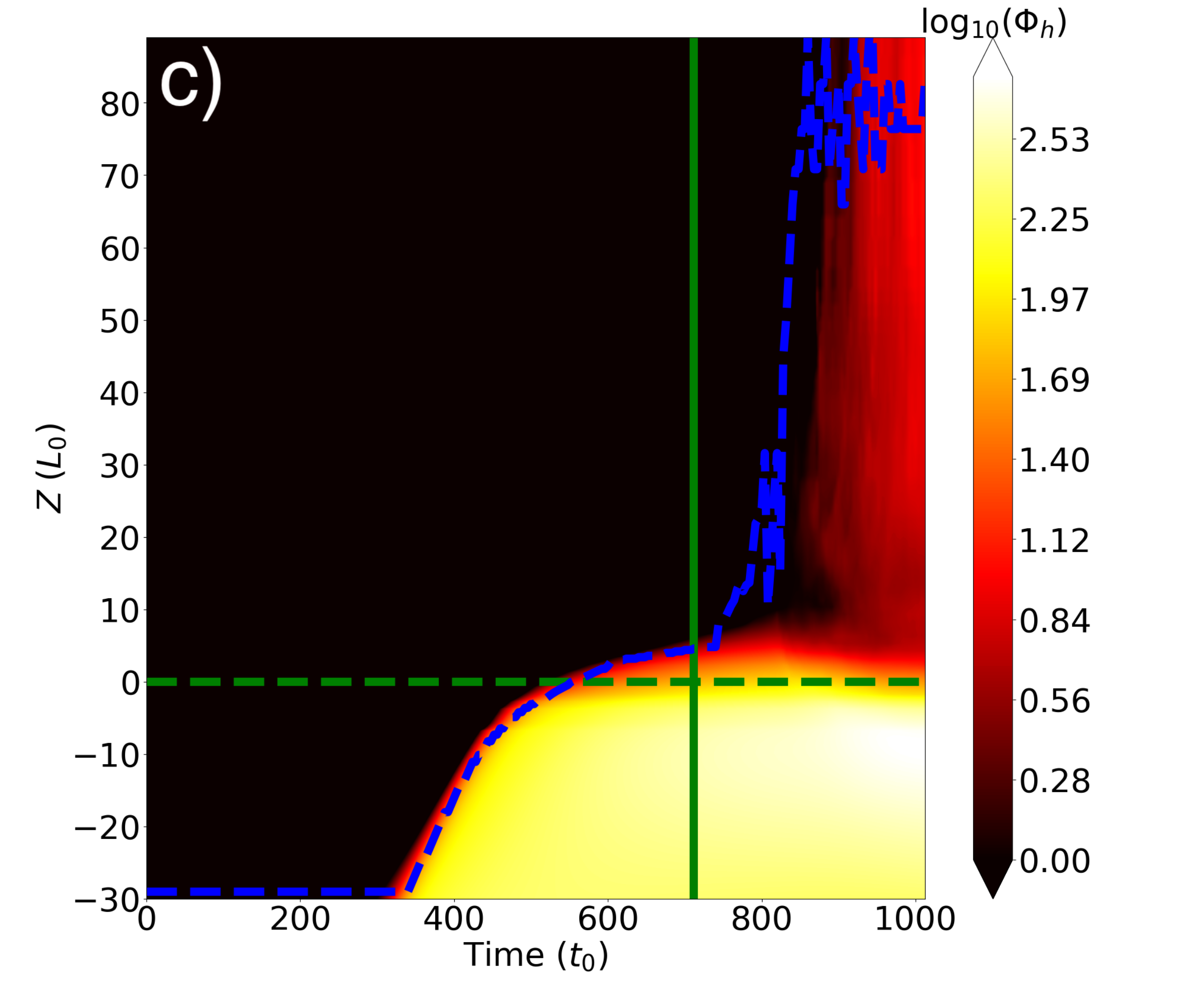}
\caption{Height-time diagrams of a) the unsigned vertical magnetic flux, b) gradient of the unsigned vertical magnetic flux, and c) the unsigned axial flux through the $y=0$ plane, showing a fraction of the simulation domain in the $z$-direction. The horizontal dashed line is at $z=0$ and the vertical solid line shows the time at which the peak photospheric field strength first exceeds $0.05B_0$. The dashed \textcolor{black}{blue} line follows the location of peak $\xi/\chi$ as a function of time.}\label{fig:heighttime}
\end{figure*}

To confirm this, we calculated the total vertical unsigned flux
\beg{Phiflux}
\Phi(z,t) = \int{dx\;dy\;|B_z(x,y,z,t)|},
\done 
at each height at each time step in our simulation, and made a height-time diagram of $\log_{10}(\Phi)$ and $d\log_{10}(\Phi)/dz$, shown in \autoref{fig:heighttime}. The flux rope is taken to emerge when the peak field strength at the photosphere (shown with the horizontal dashed line) exceeds a relatively modest $0.05B_0$, which occurs at a time marked by the vertical solid line, around $t=712\;t_0$. It is clear from the figure that a significant amount of flux appears above the photosphere before the instability criterion is exceeded. Furthermore, it appears that the flux rope decelerates as it passes near and through the photosphere, before accelerating upwards again as it expands into the corona, a process known as two-step emergence \citep{Cheung14}. The dashed blue line in the Figure tracks the temporal evolution of the height of the maximum value of $\xi/\chi$. Here it is seen that flux builds up above the location of peak $\xi/\chi$, at a height of around $z_i=5L_0$ above the photosphere --  slightly higher than the value found by \citet{Archontis13} -- before rising rapidly around $t=800$. In addition, in \autoref{fig:heighttime} we plot a height-time diagram of the flux of $B_y$ through the $y=0$ plane as a function of time:
\beg{Phihor}
\Phi_h(z,t) =  \int_{y=0}{dx\delta z(z)\;|B_y(x,z,t)|}.
\done 
This quantity shows the pile up of flux near the photosphere. Regions of strong $B_y$ pileup above the photosphere correspond, in panel b), to locations of large gradients of the vertical flux. Physically, this implies that vertical field lines are bending and turning over just above the photosphere, converting vertical flux into horizontal flux.

%As field lines rise into the corona, the density continues to drop, before the buoyancy overwhelms the downward tension on the field. 
This supports the picture of the flux piling up near the photosphere before emerging due to the undular instability. Then, reconnection between adjacent field lines allows the flux to rise rapidly into the corona. \par   
To quantitatively check that this is indeed the undular instability, we measure the wavenumber and growth rate of the instability in our simulation and compare them to theoretical values. Theoretically, the scale of the fastest growing parallel wavenumber of this instability was determined by \citet{Acheson78a,Acheson78b,Acheson79,Hughes85} and \citet{Fan01a}, the latter of whom modeled the undular instability in a high-$\beta$ regime deep in the convection zone as a mechanism to impart buoyancy to flux ropes. These studies found that the fastest growing parallel mode had a wavenumber of order $H_p^{-1}$ \citep{Acheson78a,Acheson78b,Acheson79,Hughes85} or $0.3\;H_p^{-1}$ \citep[][cf. Figure 2 in that paper]{Fan01a}. \par 
The scale of the fastest growing perpendicular mode has been studied by numerous authors. \citet{Cattaneo90} showed that the presence of shear in the field will result in a preferred value of $k_\perp$, but we detected little to no shear across the layer, so this is unlikely to set the scale of the perpendicular mode. \citet{Fan01a} argued that a large viscosity could determine the scale of perpendicular modes, since short wavelengths will be damped out. On the other hand, the analysis of \citet{Acheson78a} and \citet{Acheson78b} showed that in the presence of small but finite magnetic diffusivity the fastest growing perpendicular mode will also be of order $H_p^{-1}$, so that the parallel and perpendicular modes should have approximately equivalent wavenumbers. In essence, the fastest growing wavenumbers are those which are large enough to not be damped out by viscous or diffusive effects, but small enough that magnetic tension will not suppress their growth.\par
\begin{figure*}
\includegraphics[width=0.5\linewidth]{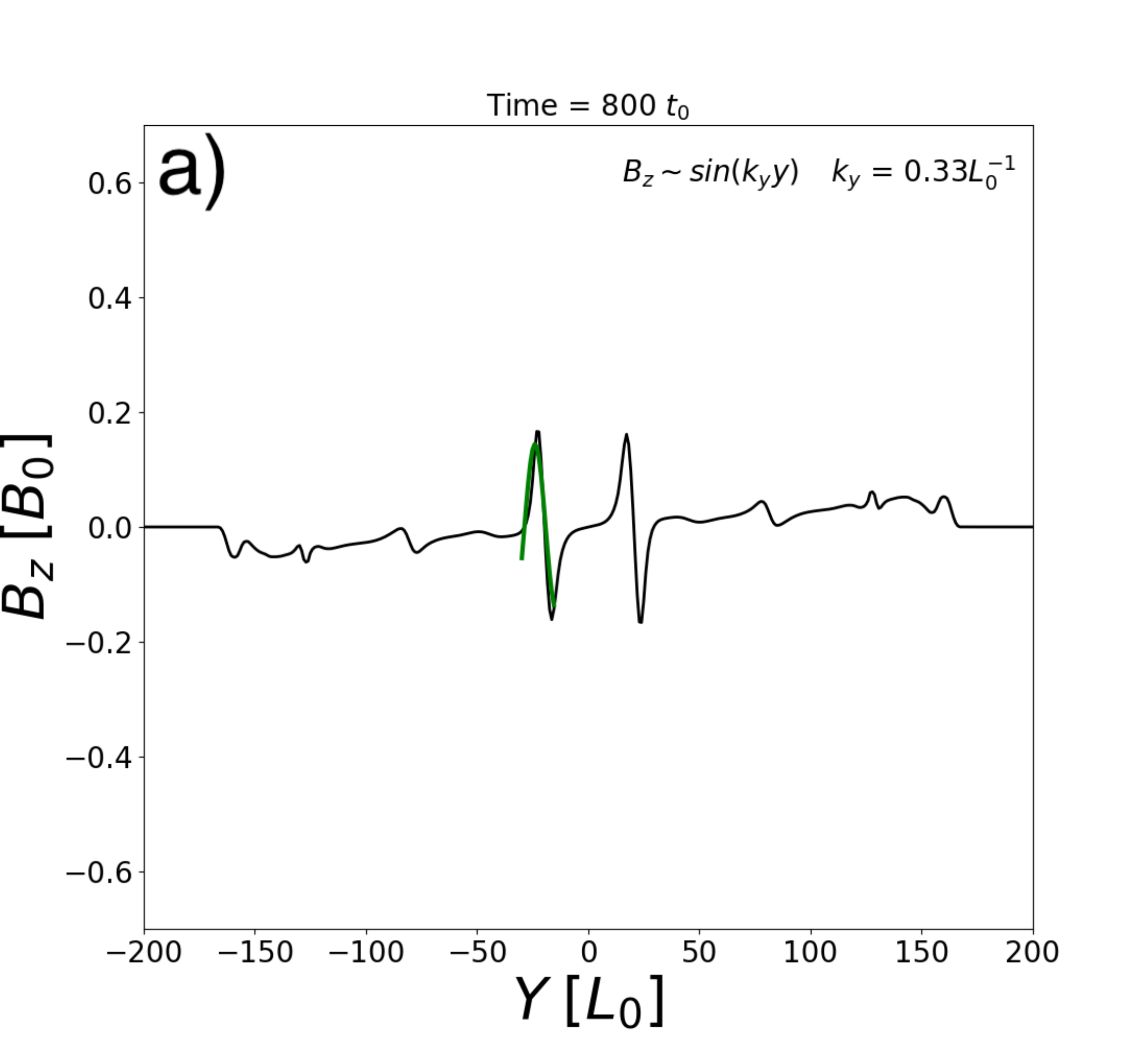}
\includegraphics[width=0.5\linewidth]{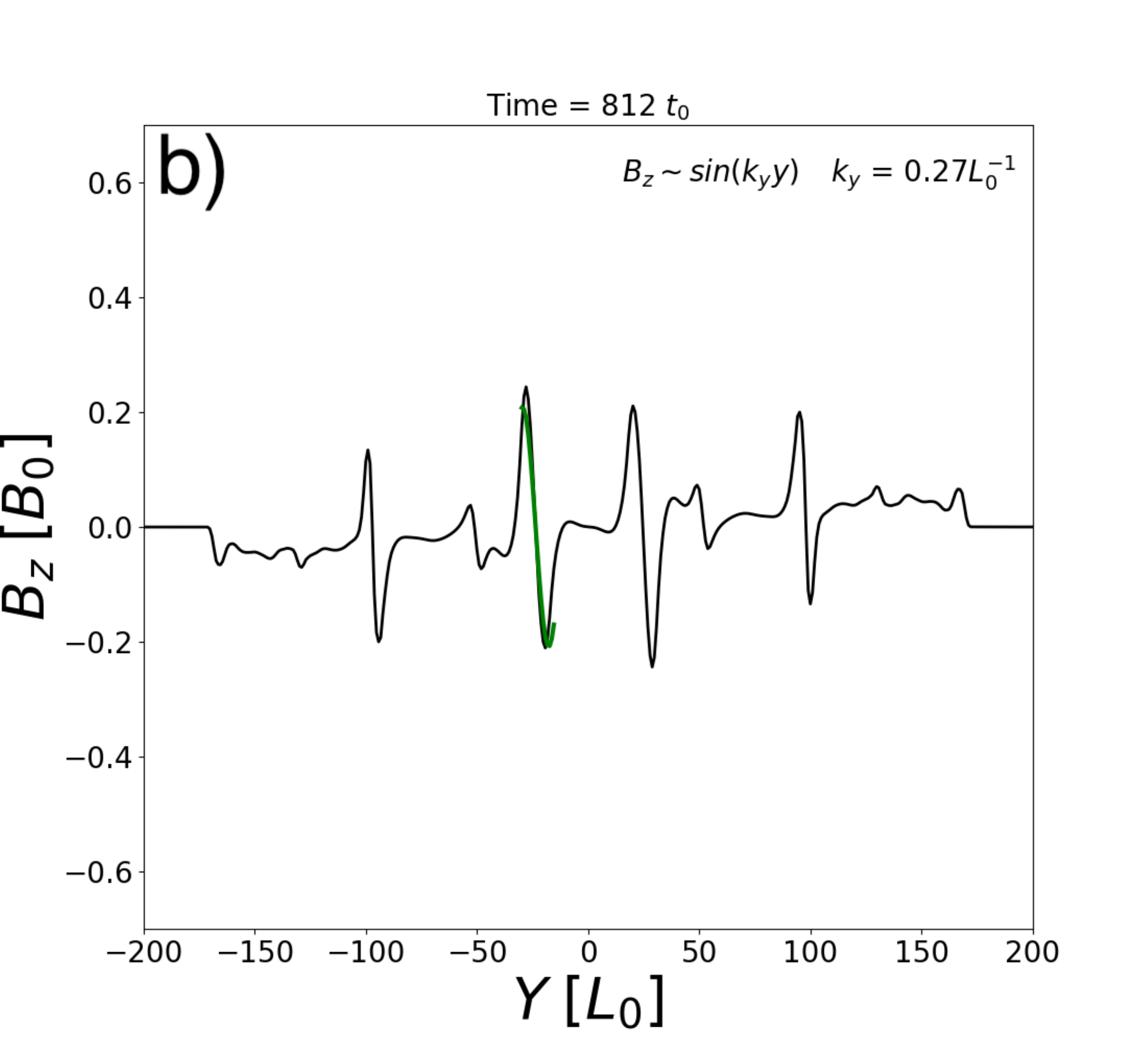}
\includegraphics[width=0.5\linewidth]{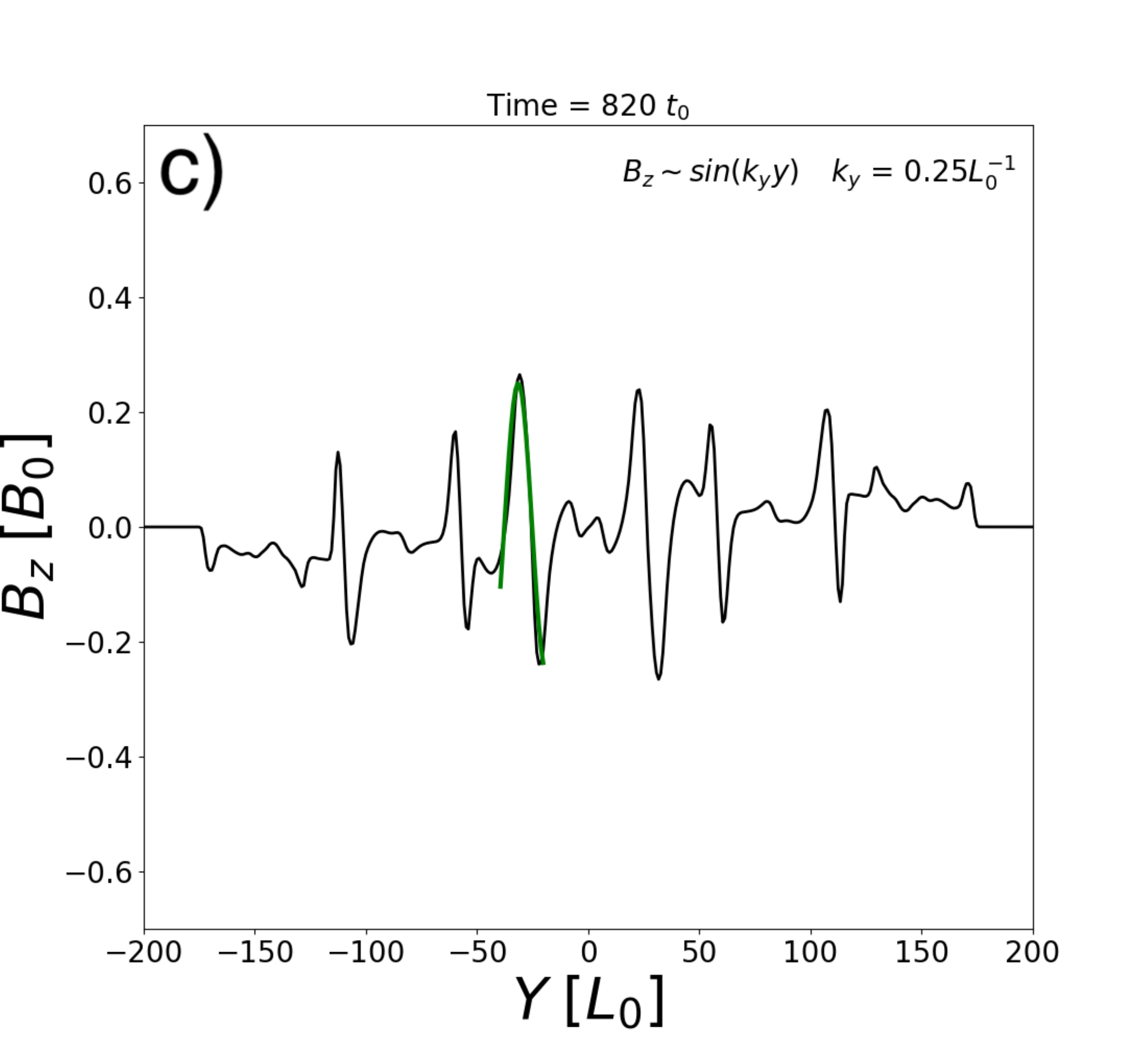}
\includegraphics[width=0.5\linewidth]{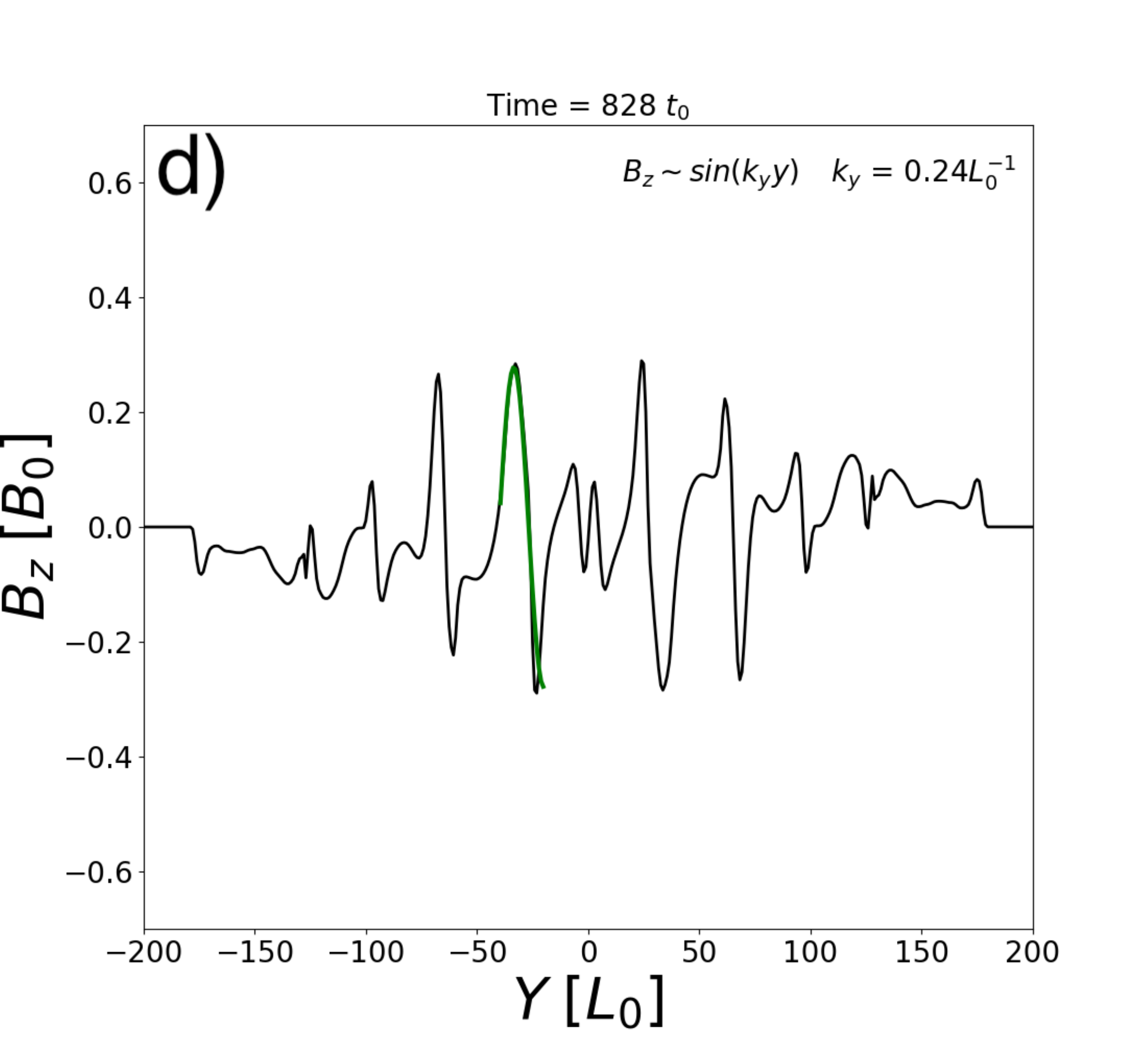}
\caption{A line cut of $B_z$ at $z=z_i$ along the $y$-direction at $x=0$ at several different times during the linear stage of the instability. The green line marks a segment of the curve that was fit to a sinusoid with wavenumber $k_y$, and its amplitude will be analyzed in \autoref{fig:growth}. For reference, the legs of the emerging flux center at $\pm y = 120\;\mathrm{L_0}$.}\label{fig:bzcut}
\end{figure*}
\begin{figure*}
\includegraphics[width=0.5\linewidth]{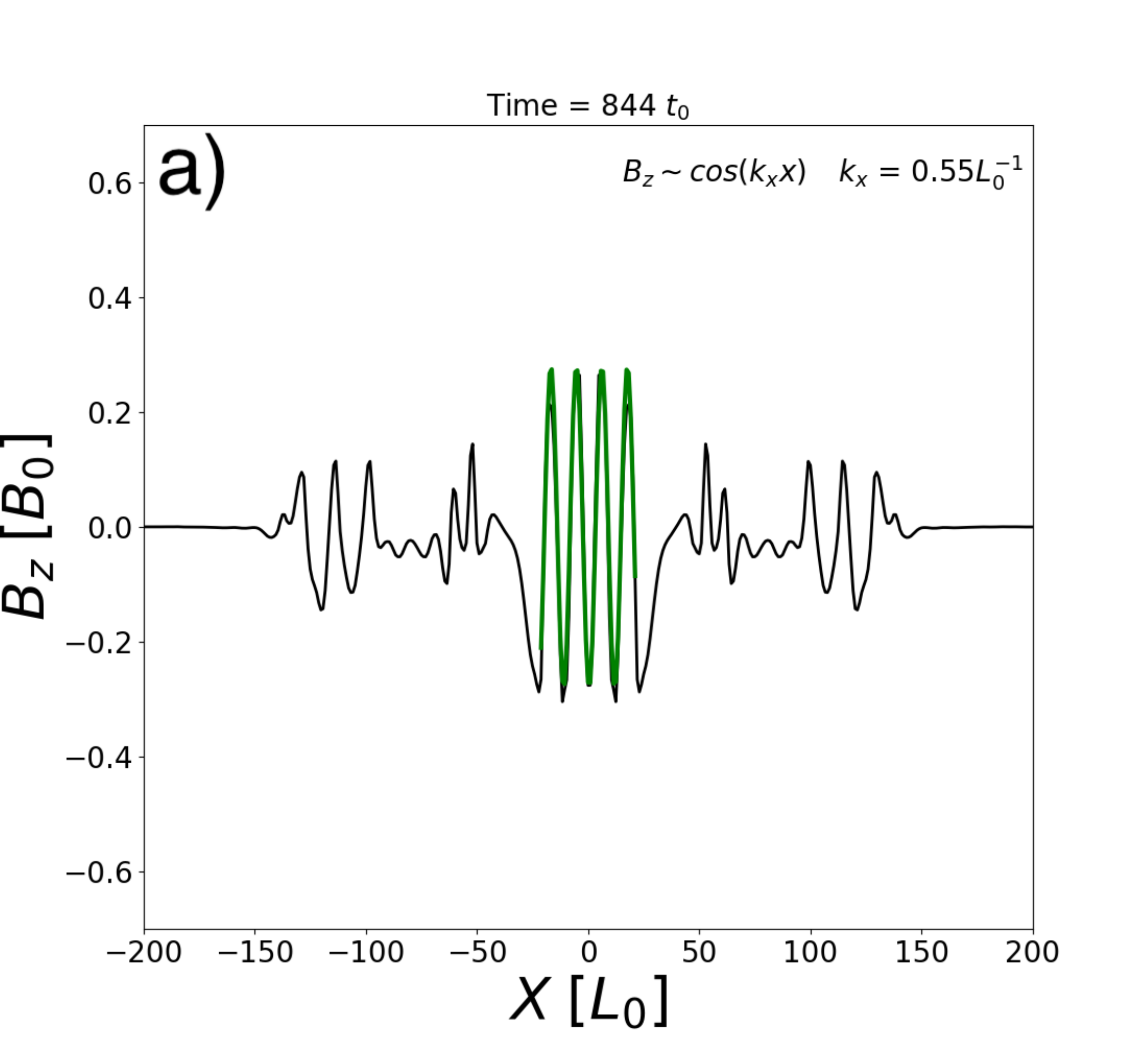}
\includegraphics[width=0.5\linewidth]{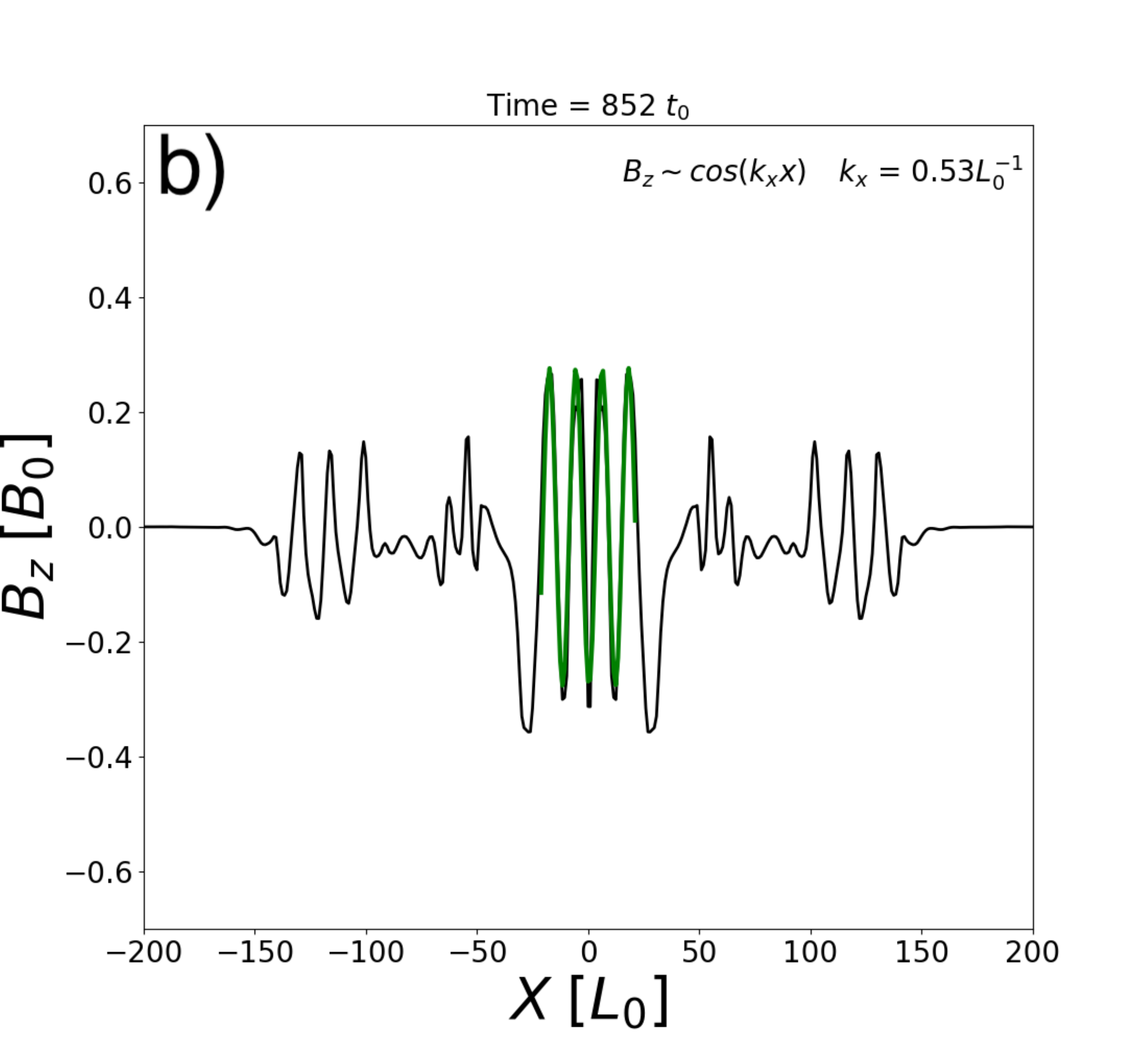}
\includegraphics[width=0.5\linewidth]{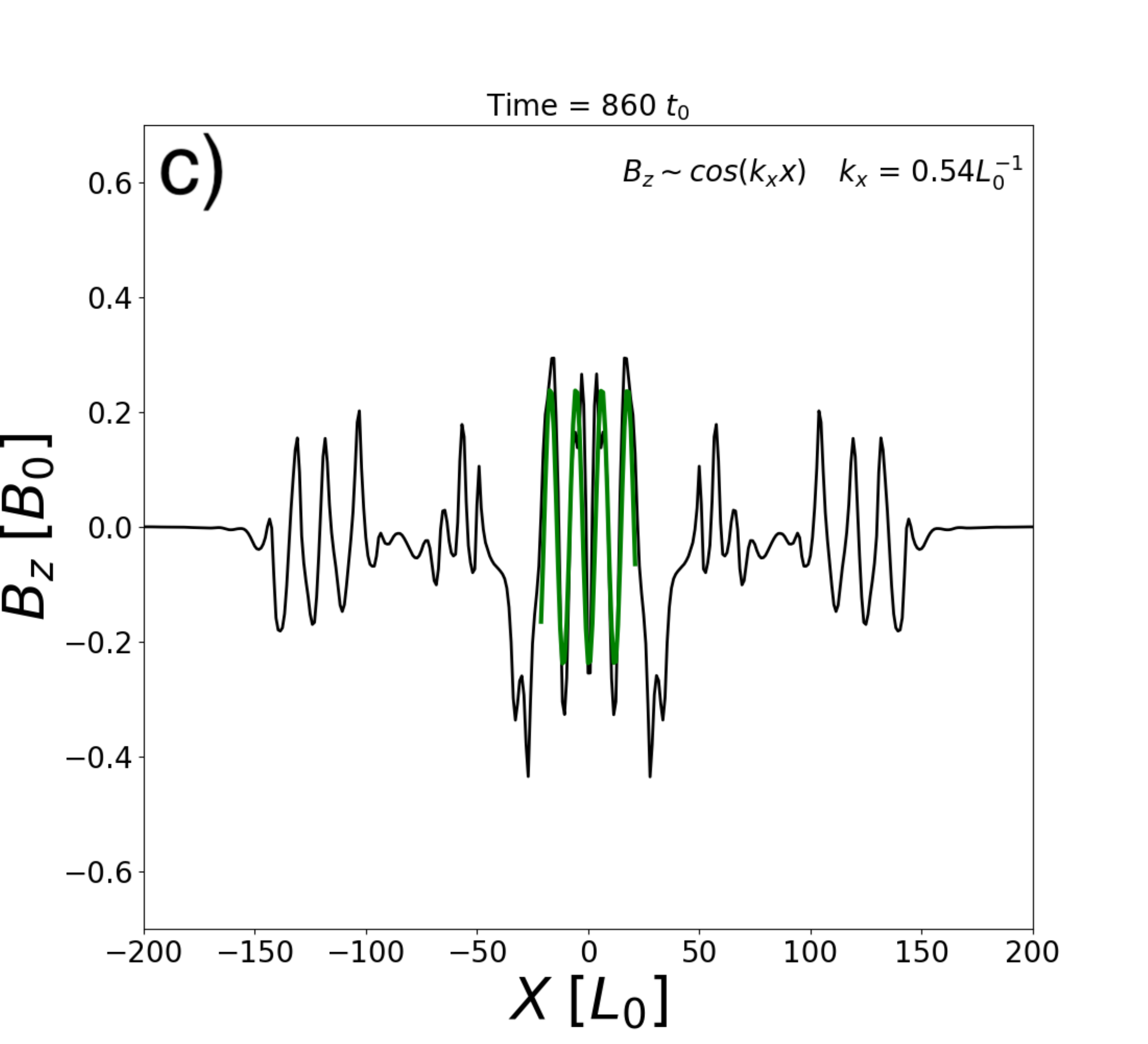}
\includegraphics[width=0.5\linewidth]{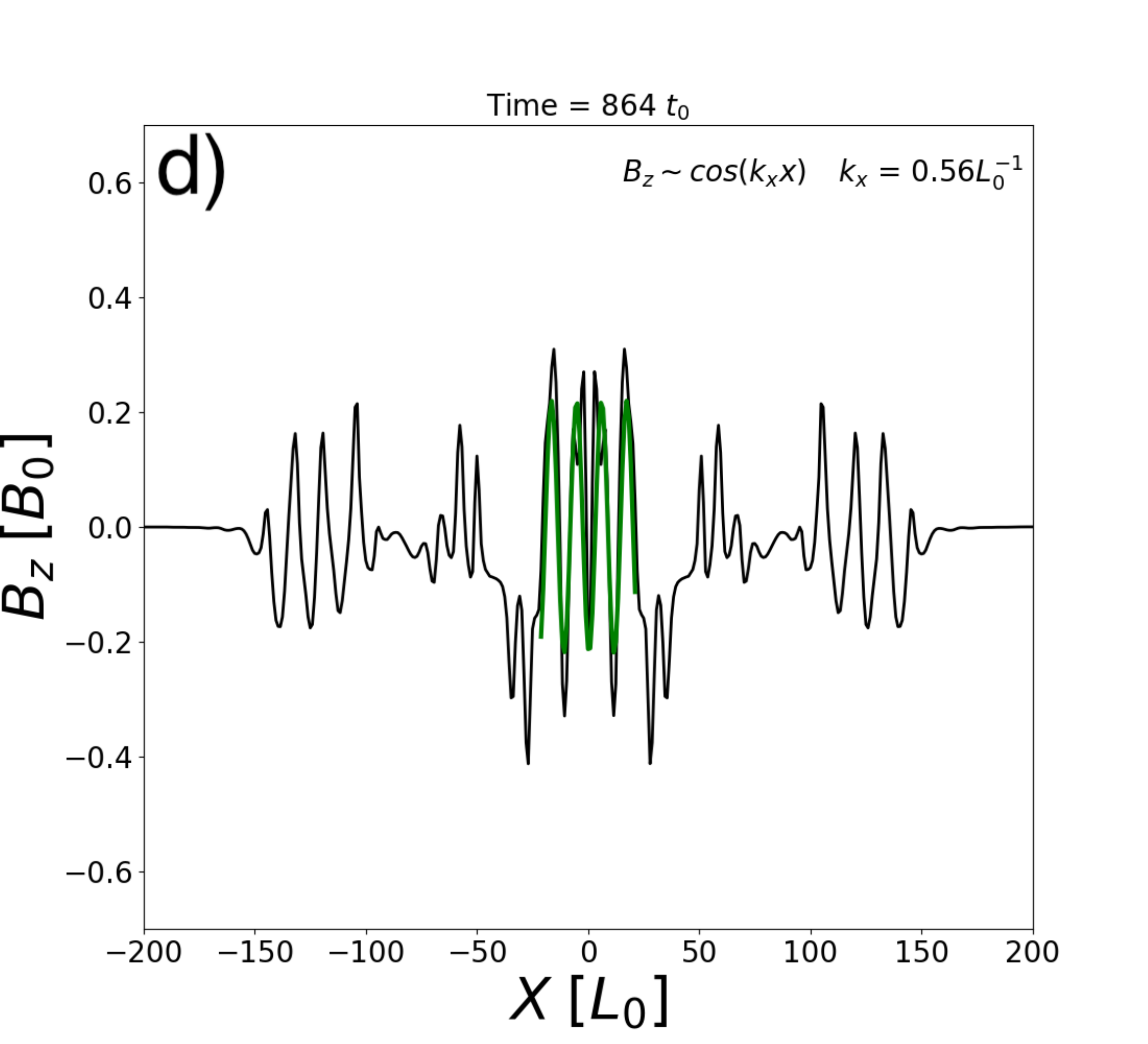}
\caption{A line cut of $B_z$ at $z=z_i$ along the $x$-direction at $y=35$ at several different times during the linear stage of the instability. The green line marks a segment of the curve that was fit to a cosinusoid with wavenumber $k_x$, and its amplitude will be analyzed in \autoref{fig:growth}.}\label{fig:perpbzcut}
\end{figure*}
\begin{figure*}
\includegraphics[width=0.5\linewidth]{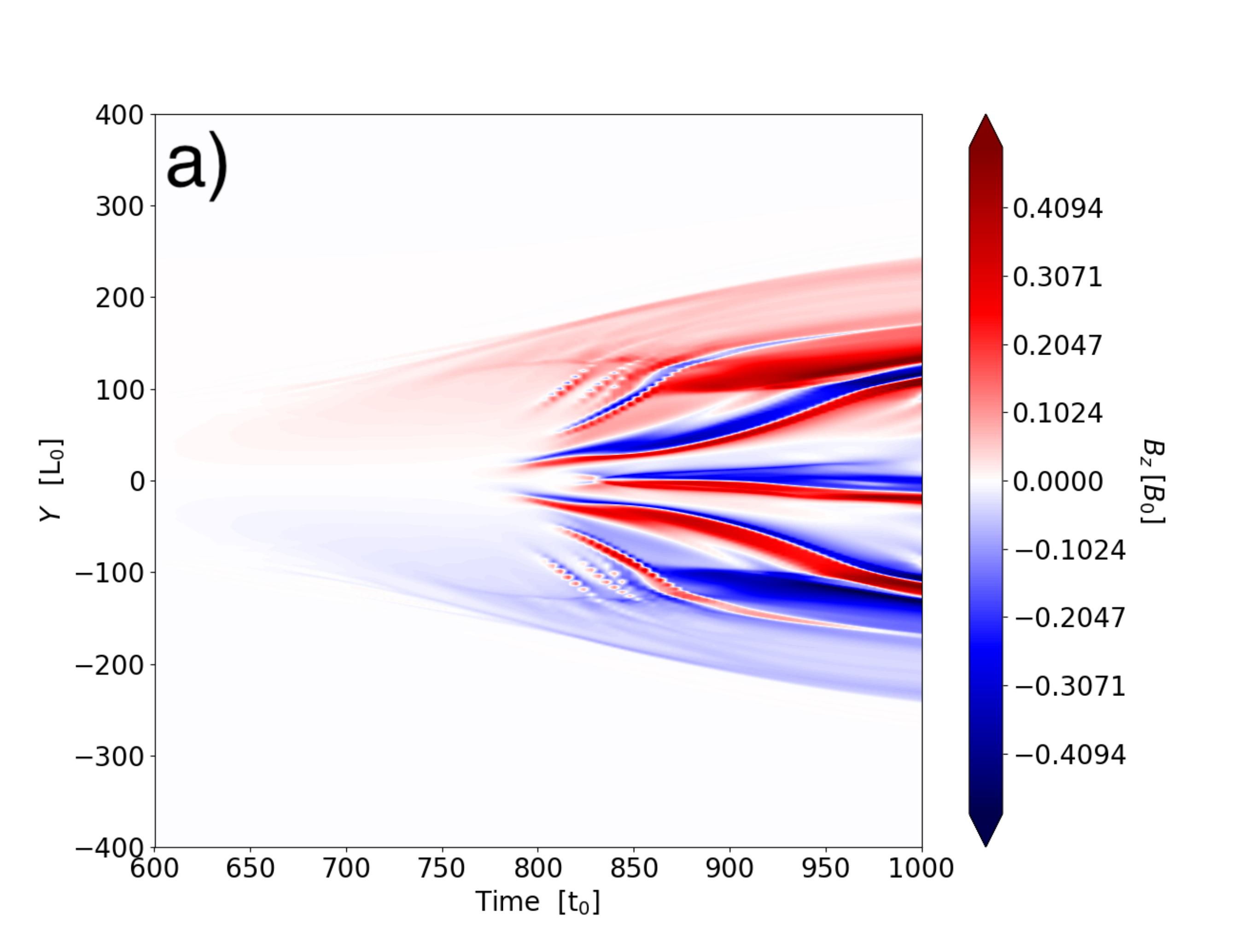}
\includegraphics[width=0.5\linewidth]{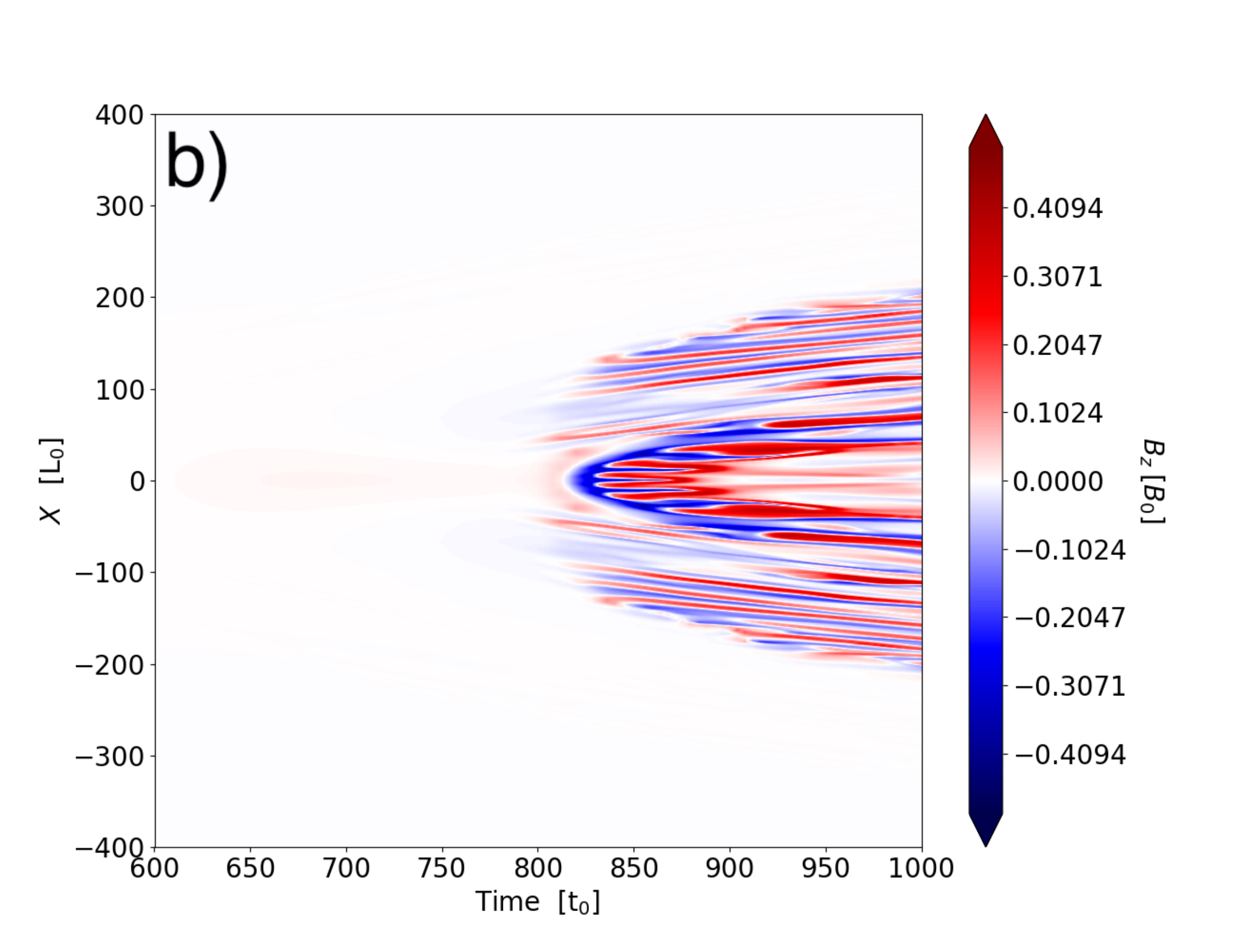}
\caption{Distance-time plot of $B_z$ along the (a) $y$- and (b) $x$- directions.}\label{fig:heighttimeBz}
\end{figure*}

%However, viscosity did not damp out the smaller spatial scales associated with the parallel mode ($k_x<k_y$), and so viscosity is unlikely to be responsible. 
To determine the fastest growing wavenumber for our simulation, we took a cut of $B_z$ along the $y$-direction at $x=0$ and $z=z_i$ for several times near the initial onset of the instability, around $t=800\;t_0$ (cf. \autoref{fig:photosphere0}b), corresponding to the parallel direction (i.e., along the flux rope axis). This cut is shown in \autoref{fig:bzcut}. Although we do not show the distribution of $B_z$ at the height $z=z_i$, it looks extremely similar to the photospheric $z=0$ distribution, just offset slightly in time. At each time step starting at $t=800\;\mathrm{t_0}$, one of the two central peaks was fit to a sinusoid (green curve) of the form $B_z\sim b_0(t) \sin(k_y y)$. Evidently, the fastest growing parallel wavenumber is of the order 
\beg{kparfastest}
k_{||}\approx 0.3\;L_0^{-1}, 
\done 
in agreement with \citet{Acheson78a, Acheson78b, Acheson79,Hughes85} and \citet{Fan01a}, and justifying \emph{a posteriori} our approach of dropping the rightmost term in \autoref{fullinstabilitycriterion}, since this term will be an order of magnitude smaller than either $\xi$ or $\chi$. This wavenumber corresponds to a fastest growing wavelength of
\beg{lambda_par}
\lambda_{||} = \frac{2\pi}{k_{||}} = \frac{2\pi}{0.3}L_0 \approx 21 L_0
\done 
(note that $L_0=H_p$ at the photosphere). This is the wavelength calculated for this instability by many previous authors \citep{Parker66,Parker69,Shibata89,Nozawa92}. The wavelength of the mode increases slightly in time as the field emerges and the `footpoints' of the field line separate. A similar analysis was performed along the perpendicular ($x$-) direction at $y=35$ and $z=z_i$ (the location of this cut is marked by the red line in \autoref{fig:photosphere0}c, though note that the height of the cut is $z=z_i$, rather than $z=0$), and fitting to a cosine function as shown in \autoref{fig:perpbzcut}. Here the perpendicular mode appears slightly later than the parallel mode, and it is seen to have a wavenumber of order 
\beg{kperpfastest}
k_\perp\approx0.55\;L_0^{-1},
\done
corresponding to a perpendicular wavelength of 
\beg{lambda_perp}
\lambda_{\perp} = \frac{2\pi}{k_{\perp}} = \frac{2\pi}{0.55}L_0 \approx 11 L_0.
\done 
We plot the amplitude of $B_z$ along the two cuts shown in Figures \ref{fig:bzcut}-\ref{fig:perpbzcut} and stack them in time, as shown in \autoref{fig:heighttimeBz}. In panel (a), two negative (positive) streaks are clearly visible in the dominant positive (negative) polarity. These are the signatures of the growing parallel mode. Here, it is obvious that the wavelength of the mode is growing in time, since each streak makes an angle with the $y=0$ axis. In panel (b), the growth of the perpendicular mode can be seen, but in this case the streaks lie almost parallel to the $x=0$ line, indicating that the wavelength of the perpendicular mode does not change substantially, in agreement with \autoref{fig:perpbzcut}.\par 
\par 

\begin{figure*}
\includegraphics[width=0.5\linewidth]{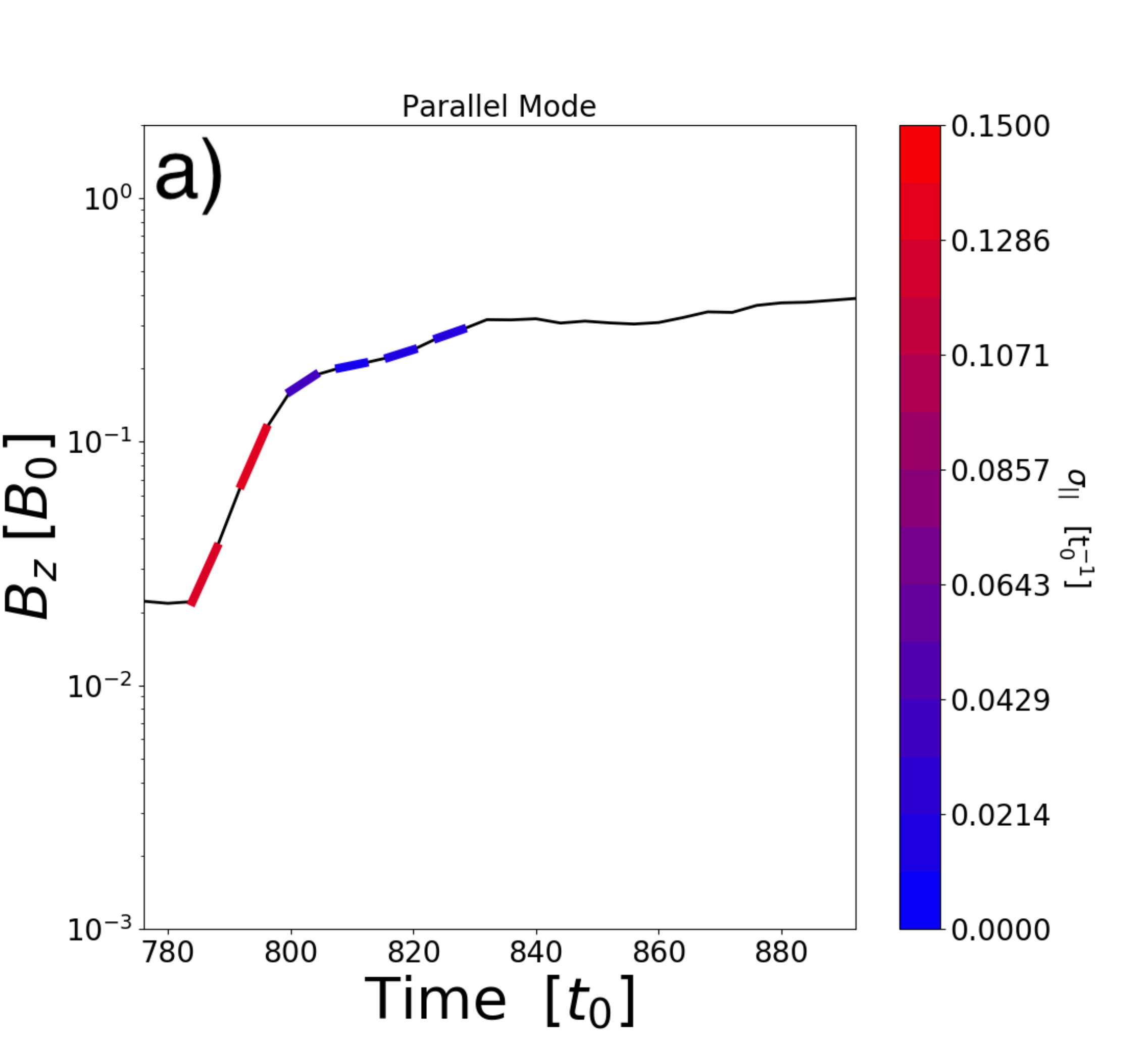}
\includegraphics[width=0.5\linewidth]{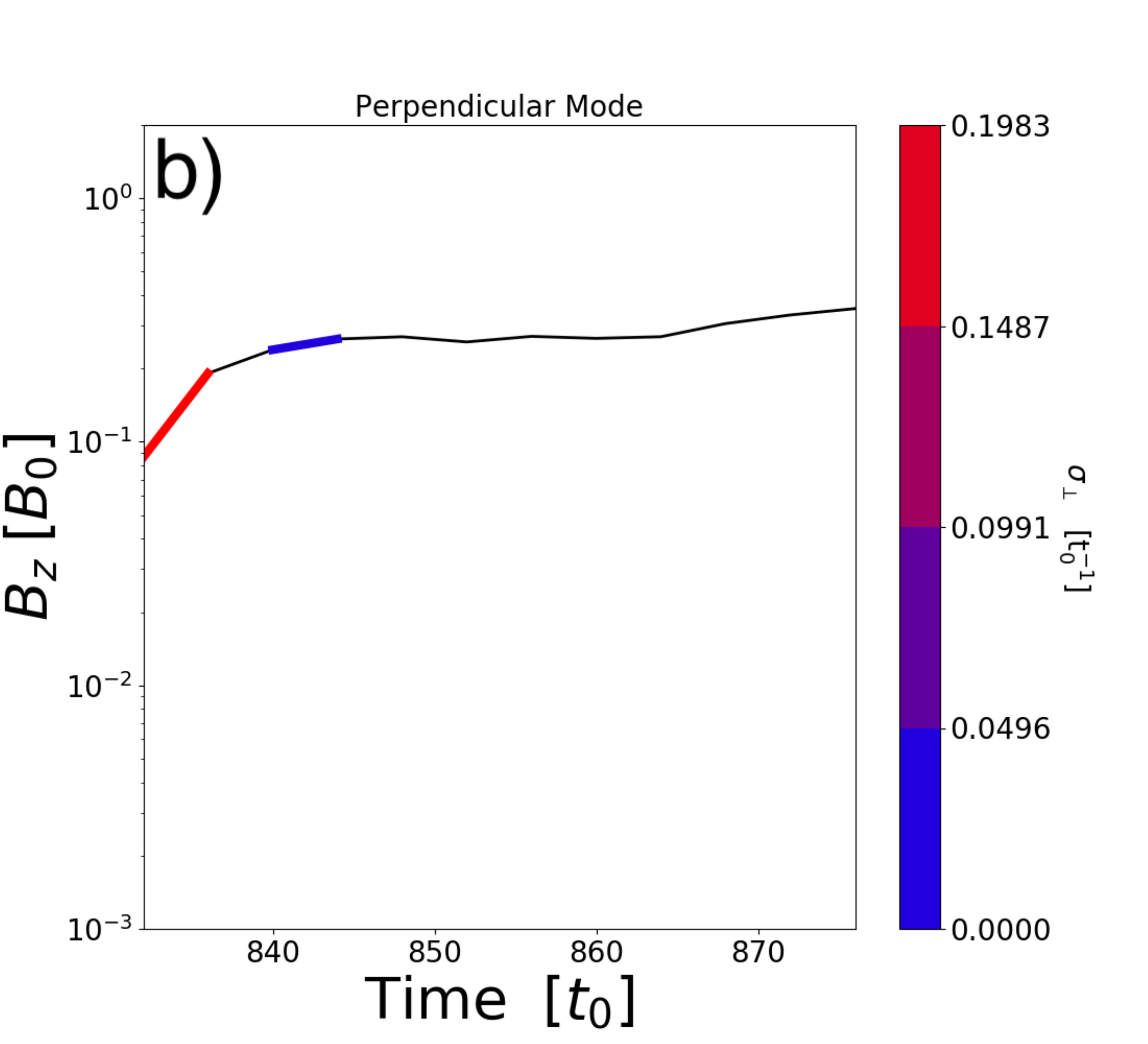}
\caption{Amplitudes of the modes shown in Figures \ref{fig:bzcut} and \ref{fig:perpbzcut}, representing the growth of the parallel (a) and perpendicular (b) modes. Each $5$ consecutive points along the curve were fit to an exponential of the form $B_z\sim b_0e^{\sigma t}$, and the color shading represents the value of $\sigma$ in the parallel (a) and perpendicular (b) directions.} \label{fig:growth}
\end{figure*}

To determine the growth rate of the instability, in \autoref{fig:growth} we plot the amplitudes of the modes in Figures \ref{fig:bzcut} and \ref{fig:perpbzcut} during the very early phase of the evolution. Each $5$ consecutive points along the curve were fit to a running exponential of the form $B_z\sim b_0e^{\sigma t}$. The color magnitude shows the value of the growth rate $\sigma$ along each direction. During the early stages of its growth, the instability has growth rates of order
\beg{gammapargrowth}
\sigma_{||} \approx 0.15 \;\mathrm{t_0^{-1}},
\done
and
\beg{gammaperpgrowth}
\sigma_{\perp} \approx 0.20 \;\mathrm{t_0^{-1}},
\done
in the parallel and perpendicular directions, respectively.
In the literature, there are several theoretical estimates for the growth rate of the undular/magnetic buoyancy instability.  \citet{Acheson79} estimated the growth rate of the instability to be of order \citep[see Equation 2.7 in][]{Acheson79}:
\beg{growthAcheson}
\sigma \sim \Big(\frac{V^2_a}{H_p^2}\xi\Big)^{1/2},
\done
with $\xi$ given in \autoref{MBIxi}. At the photosphere, $V_a\approx 1$, and $\xi\approx0.03$ when the mode amplitude rapidly increases around $t=800\;\mathrm{t_0}$ (cf. \autoref{fig:Parkerterms}) so that (with $H_p=1$ in our units), 
\beg{growthAchesonprediction}
\sigma_{Acheson} \approx 0.17\;\mathrm{t_0^{-1}}.
\done
Thus, the parallel growth rate in our simulation is of the same order of magnitude as the growth rate derived from \citet{Acheson79}. \par 
In the high-$\beta$ regime, \citet{Fan01a}, estimates a growth rate for our measured wave number of (see Fig. 2 in that paper):
\beg{growthFanprediction}
\sigma_{Fan} \approx 0.1 \;\mathrm{t_0^{-1}},
\done
which is in good agreement with our measurement during the very initial stages seen in \autoref{fig:growth}. \par 
\citet{Chandra61} used the Boussinesq approximation to predict a growth rate of the instability of 
\beg{growthChandra}
\sigma = \Bigg[\frac{g\sqrt{k_\perp^2+k_{||}^2}}{(\rho_++\rho_-)}\delta\rho_{ph} - \frac{\big(k_{||}B_y)^2}{2\pi(\rho_++\rho_-)}\Bigg]^{1/2},
\done
where $g$ is gravity, $k_{||}=k_y$ and $k_\perp=k_x$ are the wavenumbers parallel and perpendicular to the magnetic field $\vecB$, and $\rho_+$ and $\rho_-$ are the densities above and below the unstable layer, with density difference $\delta\rho_{ph}$. Although in our simulation, $\delta\rho_{ph}/\rho_{ph}\sim O(1)$ at the photosphere, and the Boussinesq approximation is not applicable, we can, nevertheless, test how close the growth rate derived from this approximation is to the one measured here. Averaging $\rho_+$, $\rho_-$, and $B_y^2$ over the interface region $x,y\in[-50,50]\;L_0$ yields
\beg{growthChandraprediction}
\sigma_{CH} \approx 0.31 \;\mathrm{t_0^{-1}},
\done 
which is in good agreement with the growth rate shown in \autoref{fig:growth}.
%In the thin tube approximation, \citet{Spruit82} find that the growth rate of each mode depends on the quantity $\beta\delta$, where $\beta\approx1$ at the photosphere, and $\delta$, defined in \autoref{defdelta}, is $\approx-0.4$. In this regime, they predict a growth rate of \citep[see Fig. 2 of ][]{Spruit82}:
%\beg{growthSpruit}
%\sigma_{SvB} \approx \Big(\frac{0.02}{2}\Big)^{1/2} = 0.1 \;\mathrm{t_0^{-1}}.
%\done 
Evidently, the growth rate for this instability derived in the Boussinesq regime, which is admittedly inapplicable in our simulation, produces values that are in good agreement with our simulation results. \par 
Thus, both the fastest growing parallel and perpendicular wave numbers and growth rates of the instability observed in our simulation are in reasonable agreement with theoretical predictions of these same quantities in the undular/magnetic buoyancy instability. \par

\subsection{Interaction of the Lobes}\label{sec:interaction}
\begin{figure}
\includegraphics[width=\linewidth,trim={0cm 4cm 0cm 4cm},clip]{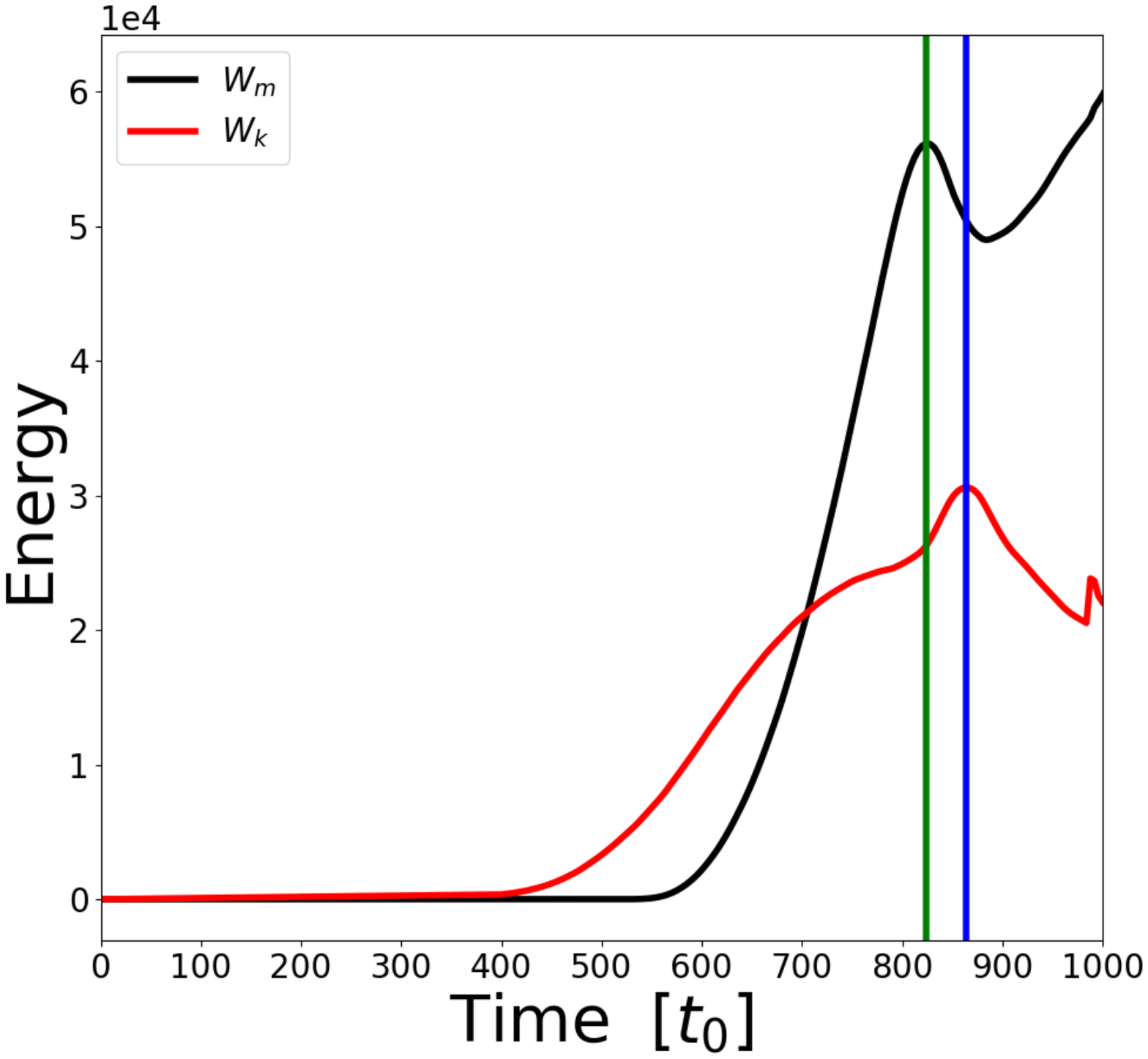}
\caption{Magnetic energy (black) and kinetic energy (red) in the corona as a function of time. The green (blue) line at $t=824\;\mathrm{t_0}$ ($t=864\;\mathrm{t_0}$) shows the local maximum of the magnetic (kinetic) energy.}\label{fig:Wm}
\end{figure}

\begin{figure*}
\includegraphics[width=0.5\linewidth]{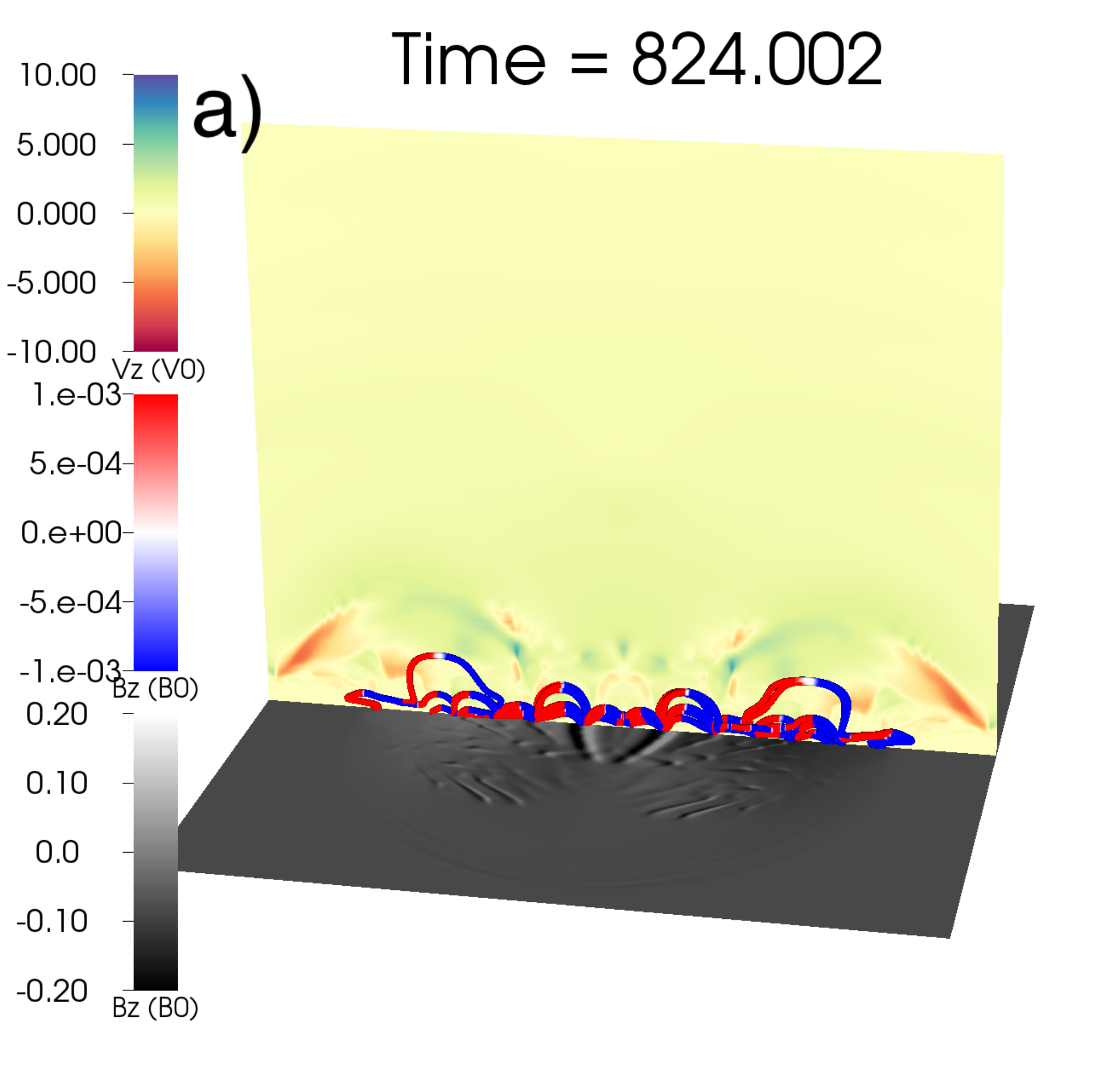}
\includegraphics[width=0.5\linewidth]{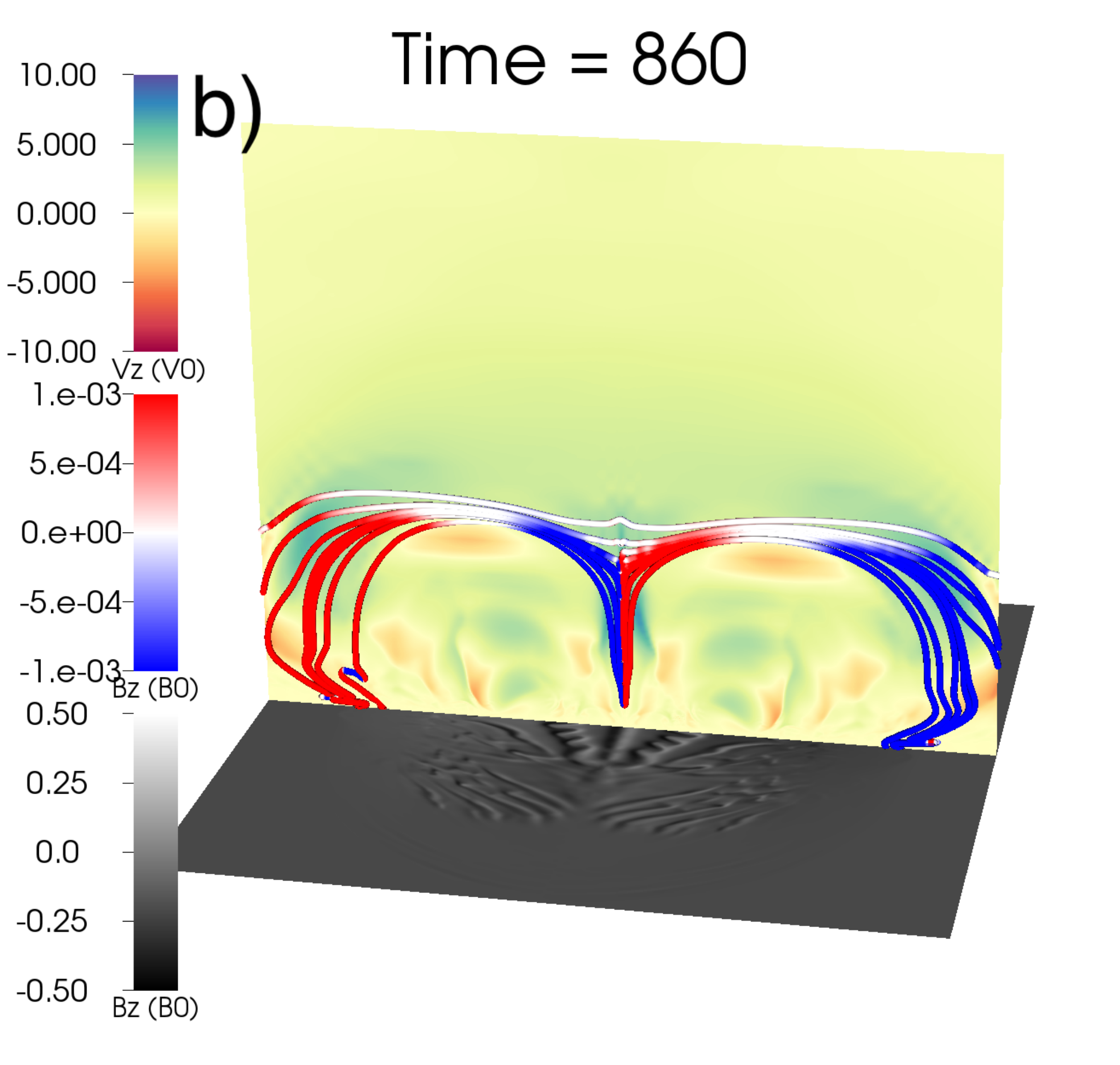}
\includegraphics[width=0.5\linewidth]{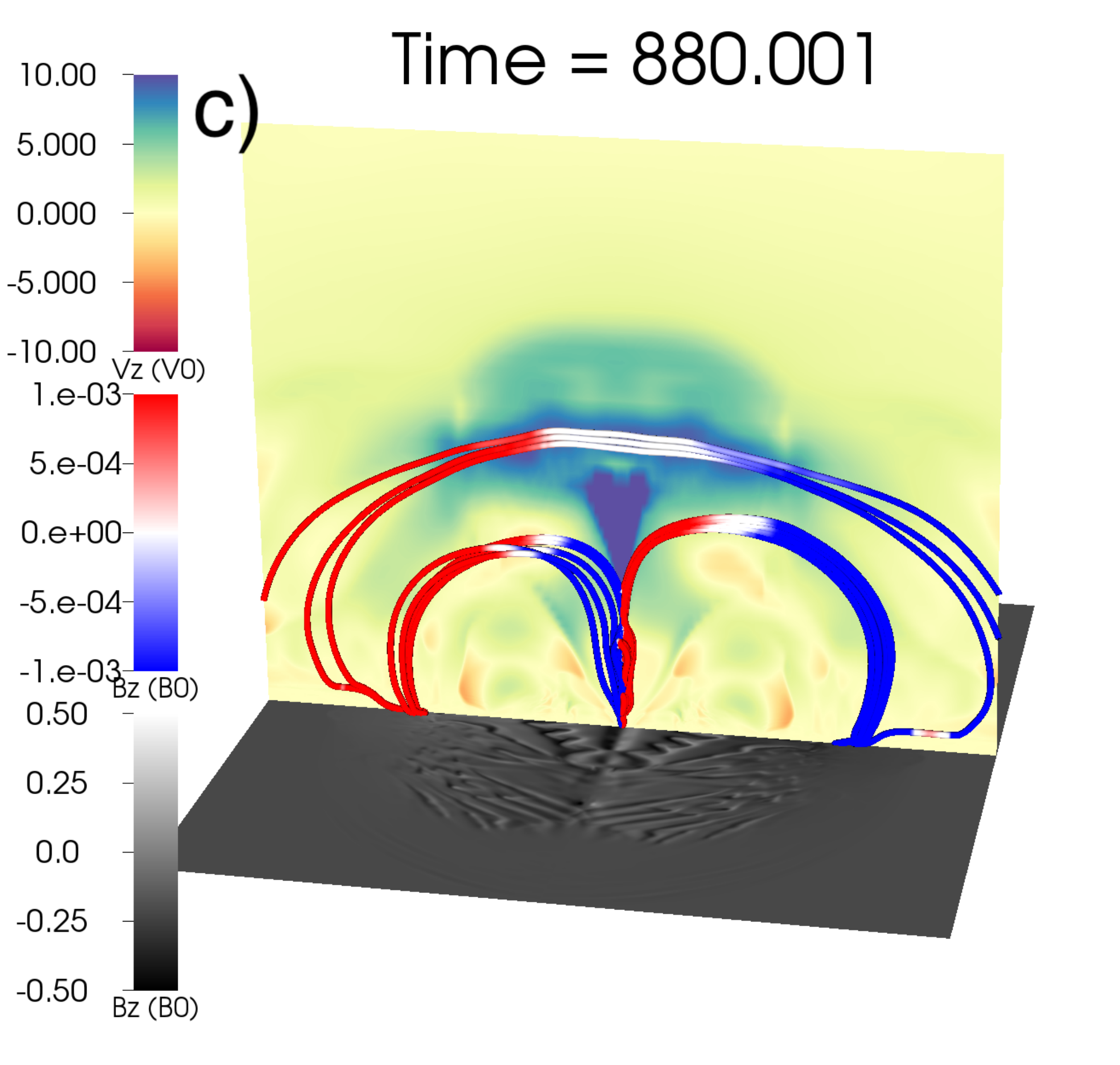}
\includegraphics[width=0.5\linewidth]{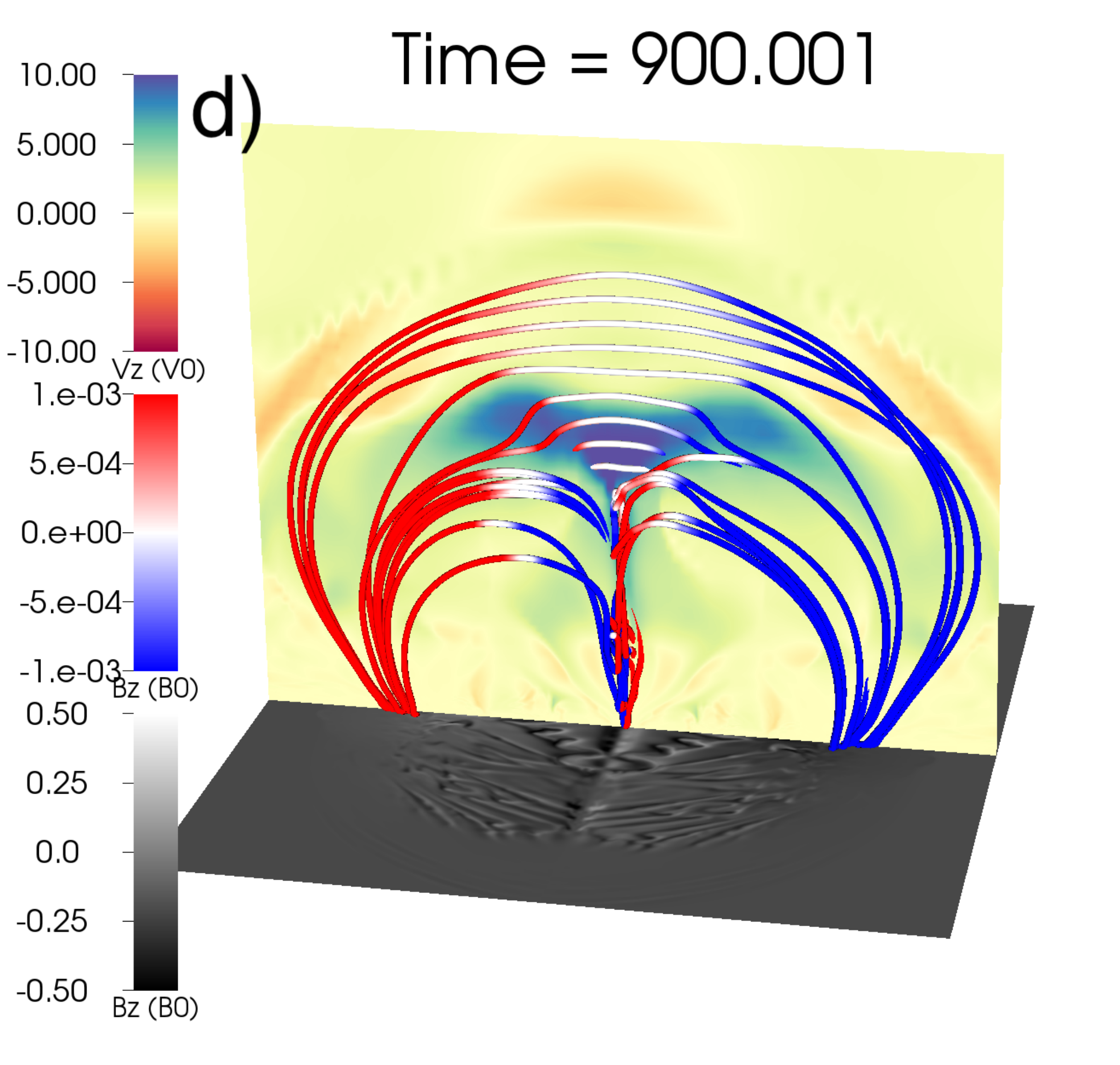}
\caption{Field lines colored by $B_z$ overplotted on a vertical plane of $V_z$ and a photospheric magnetogram at several times before and during the reconnection event.}\label{fig:rx}
\end{figure*}

The emergence of magnetic lobes was found by \citet{Archontis13} to result in significant reconnection and a jet-like disturbance in the corona. In our simulation, we see a similar event around $t=820\;\mathrm{t_0}$ that may affect the linear evolution of the undular mode, which stops around the same time. \autoref{fig:Wm} shows the magnetic energy
\beg{Wm}
W_m = \frac{1}{2\mu_0}\int_{z>0}{dV\; B^2},
\done 
and kinetic energy
\beg{Wk}
W_k = \frac{1}{2}\int_{z>0}{dV\;\rho v^2}
\done
in the corona as a function of time. The decrease in the magnetic energy that starts around $t=824\;\mathrm{t_0}$ (green line) corresponds approximately to the same magnitude increase in the kinetic energy which peaks at $t=864\;\mathrm{t_0}$ (blue line). In \autoref{fig:rx}, we plot a series of field lines, seeded along a vertical line at ($x=y=0,\;z>0$) in the corona on top of vertical cuts of the vertical velocity $V_z$. The field lines are colored by $B_z$. Early in the interaction, at $t=824\;\mathrm{t_0}$, the field lines are only barely extending above the photosphere. They slowly rise, and it is clear that by $t=860\;\mathrm{t_0}$, the two primary lobes are interacting, and field lines with opposite signs of $B_z$ begin to reconnect. This creates a significant upflow, as the newly reconnected concave up field lines release their tension. This upflow is only seen low down in the corona before $t=860\;t_0$, and with much smaller magnitude, since the field is not significantly expanding into the corona, but by $t=880\;\mathrm{t_0}$ and $t=900\;\mathrm{t_0}$ there is a clear upflow visible in the corona along with a concave-down arcade overlying the interacting loops.
%The newly reconnected flux remains constrained by overlying field, however, preventing a large scale eruption. 
This type of event likely represents an Ellerman bomb, which is associated with this type of topology \citep{Pariat04,Pariat06,Isobe07,Pariat09,Archontis09b,Danilovic17}.  

\section{Conclusions}\label{sec:Conclusions}
In this paper, we have simulated the buoyant rise and emergence of an untwisted toroidal flux rope from deep in the convection zone. The flux rope is able to rise coherently through the convection zone, emerge through the photosphere and form a relatively large ($\approx 40\;\mathrm{Mm}$ across) active region. The emergence mechanism of our flux rope is identified to be the undular instability \citep{Parker77,Acheson79,Fan01a, Archontis13}. The instability is responsible for creating an active region displaying lots of salt-and-pepper flux concentrations, which have previously been identified as the result of convective flows destroying the rising flux rope \citep{Rempel14,Dacie17}. However, the simulation presented here shows that the undular instability is a second mechanism which could account for the presence of salt-and-pepper structures on the photosphere. The structures formed in this simulation resemble those seen in observations \citep{Kitiashvili15,Dacie17}. \citet{Toriumi12} find similar salt-and-pepper structures, however their structures are due to a fluting instability, which produces more wavelengths and more regular structures than are seen here.\par
\begin{figure*}
\includegraphics[width=0.5\linewidth,trim={0cm 6cm 0cm 4cm},clip]{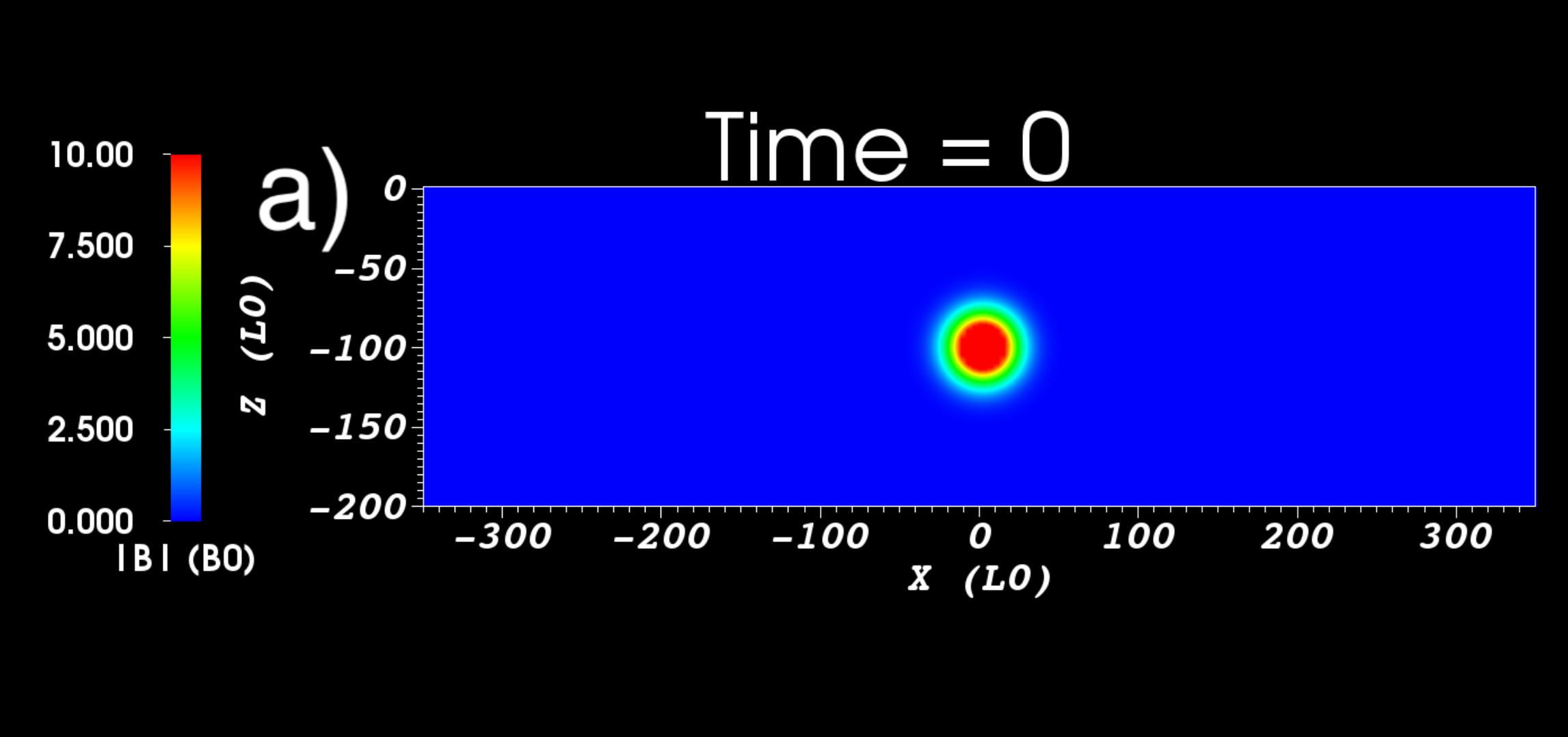}
\includegraphics[width=0.5\linewidth,trim={0cm 6cm 0cm 4cm},clip]{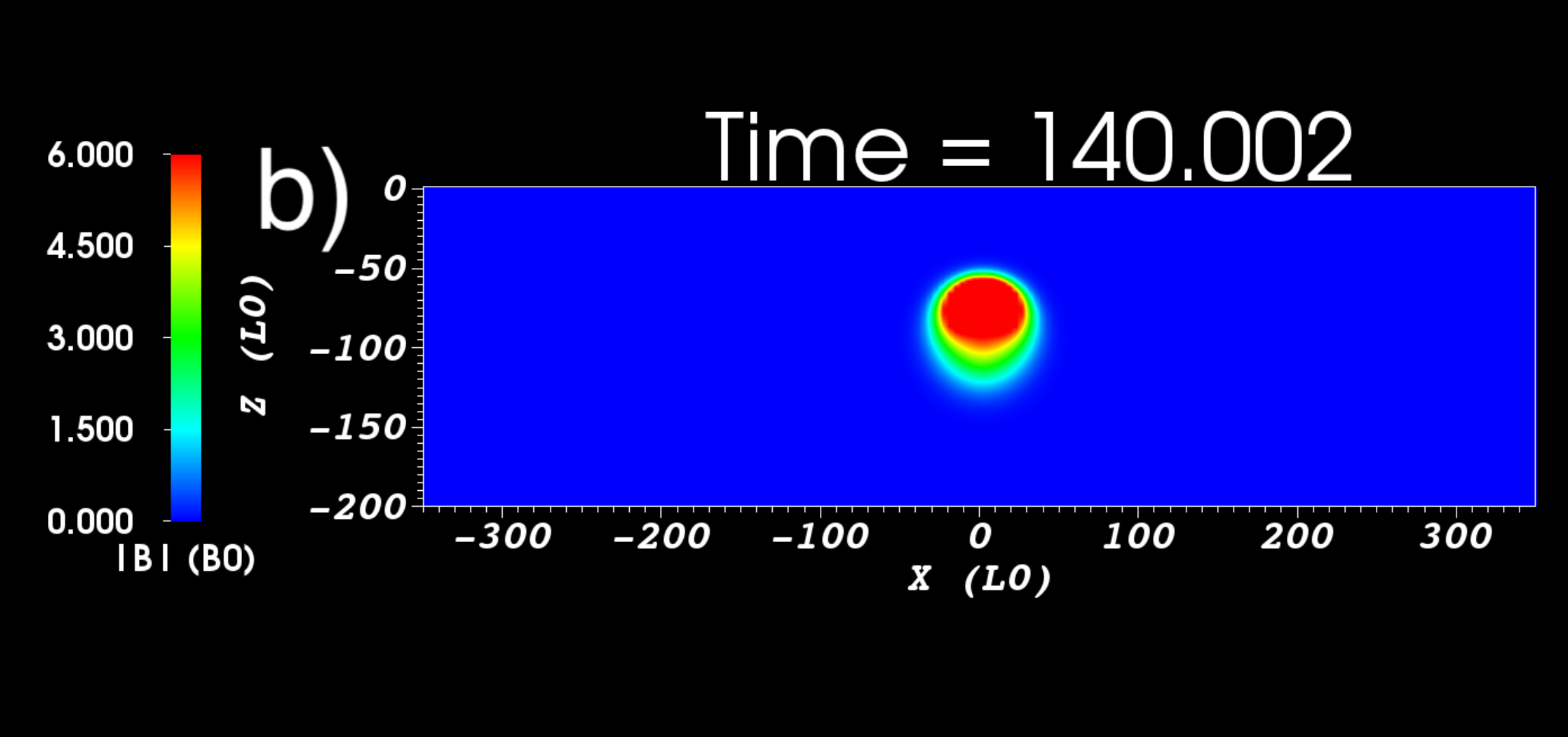}
\newline
\includegraphics[width=0.5\linewidth,trim={0cm 6cm 0cm 4cm},clip]{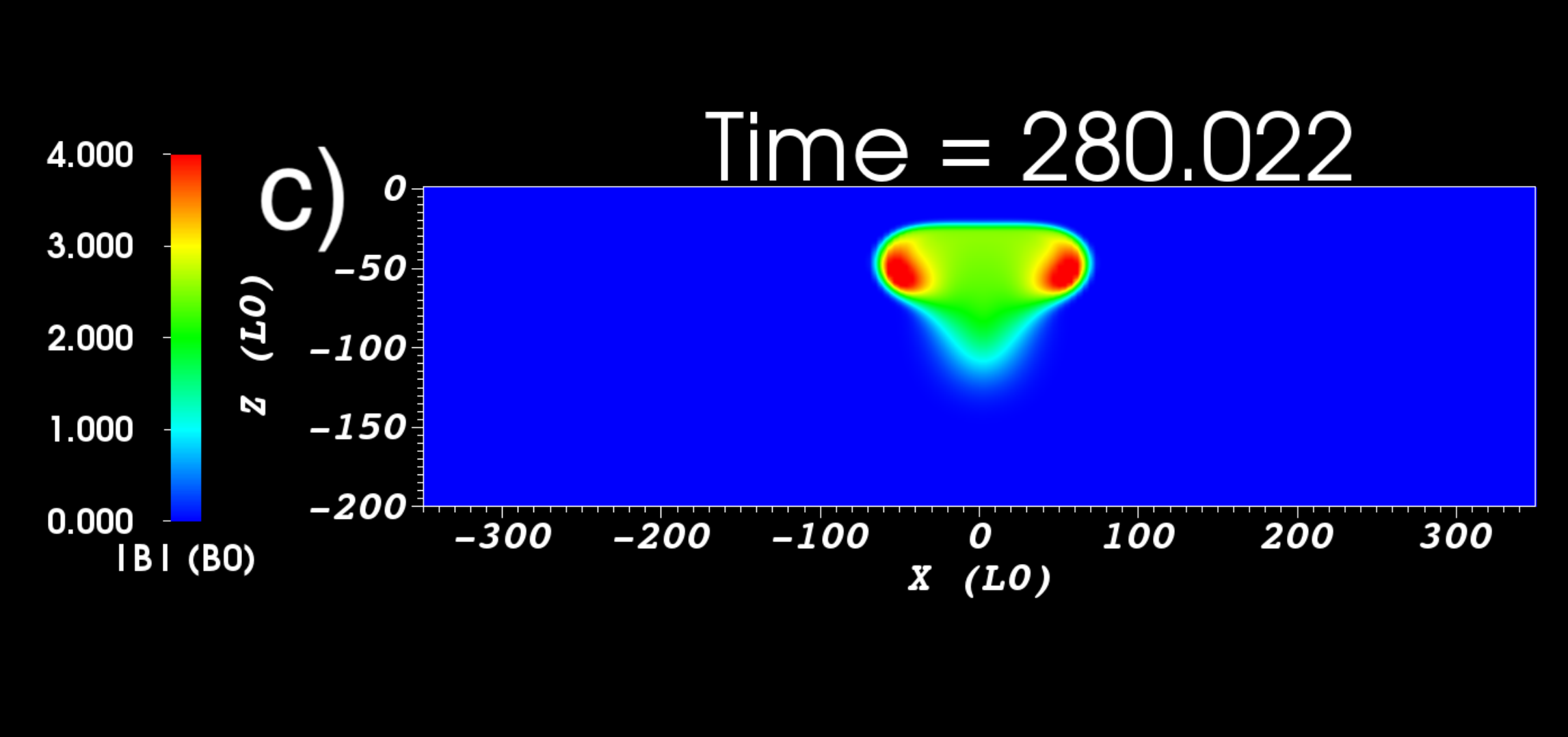}
\includegraphics[width=0.5\linewidth,trim={0cm 6cm 0cm 4cm},clip]{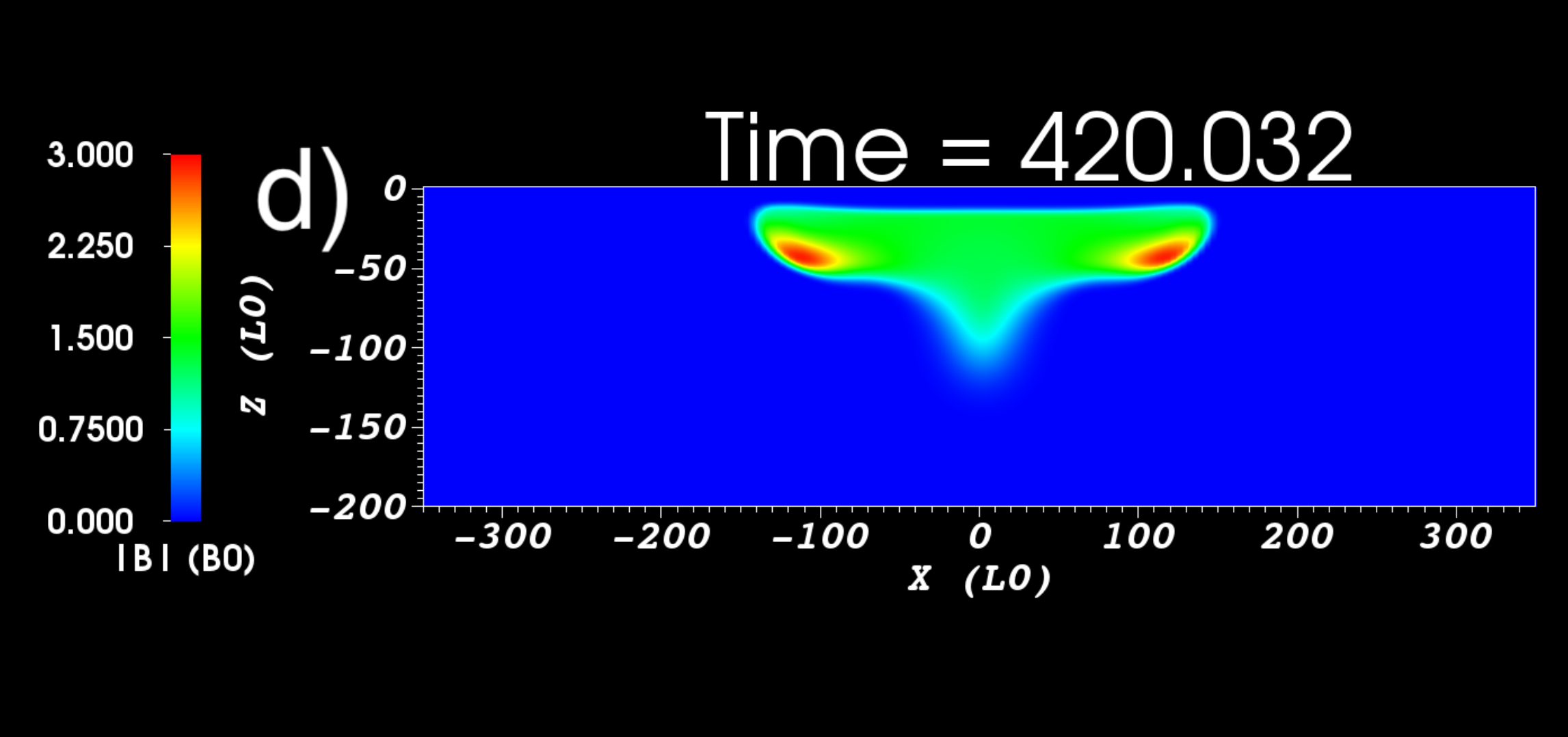}
\newline
\includegraphics[width=0.5\linewidth,trim={0cm 6cm 0cm 4cm},clip]{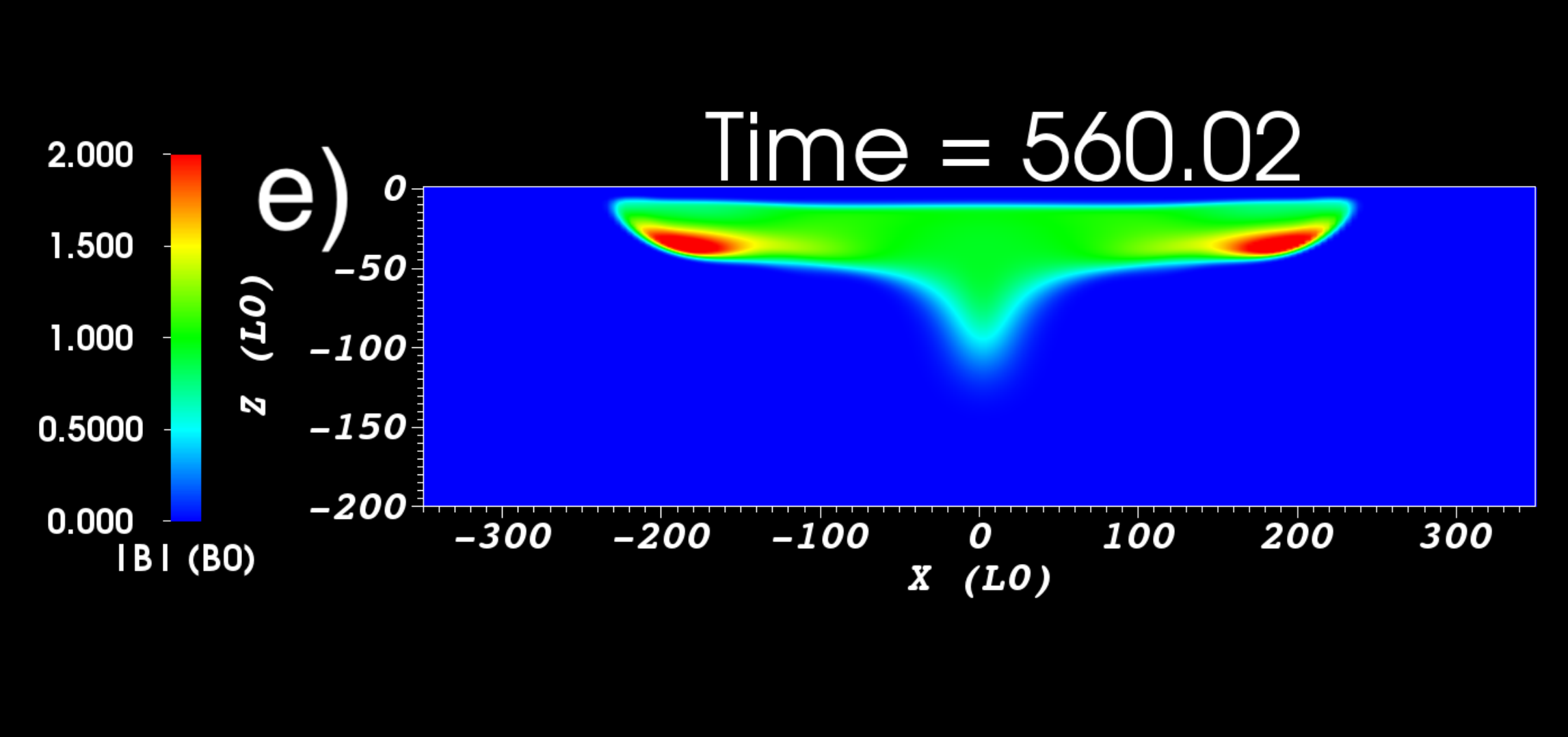}
\includegraphics[width=0.5\linewidth,trim={0cm 6cm 0cm 4cm},clip]{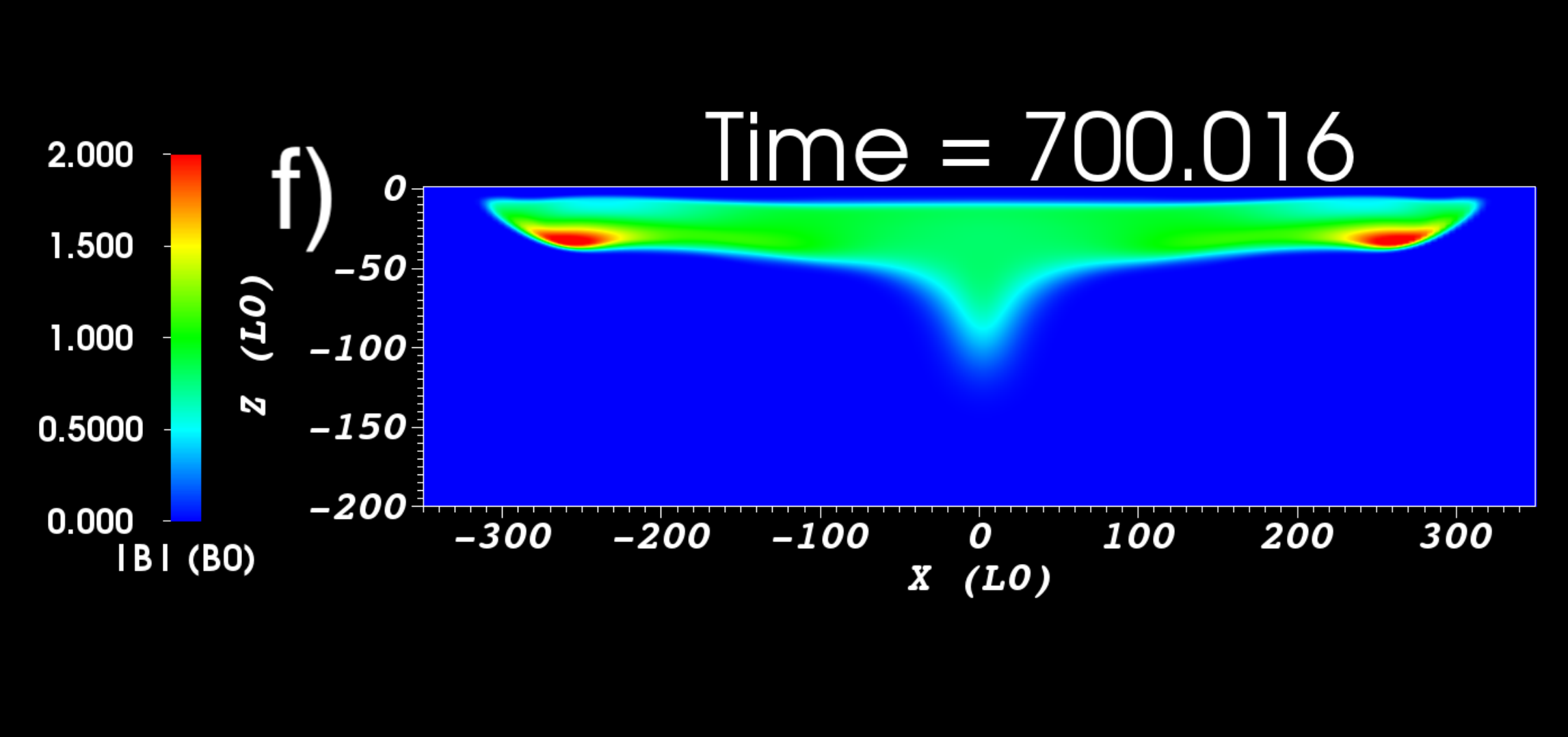}
\caption{Contours of $|B|$ at different times during the 2.5D simulation in the $x-z$ plane. \label{fig:bmag2D}}
\end{figure*}

 
Our work demonstrates that, contrary to many previous works, there is no minimum twist required for the flux rope to be able to rise and emerge from the convection zone into the corona \citep{Parker79,Schuessler79,Longcope96,MI96,Emonet98,Fan98a,Wissink00,Murray06,Toriumi11}. Our results beg the question of why the untwisted flux rope presented here was able to both rise and emerge coherently while many other twisted flux ropes, for example, the weakly twisted flux ropes of \citet{Murray06} or \citet{Toriumi11} were not able to rise and emerge. One possible answer for why this flux rope was able to rise coherently is that the toroidal shape of our flux rope plays an important role. \citet{Abbett00} argued that in three dimensions, the viscous forces that destroy the flux rope do so only over a segment of the flux rope, and that if this segment is sufficiently short (for example, if the flux rope were sufficiently curved, as in a torus), then an untwisted flux rope would not be fragmented and could rise coherently. To test this hypothesis, we performed a 2.5D simulation of an untwisted cylindrical flux rope cross section with the same simulation parameters as described above, except that the tube is a cylinder rather than a torus, and the variables along the ignorable direction ($y$, corresponding to the axial direction of the cylinder) were considered translationally invariant. Contours of $|B|$ are shown in \autoref{fig:bmag2D}. Evidently, this flux rope rises a short distance and then breaks up into separate segments on either side of $x=0$, and when it reaches the photosphere it spreads out, rather than emerging. Comparing Figures \ref{fig:bmagxz0p0} and \ref{fig:bmag2D}, it is clear that these are two fundamentally different behaviors, and clearly the curved three-dimensional structure of the toroidal flux rope plays an important role in the coherent rise of that flux rope, \textcolor{black}{either because viscous forces on a torus are important only over a very short segment, or because plasma is able effectively to drain down the legs of the rising torus}. This lends credence to the hypothesis that the toroidal shape of the flux rope is crucial to a coherent rise and emergence. \textcolor{black}{Finally, the coherence of the rising torus may also be facilitated by the line tied boundary conditions, which prevent flux from diffusing or spreading out near the bottom of the torus.}

\par 
The emergence process that we observed is in line with the findings of other studies which simulated the rise of weakly twisted flux ropes. \citet{Archontis13} demonstrated that a weakly twisted flux rope, rather than forming a single concave down overarching quasi-potential loop in the corona, formed instead two concave down loops in the corona, which then reconnected. At the photosphere, they showed that, rather than forming a pair of opposite polarity regions - as is typical for moderately twisted flux ropes - the flux rope formed two extended opposite polarities between the two primary polarities. Our toroidal simulations corroborate this result, with the additional finding that there is a significant amount of salt-and-pepper structure dispersed throughout the active region. In other words, there are many magnetic field undulations, as opposed to just the two seen in \citet{Archontis13}. \par 
A key question from this study is: why do some flux ropes emerge in convectively stable simulations with the salt-and-pepper structure indicative of the undular instability \citep[][this work]{Archontis13}, whereas other flux ropes emerge with little-to-no evidence of such structure \citep[e.g.,][]{MacTaggart09,Toriumi11,Leake13,Toriumi17b,Knizhnik18b}? 
We suggest that there are several factors that influence the presence of salt-and-pepper features at the photospheric level. One required property of flux emergence simulations to see these features is that the width of the active region should be at least several wavelengths of the undular mode. A fastest growing wavenumber of $\approx 0.3\;L_0^{-1}$, corresponds to a wavelength of $\approx 21\;L_0$, so that the resulting active region needs to be several times this size. The simulations of \citet{Knizhnik18b}, for example, had a width of only about $50\;L_0$, so that only about two wavelengths could have been seen. At this scale, salt-and-pepper features would be difficult to identify. Similarly, the simulations of \citet{Leake13} and \citet{MacTaggart09} formed an active region of approximate width $60\;L_0$, so they would be expected to have only seen about three wavelengths, not enough to see a large number of salt-and-pepper features.\par 
This cannot be the complete answer, however, since 1) two or three wavelengths of the undular mode should still have been noticed, and 2) the simulation of \citet{Archontis13} was approximately the same size as that of \citet{Knizhnik18b}, and yet the former observed the presence of undular field lines while the latter did not. Conversely, some simulations with large active regions, such as the simulation of \citet{Toriumi17b}, with an active region of width $\approx 100\;L_0$, show no sign of undulating field lines. One difference between the studies of \citet{Archontis13} and \citet{Knizhnik18b} was that the flux rope of \citet{Archontis13} was twisted only weakly. It is possible that this prevented emergence for long enough that a significant gradient in the magnetic field was able to build up, triggering the onset of the instability. On the other hand, the simulation of \citet{Toriumi17b} with no undular signatures had a smaller twist than that used by \citet{Archontis13}, so this cannot be the complete picture either. More investigation is needed into why some flux emergence simulations produce salt-and-pepper structure resulting from undulating field lines while other simulations do not.\par 
%Why our flux rope was able to emerge, while similar flux ropes rose to the surface but then did not emerge \citep[e.g.,][]{Murray06,Toriumi11} is unclear. 
Another important factor to consider in determining whether emergence occurs is the time required for a significant amplitude of the instability to set in. Since the instability criterion in \autoref{fullinstabilitycriterion} depends on $\beta$ and the vertical magnetic field gradient, it's clear that strong gradients in the magnetic field and small values of $\beta$ (corresponding to strong magnetic fields) are crucial to the emergence of flux ropes through the photosphere. In the case of \citet{Murray06} and \citet{Toriumi11}, it is possible that the simulation was not run out for long enough to allow a sufficient amount of magnetic field to pile up below the photosphere to generate a strong vertical gradient in $\vecB$, or to lower the $\beta$ sufficiently. This evidence is supported by the results of \citet{Archontis13}, who were able to obtain an emerged flux rope by extending the simulation duration. For a growth rate of order $0.2\;\mathrm{t_0^{-1}}$ (cf. \autoref{fig:growth}), a significant amplitude of the instability can be obtained within a time period of order $20-50\;\mathrm{t_0}$. While, this may explain the lack of emergence in \citet{Murray06}, \citet{Toriumi11} allowed the simulation to run for several hundred $\mathrm{t_0}$ after the flux rope neared the photosphere, so more work is required to understand why their flux rope did not emerge. The magnitude of the vertical gradient is influenced by the scale of the original flux rope as well as its initial field strength, and in turn how both of these quantities change when the flux rope reaches the photosphere. It is possible that one or both of these quantities were not sufficiently large for the instability criterion to be exceeded. \par
To summarize, we hypothesize that there are several key factors needed in order for a convection free simulation to produce the salt-and-pepper structure seen here. 1) Continual piling up of magnetic flux near the photosphere, enabling the upward magnetic pressure gradient to grow until it is large enough to lift the heavy photospheric plasma. This is a general criterion for flux emergence, and without a strong magnetic gradient no flux will appear at the photosphere. 2) The flux rope must be untwisted or twisted only weakly. Strong twist allows the flux rope to maintain its circular cross section, and the downward concavity of the rising field lines will facilitate emergence through the photosphere. On the other hand, weak twist allows the flux rope to expand and become more horizontal. 3) The active region produced by the emerged flux rope must be sufficiently large in order to observe multiple wavelengths of the undular instability.
Finally, the role of convection in flux emergence is also important to understand. Previous studies have argued that weakly twisted flux ropes, in particular, require convective upflows to emerge, since these counteract the deceleration of the flux ropes caused by their fragmentation \citep{Abbett07,Cheung07,Isobe08,Dacie17}. However, our results demonstrate that convection is not necessary for the emergence of weakly twisted flux ropes. On the other hand, convection could also act to break up a weakly twisted flux rope and produce more small scale flux, or, instead, could act to concentrate flux bundles together, which would bring the kernel density estimations at large field strengths, shown in \autoref{fig:Dacie}, more in line with observations. Many previous global-scale convective flux emergence simulations employ the anelastic approximation, which becomes invalid $20-30\;\mathrm{Mm}$ below the photosphere. As shown above, this is an important region where flux piles up, and thus parameter surveys of twist \citep[e.g., ][]{Fan08,Jouve13} cannot answer definitively whether weakly twisted flux ropes will emerge. Further work is required to determine the role that convection plays in flux emergence.

\acknowledgments{
K.J.K.\ was supported for this work by the Office of Naval Research through the National Research Council, the Jerome and Isabella Karle Distinguished Scholar Fellowship and NASA's Heliophysics Supporting Research Program. J.E.L.\ was supported by NASA LWS and ISFM programs. M.G.L.\ was supported for this work by the Office of Naval Research 6.1 program and by the NASA Living with a Star Program and Heliophysics Supporting Research programs. K.J.K.\ would like to acknowledge helpful discussions with A. Hillier. The numerical simulations were performed under a grant of computer time at the Department of Defense High Performance Computing Program. The authors are grateful to the anonymous referee who greatly improved the quality and scope of this work.}

%\printbibliography
\bibliography{bibliography}
\end{document}